\documentclass[11pt,a4paper]{article}                           

\usepackage[bookmarksopen, bookmarksnumbered, bookmarksopenlevel=2]{hyperref}
\usepackage{tikz}
\usepackage{tikz-3dplot}
\usetikzlibrary{calc}
\usetikzlibrary{decorations} %
\usepackage[UKenglish]{babel}
\usepackage[toc,page]{appendix}
\usepackage{amsmath}
\usepackage{amssymb}
\usepackage{graphicx}
\usepackage{hhline}
\usepackage[bf]{caption}
\usepackage{cite}
\usepackage[vcentermath]{youngtab}
\usepackage{geometry}
\DeclareSymbolFont{bbold}{U}{bbold}{m}{n}
\DeclareSymbolFontAlphabet{\mathbbold}{bbold}
\newcommand{\one}{\mathbbold{1}}
 
\hypersetup{
    pdftitle={Mordell-Weil Torsion and the Global Structure of Gauge Groups in F-theory},
    pdfauthor={Christoph Mayrhofer, Dave Morrison, Oskar Till, Timo Weigand},
    pdfsubject={F-theory compactifications}
}
\numberwithin{equation}{section}
\numberwithin{table}{section}

\newcommand{\bea}{\begin{eqnarray}}
\newcommand{\eea}{\end{eqnarray}}

\newcommand{\executeiffilenewer}[3]{%
 \ifnum\pdfstrcmp{\pdffilemoddate{#1}}%
 {\pdffilemoddate{#2}}>0%
 {\immediate\write18{#3}}\fi%
}

\begin{document}

\baselineskip=14pt
\parskip 5pt plus 1pt

\vspace*{-1.5cm}
\begin{flushright}    % Publication numbers
  {\small

  }
\end{flushright}

\vspace{2cm}
\begin{center}        % Main title
  {\LARGE Mordell-Weil Torsion and \\ \vspace{3mm} the Global Structure of Gauge Groups in F-theory}
\end{center}

\vspace{0.75cm}
\begin{center}        % Authors
Christoph Mayrhofer$^1$, David R.~Morrison$^2$, Oskar Till$^1$ and Timo Weigand$^1$
\end{center}

\vspace{0.15cm}
\begin{center}        % Institutes
\emph{ $^1$Institut f\"ur Theoretische Physik, Ruprecht-Karls-Universit\"at, \\
             Philosophenweg 19, 69120,
             Heidelberg, Germany\\
             $^2$Departments of Mathematics and Physics,\\
University of California Santa Barbara, Santa Barbara, USA}
\end{center}

\vspace{2cm}

%%%%%%%%%%%%%%%%%%%%%%%%%%%%%%%%%%%%%%%%%%%%%%%
%%%%%%%%%%%%%%%%%%%%%%%%%%%%%%%%%%%%%%%%%%%%%%%
%%%%%%%%%%%%%%%%%%%%%%%%%%%%%%%%%%%%%%%%%%%%%%%
%%%%%%%%%%%%%%%%%%%%%%%%%%%%%%%%%%%%%%%%%%%%%%%
%%%%%%%%%%%%%%%%%%%%%%%%%%%%%%%%%%%%%%%%%%%%%%%
%%%%%%%%%%%%%%%%%%%%%%%%%%%%%%%%%%%%%%%%%%%%%%%
%%%%%%%%%%%%%%%%%%%%%%%%%%%%%%%%%%%%%%%%%%%%%%%
%%%%%%%%%%%%%%%%%%%%%%%%%%%%%%%%%%%%%%%%%%%%%%%

\begin{abstract}
\noindent We study the global structure of the gauge group $G$ of F-theory compactified on an elliptic fibration $Y$. The global properties of $G$ are encoded in the torsion subgroup of the Mordell-Weil group of rational sections of $Y$.
Generalising the Shioda map to torsional sections we construct a specific integer divisor class on $Y$ as a fractional linear combination of the resolution divisors associated with the Cartan subalgebra of $G$. This divisor class can be interpreted as an element of the refined coweight lattice of the gauge group.
As a result, the spectrum of admissible matter representations is  strongly constrained and the gauge group is non-simply connected.
We exemplify our results by a detailed analysis of the general elliptic fibration with Mordell-Weil group $\mathbb Z_2$ and $\mathbb Z_3$ as well as a further specialization to $\mathbb Z \oplus \mathbb Z_2$. 
Our analysis exploits the representation of these fibrations as hypersurfaces in toric geometry.

\end{abstract}

\thispagestyle{empty}
\clearpage
\setcounter{page}{1}

%%%%%%%%%%%%%%%%%%%%%%%%%%%%%%%%%%%%%%%%%%%%%%%
%%%%%%%%%%%%%%%%%%%%%%%%%%%%%%%%%%%%%%%%%%%%%%%
%%%%%%%%%%%                 %%%%%%%%%%%%%%%%%%%
%%%%%%%%%%%  DOCUMENT BODY  %%%%%%%%%%%%%%%%%%%
%%%%%%%%%%%                 %%%%%%%%%%%%%%%%%%%
%%%%%%%%%%%%%%%%%%%%%%%%%%%%%%%%%%%%%%%%%%%%%%%
%%%%%%%%%%%%%%%%%%%%%%%%%%%%%%%%%%%%%%%%%%%%%%%
%%%%%%%%%%%%%%%%%%%%%%%%%%%%%%%%%%%%%%%%%%%%%%%

\newpage

\tableofcontents
%%%%%%%%%%%%%%%%%%%%%%%%%%%%%%%%%%%%%%%%%%%%%%%%%%%%%%%%%%%%%%%%%%%%%%%%%%%%%%%%%%%%%%%%%%%%%%%%%%%%%%%%%%%%%%%%%%%%%%%%%%%%%%%%%%%%%%%%%%%%%%%
\section{Introduction}
%%%%%%%%%%%%%%%%%%%%%%%%%%%%%%%%%%%%%%%%%%%%%%%%%%%%%%%%%%%%%%%%%%%%%%%%%%%%%%%%%%%%%%%%%%%%%%%%%%%%%%%%%%%%%%%%%%%%%%%%%%%%%%%%%%%%%%%%%%%%%%%

The Mordell-Weil group of rational sections of an elliptic fibration has attracted a great deal of interest in the recent F-theory literature. 
The free part of the Mordell-Weil group encodes information about the 
abelian sector of an F-theory model \cite{oai:arXiv.org:hep-th/9602114,oai:arXiv.org:hep-th/9603161,Klemm:1996hh}. Explicit realizations of $U(1)$ gauge groups via such rational sections in F-theory have been studied in detail \cite{Grimm:2010ez,Braun:2011zm,Krause:2011xj,Grimm:2011fx,oai:arXiv.org:1202.3138,Morrison:2012ei,oai:arXiv.org:1210.6034,Mayrhofer:2012zy,Braun:2013yti,Borchmann:2013jwa,Cvetic:2013nia,Braun:2013nqa,Cvetic:2013uta,Borchmann:2013hta,Cvetic:2013jta,Cvetic:2013qsa,Morrison:2014era,Martini:2014iza,Bizet:2014uua}.\footnote{This has to a large extent been motivated from model building and the need for abelian selection rules in GUT models, see \textit{e.g.}\ \cite{Marsano:2009wr,Hayashi:2010zp,Dolan:2011iu,Dolan:2011aq,Marsano:2011nn,Maharana:2012tu} for a rather incomplete list of references.} In this paper we investigate the role of the torsion part of the Mordell-Weil group and its relation to global properties of the non-abelian gauge sector of the F-theory vacuum. \\ \\
Non-abelian gauge symmetries in F-theory have their origin in the codimension-one singularity structure. The by now algorithmic procedure to engineer gauge theories takes the Kodaira classification of singular fibers as the starting point. However, the resolution of codimension-one singularities provides only information on the gauge algebra, and not on the gauge group. The Lie group whose Lie algebra is given by the geometric data might be simply connected, making the lift of the Lie algebra and its representations trivial. If it is non-simply connected only a subset of the matter representations will be present in the gauge theory. The difference between such theories is in particular measured by non-local operators, see \textit{e.g.}~\cite{Aharony:2013hda}. For the example of the Standard Model, the gauge group is presumably not the simply connected $SU(3)_c \times SU(2)_W \times U(1)_Y$ but really $(SU(3)_c \times SU(2)_W \times U(1)_Y)/\mathbb Z_6$, where $\mathbb Z_6$ is a subgroup of the center $\mathbb Z_3 \oplus \mathbb Z_2 \oplus U(1)$, and only matter multiplets invariant under the action of $\mathbb Z_6$ are present \cite{Mccabe:2007}. Indeed, embedding the Standard Model into $SU(5)$ amounts to choosing a block diagonal decomposition $S(U(2) \times U(3)) \subset SU(5)$ such that its determinant is unity, and $S(U(2)\times U(3))$ is isomorphic to $(SU(3)_c \times SU(2)_W \times U(1)_Y)/\mathbb Z_6$, see \textit{e.g.}\ \cite{Baez:2009dj}. \\ \\
The global properties of a gauge group are related to torsion elements of the Mordell-Weil group.
The study of torsional sections in F-theory fibrations (\textit{i.e.}\ sections of the fibration which induce torsion elements in the Mordell-Weil group) was initiated in \cite{Aspinwall:1998xj}. By utilizing the duality between F-theory and heterotic theory in eight dimensions it was shown that the fundamental group of the gauge group is isomorphic to the torsion subgroup of the Mordell-Weil group, and it was conjectured that the same result holds for six-dimensional compactifications. The general framework relating the Mordell-Weil group of the fibration to the gauge group of F-theory was laid out in \cite{LieF}.  The Mordell-Weil group has also been studied via string junctions and configurations of $(p,q)$-branes \cite{Fukae:1999zs}. This approach was eight-dimensional and reproduces the classification of Mordell-Weil lattices for elliptic surfaces \cite{Oguiso:1991,MR1813537}. Subsequent work addressed the same problem for elliptic threefolds \cite{Guralnik:2001jh}.  \\ \\ %We will not comment further on this approach in this note. \\ 
In this article, we show how the global structure of the gauge theory manifests itself for F-theory in any dimension. Rather than relying on heterotic duality or the physics of string junctions, we directly study the effect of a torsional Mordell-Weil
subgroup on the physics of the F-theory compactification with geometric means. 
Our starting point is a generalization of the Shioda map \cite{Shioda:1989,Wazir:2001,Park:2011ji}  to torsional sections. Unlike for non-torsional sections, this map defines a trivial divisor class on the elliptic variety. We use this class to construct an element in the coweight lattice which takes integer values on any charged matter representation that can occur in the compactification. The coweight in question is associated with a fractional linear combination of the resolution divisors which correspond to the Cartan generators of the gauge algebra.
The requirement that this fractional linear combination must have integer pairing with the matter representations strongly constrains the set of admissible representations. As a result, the center of the gauge group is smaller compared to naive expectations and the gauge group acquires a non-trivial first fundamental group. % These are precisely the representations invariant under the center of the gauge group. 
The divisor associated with the coweight is a torsional element of $H^{1,1}(\hat Y,\mathbb Z)$ modulo the resolution divisors associated with the gauge algebra realized on the elliptic fibration $\hat Y$. This clarifies the relation between torsion in the Mordell-Weil group and torsion in the cohomology group of the elliptic fibration.\\ \\
To exemplify this general structure we explicitly analyze F-theory compactifications on elliptic fibrations whose fiber can be realized as a hypersurface in a toric ambient space. 
Out of the 16 possible toric realizations of such elliptic fibrations, three are known to have torsional Mordell-Weil group $\mathbb Z_2$, $\mathbb Z_3$ and $\mathbb Z \oplus \mathbb Z_2$ \cite{Braun:2013nqa}. We will show that the first two correspond, in fact, to the most general elliptic fibrations with Mordell-Weil group $\mathbb Z_2$ and $\mathbb Z_3$ in the list presented in \cite{Aspinwall:1998xj}, while the $\mathbb Z \oplus \mathbb Z_2$ model is a restriction of the $\mathbb Z_2$ fibration. Certain blow-downs of these fibrations have also been considered previously in \cite{Berglund:1998va} as examples of elliptic fibrations with restricted $SL(2,\mathbb Z)$-monodromy.
The fibrations we consider allow for a representation as a global Tate model and can be obtained as a special case of the $U(1)$ restricted Tate model \cite{Grimm:2010ez}.
The restriction of the complex structure of the fibration necessary to implement torsion in the  Mordell-Weil group automatically induces non-abelian singularities in codimension one, which we resolve and study in detail. 
The associated gauge group factor can be viewed as the non-abelian enhancement of the $U(1)$ gauge group in the underlying $U(1)$ restricted Tate model, to which the geometries are consequently related by a chain of (un)Higgsings.
Furthermore, we exemplify the construction of extra non-abelian gauge group factors via toric tops \cite{Candelas:1997pv,Bouchard:2003bu}. The possible extra gauge group factors follow a specific pattern dictated by the torsional sections. As predicted by our general analysis of Mordell-Weil torsion, only a subset of typically realized matter representations is present in the geometry. \\\\
In section \ref{sec_2} we begin with a brief review of the Mordell-Weil group with special emphasis on its torsion subgroup. In section \ref{sec_3} we outline the general picture of our geometric construction of the coweight lattice and elucidate the relation between the Mordell-Weil group and global properties of the gauge group in F-theory. Our exemplification of these general results for elliptic fibrations with Mordell-Weil torsion $\mathbb Z_2$, $\mathbb Z \oplus \mathbb Z_2$ and $\mathbb Z_3$ follows in sections \ref{sec_4}, \ref{sec_5} and \ref{sec_6}, respectively. Some computational details are relegated to the appendix.

%%%%%%%%%%%%%%%%%%%%%%%%%%%%%%%%%%%%%%%%%%%%%%%%%%%%%%%%%%%%%%%%%%%%%%%%%%%%%%%%%%%%%%%%%%%%%%%%%%%%%%%%%%%%%%%%%%%%%%%%%%%%%%%%%%%%%%%%%%%%%%%
\section{The arithmetic of elliptic fibrations} \label{sec_2}
%%%%%%%%%%%%%%%%%%%%%%%%%%%%%%%%%%%%%%%%%%%%%%%%%%%%%%%%%%%%%%%%%%%%%%%%%%%%%%%%%%%%%%%%%%%%%%%%%%%%%%%%%%%%%%%%%%%%%%%%%%%%%%%%%%%%%%%%%%%%%%%
In this section, we give a brief review of the Mordell-Weil group of a family of elliptic curves. We describe how meromorphic sections naturally come with a group structure and comment in particular on the finite part of this group, the part associated to ``torsional sections.'' This is a classic topic in mathematics and for more extensive treatments see \textit{e.g.}~\cite{Silverman:2008,MR715605}. 
%%%%%%%%%%%%%%%%%%%%%%%%%%%%%%%%%%%%%%%%%%%%%%%%%%%%%%%%%%%%%%%%%%%%%%%%%%%%%%%%%%%%%%%%%%%%%%%%%%%%%%%%%%%%%%%%%%%%%%%%%%%%%%%%%%%%%%%%%%%%%%%
\subsection{The Mordell-Weil group}
%%%%%%%%%%%%%%%%%%%%%%%%%%%%%%%%%%%%%%%%%%%%%%%%%%%%%%%%%%%%%%%%%%%%%%%%%%%%%%%%%%%%%%%%%%%%%%%%%%%%%%%%%%%%%%%%%%%%%%%%%%%%%%%%%%%%%%%%%%%%%%%
An elliptic curve $E$ is a smooth complex curve of genus one with a marked point. Such a curve may be given in Weierstrass form
\begin{equation}\label{eq:weierstrass_equation}
y^2 = x^3 + f x z^4 + g z^6
\end{equation}
with coordinates $[x:y:z] \in \mathbb P^2_{2,3,1}$ and  $f$, $g$ valued in some field $K$. For fixed values of $f$ and $g$ this genus one curve is the flat torus given by the quotient
\begin{equation}
E = \frac{\mathbb C}{\Lambda}
\end{equation}
of the complex plane $\mathbb C$ by $\Lambda = \langle 1, \tau \rangle$, \textit{i.e.}\ the lattice generated by $1$ and $\tau$. These two descriptions are equivalent and for $z \in \mathbb C$ the isomorphism is given by\footnote{Actually, this parametrisation corresponds to the Weierstrass equation $y^2 = 4x^3 + fx + g$, but this difference will not be of interest for our purposes.}
\begin{equation}\label{eq:weierstrass_isomorphism}
z \mapsto [\wp(z): \wp'(z):1]\, ,
\end{equation}
where $\wp$ is the doubly periodic Weierstrass function. The complex structure parameter $\tau$ is related to the Weierstrass equation via the modular $j$-function
\begin{equation}
j(\tau) \sim \frac{f^3}{4f^3 + 27 g^2}.
\end{equation}
Because of the isomorphism \eqref{eq:weierstrass_isomorphism} the addition of complex numbers in $\mathbb C/\Lambda$ induces an addition of points on the curve \eqref{eq:weierstrass_equation}. The set of rational points on $E$, \textit{i.e.}\ points given by rational expressions in the field $K$, is closed under this addition and thus forms an abelian group. This group is often denoted by $E(K)$ and the abelian structure makes elliptic curves examples of abelian varieties. The original Mordell-Weil theorem states that this group is finitely generated when $K$ is a ``number field'', \textit{i.e.}\ a finite extension of the rational numbers.  In this case,
\begin{equation}
E(K) = \mathbb Z^r \oplus \mathbb Z_{k_1} \oplus \dots \oplus \mathbb Z_{k_n} \, .
\end{equation}
The rank $r$ of this group is the number of generators of the free subgroup and the finite part is called the torsion subgroup $E(K)_\textmd{tors}$.  A theorem by Mazur states that for a curve over the rationals, the torsion subgroup $E(\mathbb Q)_\textmd{tors}$ is either $\mathbb Z_k$ for $k = 1, \dots, 10,12$ or $\mathbb Z_2 \oplus \mathbb Z_k$ for $k = 2,4,6,8$. The converse statement also holds, \textit{i.e.}\ all possibilities are realised.

%%%%%%%%%%%%%%%%%%%%%%%%%%%%%%%%%%%%%%%%%%%%%%%%%%%%%%%%%%%%%%%%%%%%%%%%%%%%%%%%%%%%%%%%%%%%%%%%%%%%%%%%%%%%%%%%%%%%%%%%%%%%%%%%%%%%%%%%%%%%%%%
\subsection{Elliptic fibrations with torsion Mordell-Weil group}\label{section:fibrations_with_torsion}
%%%%%%%%%%%%%%%%%%%%%%%%%%%%%%%%%%%%%%%%%%%%%%%%%%%%%%%%%%%%%%%%%%%%%%%%%%%%%%%%%%%%%%%%%%%%%%%%%%%%%%%%%%%%%%%%%%%%%%%%%%%%%%%%%%%%%%%%%%%%%%%
The notion of the Mordell-Weil group also applies to families of elliptic curves, \textit{i.e.}\ fibrations
\begin{equation}
\pi: Y \rightarrow \mathcal{B}
\end{equation}
with a distinguished zero-section $\sigma_0$ such that the fiber $\pi^{-1}(b)$ for a generic point $b \in \mathcal{B}$ is an elliptic curve. We can regard the coefficients of the Weierstrass equation \eqref{eq:weierstrass_equation} as taking  values in the field $K$ of meromorphic functions on the base $B$.
Each meromorphic section of the fibration determines an element of $E(K)$, because it detemines $x=x(b)$ and $y=y(b)$, the ``coordinates'' of the point, as elements of the field $K$ of meromorphic functions.  The zero-section $\sigma_0$ maps to the identity element in the group $E(K)$, and the group structure
is given by fiberwise addition of points.\\\\
The ``Mordell-Weil theorem for function fields'' (proved by Lang and N\'eron \cite{MR0102520}) says that in this situation, $E(K)$ is also finitely generated
unless the fibration is ``split'', \textit{i.e.}\ unless $Y$ is birational to
a product $E\times \mathcal{B}$. Note that the zero-section does not serve as one of the generators of the group.  In particular,
the Mordell-Weil group is trivial when  the zero section is the only section of the fibration, and extra rational sections are needed to have a non-trivial group.  
For certain elliptic surfaces the possible groups $E(K)$ have been classified analogously to the Mazur theorem for elliptic curves. For instance, for a rational elliptic surface the non-trivial possibilities for the Mordell-Weil group are
\begin{equation}
\begin{aligned}
&\mathbb Z^r\, \,(1 \leq r \leq 8), & \qquad & \mathbb Z^r \oplus \mathbb Z_2\,\, (1\leq r \leq 4),&\qquad & \mathbb Z^r \oplus \mathbb Z_3\,\, (1\leq r \leq 2), \\ 
&\mathbb Z^r \oplus \mathbb Z_2 \oplus \mathbb Z_2\,\, (1\leq r \leq 2),&\qquad &\mathbb Z \oplus \mathbb Z_4, &\qquad& \mathbb Z_2 \oplus \mathbb Z_4, \\
&\mathbb Z_2 \oplus \mathbb Z_2, &\qquad & \mathbb Z_3 \oplus \mathbb Z_3, &\qquad& \mathbb Z_k \, \,(2 \leq k \leq 6)
\end{aligned}
\end{equation}
and in particular the Mordell-Weil group for any rational elliptic surface is torsion-free if its rank is greater than 4 \cite{Oguiso:1991}. For elliptic K3 surfaces the list is more complicated, but completely known \cite{MR1813537}; in particular, the possibilites for non-trivial torsion in the Mordell-Weil group are
\begin{equation}
\begin{aligned}
& \mathbb Z_k \, \,(2 \leq k \leq 8), & \qquad
&\mathbb Z_{2} \oplus \mathbb Z_{2k} \,\, (1\leq k \leq 3),
&\qquad
& \mathbb Z_3 \oplus \mathbb Z_3, &\qquad
& \mathbb Z_4 \oplus \mathbb Z_4.
\end{aligned}
\end{equation}
 The general situation for higher-dimensional fibrations, \textit{e.g.}\  three- and fourfolds, is not as well understood and classifications only exist in special cases such as \cite{Kloosterman:2011}. \\\\
A useful tool to study in particular higher-dimensional examples of elliptic fibrations is toric geometry.
In toric geometry an elliptic curve may be realized as a hypersurface or a complete intersection in a toric ambient space. The possible realizations as hypersurfaces are classified by the 16 reflexive polygons in two dimensions. The associated toric ambient spaces are $\mathbb P^2_{1,1,2}$, $\mathbb P^1 \times \mathbb P^1$, $\mathbb P^2$ or blow-ups thereof. Three of these polygons admit torsional sections given as the intersection of an ambient toric divisor with the elliptic curve. According to the enumeration of polygons in \cite{Bouchard:2003bu}, the elliptic curves in the ambient spaces defined by polygon 13, 15 and 16 have  toric Mordell-Weil groups $\mathbb Z_2$, $\mathbb Z \oplus\mathbb Z_2$ and $\mathbb Z_3$, respectively  \cite{Braun:2013nqa} (see also \cite{GrassiPerduca}). These cases will be studied in detail in this paper including  the toric implementation of further non-abelian gauge groups via tops. \\\\
An important ingredient in our analysis is the correspondence between rational sections and certain divisor classes on the fibration, more precisely elements of the N\'eron-Severi group of divisors modulo algebraic equivalence.
Note that the N\'eron-Severi group coincides with the Picard group of divisors modulo linear equivalence for spaces with vanishing first cohomology group, which is the situation of relevance throughout this paper.\footnote{For this reason, we will systematically restrict our notation to refer to the N\'eron-Severi group rather than the Picard group.}
Let $E$ be a general fiber of $\pi$.  Each divisor $D$ on $Y$ can be restricted
to a divisor $D|_E$ on $E$ which has a specific degree  $D\cdot E$.  For example,
sections restrict to divisors of degree $1$.  Now for an arbitrary divisor $D$,
the linear combination $D-(D\cdot E)\sigma_0$ restricts to a divisor of
degree $0$ on $E$.  But the set of divisors of degree $0$ on $E$ is just
$E$ itself.\\\\
In this way, we get a surjective homomorphism of groups
\begin{equation} \label{eq:psi}
\psi:NS(Y) \to E(K)
\end{equation}
which sends $[D]$ to the $K$-valued point of $E$ determined by restricting
the divisor $D-(D\cdot E)\sigma_0$ to $E$.  (It is surjective because
every element of $E(K)$ arises from a rational section $\sigma$.)
The kernel of this homomorphism is generated by the zero section 
and by divisors whose restriction to the general fiber $E$ is trivial.\\\\
Recall that the elliptic fiber degenerates when the discriminant $\Delta = 4 f^3 + 27 g^2$ vanishes. The singularities, if any, in the total space of $Y$ can always be resolved,
and $\cal{B}$ can be further blown up if, necessary, to ensure a birational
model $\hat \pi:\hat Y\to \hat{\cal B}$ of our fibration with a nonsingular
total space $\hat Y$ and a flat fibration, \textit{i.e.}\ a fibration in which all
of the fibers are one-dimensional.  In the sequel we assume that our original
base $\cal{B}$ allows for a resolution $\hat Y$ which is nonsingular
and has a flat family.
The resolution process introduces a set $F_i$ of resolution divisors which are $\mathbb P^1$-fibrations over the codimension-one locus in the base ${\cal B}$ over which the singularity was located. 
Let $\cal{T}$ denote the subgroup of $NS(\hat Y)$ generated by the zero-section
$[\sigma_0]$, the resolution divisors $F_i$, and divisors of the form
$\pi^{-1}(\delta)$ for $\delta \in NS({\cal B})$.
The
Shioda-Tate-Wazir theorem \cite{Shioda:1972, Wazir:2001} asserts that
the kernel of the map $\psi$ in \eqref{eq:psi} is $\cal{T}$.  In particular,
\begin{equation} \label{STW}
\text{rank } NS(\hat Y) = 1+ \text{rank } NS(\mathcal B)+ \text{rank } E(K)  + \sum_{w\in \Delta} (n_w-1),
\end{equation}
where $n_w$ is the number of irreducible components of the resolved fiber over the codimension-one loci $w \in \Delta \subset \mathcal B$ over which the fiber degenerates. \\\\
The divisors on $\hat Y$ are thus generated
%, as a vector space over $\mathbb{Q}$, 
by the class of the zero section $Z = [\sigma_0]$, the pullback of divisors in $\mathcal B$, the divisor classes $S_i - Z = [\sigma_{i}] - [\sigma_0]$ from the free generators of $E(K)$ and the irreducible fiber resolution divisors $F_i$.
On the other hand the divisor class  $R-Z = [\sigma_{r}] - [\sigma_0] \in NS(\hat Y)$ associated with a torsional section $\sigma_r$ has the property
that $k(R-Z)$ can be expressed in terms of the generators of $\cal{T}$, where
$k$ is the order of the torsional element of the Mordell-Weil group.
It follows that $R-Z$ can be expressed in terms of these generators using
$\mathbb{Q}$-coefficients.
As described in the next section, this expression for $R-Z$ is closely related to the so-called Shioda map  \cite{Shioda:1989}, \cite{Wazir:2001,Park:2011ji}. This is in line with the result for elliptic surfaces in \cite{Cox:1979}, where a trivial class on the hypersurface is obtained by adding a certain rational linear combination of resolution divisors to $R-Z$.

%%%%%%%%%%%%%%%%%%%%%%%%%%%%%%%%%%%%%%%%%%%%%%%%%%%%%%%%%%%%%%%%%%%%%%%%%%%%%%%%%%%%%%%%%%%%%%%%%%%%%%%%%%%%%%%%%%%%%%%%%%%%%%%%%%%%%%%%%%%%%%%
\section{F-theory fibrations with non-trivial Mordell-Weil group} \label{sec_3}
%%%%%%%%%%%%%%%%%%%%%%%%%%%%%%%%%%%%%%%%%%%%%%%%%%%%%%%%%%%%%%%%%%%%%%%%%%%%%%%%%%%%%%%%%%%%%%%%%%%%%%%%%%%%%%%%%%%%%%%%%%%%%%%%%%%%%%%%%%%%%%%
After a brief review of the physics of the free Mordell-Weil group and abelian gauge symmetries, a subject treated in great detail in the recent F-theory literature, we outline the general picture of torsional sections and the global structure of the gauge theory. \\ \\
In the sequel we denote by $G$ the non-abelian part of the gauge group of an F-theory compactification on an elliptically fibered Calabi-Yau 4-fold $Y_4$ over the base manifold $\mathcal{B}$ and denote its Cartan subgroup by $H$.
Let us assume that the singularities of $Y_4$ responsible for the appearance of a non-abelian gauge group $G$ in codimension-one admit a crepant resolution $\hat Y_4$. Expanding the M-theory 3-form $C_3$ as $C_3 = \sum_i A_i \wedge F_i$ with $F_i$ the resolution divisors gives rise to the Cartan $U(1)$ gauge fields $A_i$. Therefore the resolution divisors $F_i$ span the coroot lattice $Q^\vee$ of the Cartan subalgebra $\mathfrak{h}$.
%%%%%%%%%%%%%%%%%%%%%%%%%%%%%%%%%%%%%%%%%%%%%%%%%%%%%%%%%%%%%%%%%%%%%%%%%%%%%%%%%%%%%%%%%%%%%%%%%%%%%%%%%%%%%%%%%%%%%%%%%%%%%%%%%%%%%%%%%%%%%%%
\subsection{The free Mordell-Weil group and the Shioda map}
%%%%%%%%%%%%%%%%%%%%%%%%%%%%%%%%%%%%%%%%%%%%%%%%%%%%%%%%%%%%%%%%%%%%%%%%%%%%%%%%%%%%%%%%%%%%%%%%%%%%%%%%%%%%%%%%%%%%%%%%%%%%%%%%%%%%%%%%%%%%%%%

%
Since the group homomorphism  \eqref{eq:psi} is surjective, there is an
injective homomorphism in the other direction after tensoring with $\mathbb{Q}$.  In the case of elliptic surfaces,
Shioda \cite{Shioda:1989} introduced such a homomorphism with a specific additional property, which was extended in \cite{Wazir:2001,Park:2011ji} to
a Shioda map for elliptic fibrations of arbitrary dimension.  For an
elliptic fourfold $\hat Y_4$, the Shioda map
\begin{equation}
\varphi: E(K) \rightarrow NS(\hat Y_4)\otimes \mathbb{Q} \, 
\end{equation}
satisfies the property that $\langle \varphi(\sigma), T \rangle = 0$ for any
divisor $T\in \mathcal{T}$, where the pairing $\langle\ ,\ \rangle$ is the 
{\em height pairing}
\begin{equation}
\langle D_1, D_2 \rangle := \pi( D_1 \cap D_2),
\end{equation}
which  projects the intersection of two divisors to the base.  It is well defined modulo linear equivalence, and so defines a pairing on the N\'eron-Severi group.
% of the intersection of two divisors.
%which is a homomorphism from the Mordell-Weil group $E(K)$ to the N\'{e}ron-Severi group of divisors modulo algebraic equivalence and was constructed for elliptic surfaces in \cite{Shioda:1989}. When $\sigma \in E(K)/E(K)_\textmd{tors}$ is different from the zero section, \textit{i.e.}\ when $\text{rank }E(K) \geq 1$ this gives a non-trivial divisor class ${\cal S} = \varphi(\sigma)$. 
%By construction, its Poincar\'{e} dual harmonic two-form, which we denote by the same symbol, 
%satisfies
%\begin{equation}
%\int_{\hat Y_4} {\cal S} \wedge F_i \wedge D_a \wedge D_b = \int_{\hat Y_4} {\cal S} \wedge Z \wedge D_a \wedge D_b = \int_{\hat Y_4} {\cal S} \wedge D_a \wedge D_b \wedge D_c = 0
%\end{equation}
%for $\{D_{a,b,c}\}$ the pullback of generic divisors in $\mathcal{B}$ to the fourfold. 
For example, given any section $S$ defining an element $S-Z$ of the
Mordell-Weil group,  we have
\bea \label{Shioda1}
\varphi(S-Z) = S - Z - \pi^{-1}(\delta) + \sum l_i F_i  
\qquad 
% {\rm with} \quad \bar{\mathcal{K}} = \pi^{-1} \bar{\mathcal{K}}_\mathcal{B}, \quad 
%l_i \in \mathbb Q
\eea
for some divisor $\delta$ on ${\cal B}$ and some rational numbers $l_i \in \mathbb Q$,
which is constructed so that for every $T\in \mathcal{T}$ we have
\begin{equation} \label{pi-condition}
\pi\left(T \cap (S - Z - \pi^{-1}(\delta) + \sum l_i F_i)\right)
\end{equation}
is linearly equivalent to zero on the base $\cal{B}$.\\\\
Let us denote by $\cal S$ the harmonic 2-form representative of the cohomology class associated with $\varphi(S-Z)$.
Expanding the M-theory 3-form as $C_3 = A_{\cal S} \wedge {\cal S}$ gives $A_{\cal S}$ as a massless $U(1)$ one-form gauge field in three dimensions.\footnote{By contrast, \emph{massive} $U(1)$s in F-theory can be understood along the lines of \cite{Grimm:2010ez,Grimm:2011tb,Braun:2014nva,Douglas:2014ywa}; see also
\cite{Jockers:2005zy,Buican:2006sn} for a similar mechanism at work in a different context.} 
The details of the map assert that this generator does not lie in the Cartan of any non-abelian gauge symmetry, and that it has `one leg in the fiber', ensuring that the gauge field $A_{\cal S}$ lifts to a one-form field in four dimensions under M-/F-theory duality \cite{oai:arXiv.org:hep-th/9602114,oai:arXiv.org:hep-th/9603161}.
The geometric realisation and the physics of extra sections has been studied extensively in the recent literature \cite{Grimm:2010ez,Braun:2011zm,Krause:2011xj,Grimm:2011fx,oai:arXiv.org:1202.3138,Morrison:2012ei,oai:arXiv.org:1210.6034,Mayrhofer:2012zy,Braun:2013yti,Borchmann:2013jwa,Cvetic:2013nia,Braun:2013nqa,Cvetic:2013uta,Borchmann:2013hta,Cvetic:2013jta,Cvetic:2013qsa,Morrison:2014era,Martini:2014iza,Bizet:2014uua}.      
%This has to a large extent been motivated by the need for $U(1)$ selection rules in GUT models, forbidding dimension 4 and 5 proton decay operators (see e.g.\ \cite{Marsano:2009wr,Hayashi:2010zp,Grimm:2010ez,Dolan:2011iu,Dolan:2011aq,Marsano:2011nn,Maharana:2012tu} and references therein). %Likewise doublet-triplet splitting of the Higgs $\mathbf{5_H}$ in the SU(5) GUT is realised by different $U(1)$ charges of the Higgs multiplets \cite{Marsano:2009wr,}. 
%%%%%%%%%%%%%%%%%%%%%%%%%%%%%%%%%%%%%%%%%%%%%%%%%%%%%%%%%%%%%%%%%%%%%%%%%%%%%%%%%%%%%%%%%%%%%%%%%%%%%%%%%%%%%%%%%%%%%%%%%%%%%%%%%%%%%%%%%%%%%%%
\subsection{Torsional sections and divisor classes} \label{tors-sec-gen}
%%%%%%%%%%%%%%%%%%%%%%%%%%%%%%%%%%%%%%%%%%%%%%%%%%%%%%%%%%%%%%%%%%%%%%%%%%%%%%%%%%%%%%%%%%%%%%%%%%%%%%%%%%%%%%%%%%%%%%%%%%%%%%%%%%%%%%%%%%%%%%%

Let us now consider the divisor class $R$ of a torsional meromorphic section of order $k$ such that $R-Z$ is a generator of the torsional part of the Mordell-Weil group of $\hat Y_4$.
Combining the  theory outlined in section \ref{section:fibrations_with_torsion}   with the properties of the Shioda map one can conclude that there 
 exists now a fractional linear combination 
of resolution divisors $F_i$ such that 
\begin{equation} \label{Sigmadef}
 \Sigma := R - Z - 
%\bar{\mathcal{K}} 
\pi^{-1}(\delta)
+ \frac{1}{k}\sum a_i F_i     \quad  \quad  {\rm with} 
%\quad \bar{\mathcal{K}} = \pi^{-1} \bar{\mathcal{K}}_\mathcal{B}, 
\quad a_i \in  \mathbb Z
\end{equation}
is trivial in $NS(\hat Y_4) \otimes \mathbb Q$ and thus in particular in $H^{2}(\hat Y_4, \mathbb R)$. Indeed, as described in  section \ref{section:fibrations_with_torsion}, it is guaranteed that $R-Z$ can be expressed as a linear combination with $\mathbb Q$ coefficients of the generators of ${\cal T}$, the subgroup of $NS(\hat Y_4)$ generated by $[\sigma_0]$, the resolution divisors $F_i$ and $\pi^{-1}(\delta)$ for some divisor class $\delta$ on ${\cal B}$.
Thus, $R-Z$ minus this linear combination is trivial in $NS(\hat Y_4) \otimes {\mathbb Q}$. 
On the other hand, the Shioda map gives a specific such linear combination of the form (\ref{Shioda1}) as
\bea
\varphi(R-Z) = R - Z - \pi^{-1}(\delta) + \sum_i l_i F_i.
\eea
The rational numbers $l_i$ are in fact of the form $\frac{a_i}{k}$ with $a_i \in \mathbb Z$.
Since $\varphi$ is a homomorphism, $\varphi(k (R-Z)) = k (R - Z - \pi^{-1}(\delta) + \sum_i l_i F_i)$ and this must be trivial in $NS(\hat Y_4) \otimes {\mathbb Q}$ because $R-Z$ is $k$-torsion. Furthermore, since $NS(\hat Y_4) \otimes {\mathbb Q}$ is torsion-free, this implies that $R - Z - \pi^{-1}(\delta) + \sum_i l_i F_i$ is trivial in  $NS(\hat Y_4) \otimes {\mathbb Q}$, as claimed above.\\\\
We will exemplify this general fact for situations in which 
 $\hat Y_4$ is a hypersurface in a toric ambient space.
 In our examples, $-k \Sigma$ turns out to be a toric divisor on the toric ambient space which does not intersect the Calabi-Yau hypersurface $\hat Y_4$.
 Furthermore, in the toric examples we will consider the base divisor $\delta$ will be given by ${\bar{\cal K}}_{\cal B}$, the anti-canonical divisor of ${\cal B}$  \footnote{This is due to the fact that we are only considering fibrations which are blow-ups of the Weierstrass type or have at least one holomorphic section.}.\\\\
Since $[\Sigma]$ is trivial as an element of $H^{2}(\hat Y_4, \mathbb R)$, it does not give rise to an extra $U(1)$ factor as would be the case if $R$ were a non-torsional rational section. We may use the triviality of $\Sigma$ in $NS(\hat Y_4) \otimes \mathbb Q$   to write  %$H^{2}(\hat Y_4, \mathbb R)$ 
\begin{equation} \label{def-Xi}
\Xi_k \equiv R-Z-\pi^{-1}(\delta) = -\frac{1}{k} \sum_i a_i F_i, \quad a_i \in \mathbb Z, \,
\end{equation} 
which by construction defines an element in $H^2(\hat Y_4,\mathbb Z)$.
One may be forgiven for thinking that the existence of a $k$-torsional point on the elliptic fiber induces a $k$-torsional element in $H^2(\hat Y_4,\mathbb Z)$. 
This is almost true but misses possible complications in the degenerate fibers at codimension-one singular loci whose resolution introduces the extra divisor classes $F_i$.
Indeed, from (\ref{def-Xi}) we see that while 
the class $[\Xi_k]$ is not torsion in the cohomology $H^{1,1}_{\mathbb Z}(\hat{Y}_4) = H^{2}(\hat{Y}_4, \mathbb Z) \cap H^{1,1}(\hat{Y}_4)$, 
it does represent a $k$-torsional element in the quotient cohomology $H^{1,1}_{\mathbb Z}(\hat{Y}_4)/ \langle [F_i] \rangle_{\mathbb Z} $ of classes modulo integer linear combinations of resolution classes. Namely, 
\begin{equation}
k \cdot [\Xi_k] = -\sum_i a_i [F_i] = 0 \,\, \text{mod } f \in \langle [F_i] \rangle_{\mathbb Z},
\end{equation}
which establishes $[\Xi_k]$  as $k$-torsion up to resolution divisors. We will give an intuitive explanation for the appearance of such a torsional element from the geometry of the elliptic fibration in the examples below - see section \ref{subsec_freequotient}.
%

%%%%%%%%%%%%%%%%%%%%%%%%%%%%%%%%%%%%%%%%%%%%%%%%%%%%%%%%%%%%%%%%%%%%%%%%%%%%%%%%%%%%%%%%%%%%%%%%%%%%%%%%%%%%%%%%%%%%%%%%%%%%%%%%%%%%%%%%%%%%%%%
\subsection{The global structure of the gauge group in presence of Mordell-Weil torsion} \label{sec_global}
%%%%%%%%%%%%%%%%%%%%%%%%%%%%%%%%%%%%%%%%%%%%%%%%%%%%%%%%%%%%%%%%%%%%%%%%%%%%%%%%%%%%%%%%%%%%%%%%%%%%%%%%%%%%%%%%%%%%%%%%%%%%%%%%%%%%%%%%%%%%%%%

While, as described, the existence of a torsional section does not give rise to any new $U(1)$ groups, it does have profound consequences on the physical properties of the F-theory compactification by restricting the matter spectrum and, equivalently, the global structure of the gauge group.\\\\
In F-theory, the non-abelian gauge algebra $\mathfrak g$ is dictated entirely by the singularity structure of the elliptic fibration $\hat Y_4$ in codimension one. The resolution divisors $F_i$ correspond to the generators of the Cartan subalgebra $\mathfrak h$ of  $\mathfrak g$.
The Cartan generators, or equivalently the resolution divisors, span the coroot lattice $Q^\vee = \langle F_i \rangle_\mathbb Z$.
%Let us denote by $Q^\vee$ the coroot lattice spanned by these $F_i$.
On the other hand, the information about the global structure of the non-abelian gauge group $G$ with Lie algebra $\mathfrak g$  is reflected in the representation content.
In F-theory localised  charged massless matter states in representation $\rho$ of the full gauge group $G$ arise from M2-branes wrapping suitable fiber components $\mathbb P^1_\rho$ over codimension-two loci on ${\cal B}$ corresponding to the intersection of several components of the discriminant locus, or to self-intersections of its components. The fiber components in question can be identified with the weights of the representation $\rho$.
The weights of all representations which are realized in the geometry span the weight  lattice $\Lambda$.
The coweight lattice $\Lambda^\vee $ is the dual lattice, defined by the integer pairing with the weight lattice $\Lambda$, 
\begin{equation} \label{pairing}
\Lambda ^\vee \times \Lambda  \rightarrow \mathbb Z.
\end{equation}
Geometrically, the coroot lattice $Q^\vee \subseteq \Lambda^\vee$ is spanned by the resolution divisors $F_i$, and the pairing is the intersection with the fiber components $\mathbb P^1_\rho$ associated with the matter representations.\\ \\
The relation between the representation data and the global structure of the gauge group   be understood as follows: For definiteness consider a semi-simple Lie group $G$. For such $G$ recall, \textit{e.g.}\ from  \cite{Bump:2013,DiFrancesco:1997}, that
\begin{equation}
\pi_1(G) \approx \frac{\Lambda^\vee}{Q^\vee}.
\end{equation}
It will be useful to compare $G$ to its \emph{universal cover} $G_0$, which has the same Lie algebra $\mathfrak g$ and whose coweight lattice is by definition 
$\Lambda_0^\vee = \langle F_i \rangle_{\mathbb Z}$. The dual weight lattice $\Lambda_0$ then contains all information about the representations that occur in a gauge theory with gauge group $G_0$.
Since by assumption $\Lambda_0^\vee = Q^\vee$, the group $G_0$ is simply-connected.

\noindent  Now, for definiteness suppose that the F-theory compactification gives rise to gauge algebra $\mathfrak{g} \oplus \mathfrak{g'}$, where $\mathfrak{g}$ and $\mathfrak{g'}$ are both semi-simple and whose Cartan subgroups are spanned by two sets of resolution divisors $F_i$ and $F_i'$. The gauge algebra $\mathfrak{g'}$ and its gauge group $G'$ will be mere spectators in what follows, but we include them to be more general.
We are interested in the structure of the global gauge group $G \times G'$. 
Suppose furthermore that the Mordell-Weil group has $k$-torsion and that the class $\Xi_k$ defined in (\ref{def-Xi}) involves only the Cartan generators $F_i$ of $\mathfrak{g}$, but not the generators $F_i'$ of 
$\mathfrak{g'}$. 
The class $\Xi_k$  is integer and therefore its intersection with the split fiber components $\mathbb P^1_{\rho}$ is integer as well. 
Group theoretically this implies that we can identify $\Xi_k$
 with a coweight of $G$.
Having fractional coefficients in $\frac{1}{k}\mathbb Z$ with respect to the $F_i$, the class $\Xi_k$ corresponds to a coweight in a  coweight lattice $\Lambda^\vee$ which is finer (by order $k$) compared to the sublattice $\Lambda^\vee_0= \langle F_i \rangle_{\mathbb Z}$ spanned by the $F_i$ alone. 
Therefore $\pi_1(G) \approx \frac{\Lambda^\vee}{Q^\vee}$ acquires a $\mathbb Z_k$ component compared to the first fundamental group of $G_0$. This leads to non-simply connected gauge groups. Since the universal covering group $G_0$ is simply connected, the gauge group $G \times G'$ in such an F-theory compactification with Mordell-Weil torsion $\mathbb Z_k$ has in fact first fundamental group
\bea
\pi_1(G) \times \pi_1(G')  =   \mathbb Z_k\  \times \pi_1(G'),
\eea 
where the spectator group $G'$ is unaffected by the Mordell-Weil torsion.

\begin{table}[t]
\begin{center}
\begin{tabular}{cc}
 \textit{Lie algebra} & \textit{Center of universal covering group} \tabularnewline 
$A_{n\,\geq\, 1}$ &  $\mathbb Z_{n+1}$ \tabularnewline
$B_{n\,\geq\, 2}$ &  $\mathbb Z_{2}$  \tabularnewline
$C_{n\,\geq \,3}$ &  $\mathbb Z_{2}$  \tabularnewline
$D_{2n+ 1\, \geq \, 4}$ &  $\mathbb Z_{4}$ \tabularnewline
$D_{2n \,\geq \,4}$ &  $\mathbb Z_{2} \oplus \mathbb Z_{2}$  \tabularnewline
$E_6$ &  $\mathbb Z_{3}$ \tabularnewline
$E_7$ &  $\mathbb Z_{2}$  \tabularnewline
$E_8$ &   -  \tabularnewline
$F_4$ &   -  \tabularnewline
$G_2$ &  -  \tabularnewline
\end {tabular}
\caption{Simple Lie algebras and the center of their universal covering groups.} \label{tab:lie_algebra_centers} 
\end{center}
\end {table}

\noindent At the same time, the integer pairing (\ref{pairing})
of coweights and weights forces the weight lattice $\Lambda$ to be coarser compared to the weight lattice $\Lambda_0$ dual to $\Lambda_0^\vee$, and the weights realized in the geometry become a subset of all weights that would be possible on the basis  of the Lie algebra alone.
Not only can one verify, as is clear by construction, that the geometrically realised  representations all have integer pairing with the coweight $-\frac{1}{k} \sum_i a_i F_i$ (appearing on the right of (\ref{def-Xi})), but also other representations which would be present in more generic fibrations without torsional sections have only fractional such pairing and are therefore `forbidden'.\\\\
Equivalently, we can think of Mordell-Weil torsion as affecting the center $Z_G$ of the gauge group $G$.
The center $Z_G$ of a semi-simple Lie group $G$  is given by  \cite{Bump:2013,DiFrancesco:1997} 
\begin{equation}
 Z_G \approx \frac{\Lambda}{Q} \,,
\end{equation}
where $Q \subset \Lambda$  the root lattice (see Figure \ref{tab:lie_algebra_centers} for a list of the center of the universal covering groups of the simple Lie algebras). Geometrically $Q$ is spanned by the fiber components associated with the adjoint representation of $G$ localised in codimension one.
As a reference consider again   the universal cover group $G_0$ introduced above  with center $Z_{G_0}$.
Since Mordell-Weil torsion $\mathbb Z_k$ renders $\Lambda$ coarser by a factor of $\mathbb Z_k$ compared to $\Lambda_0$, the center of  $G$ is smaller by the same amount, 
\bea
Z_G = Z_{G_0}/\mathbb Z_k.
\eea 
Note that this requires that $\mathbb Z_k$ be a subgroup of the center of $G_0$, which constrains the possible gauge algebra $\mathfrak g$ that can possibly appear. 
By contrast, any extra spectator Lie algebra $\mathfrak{g'}$ whose generators do not enter $\Xi_k$ is unconstrained.
For example, if the Mordell-Weil torsion is $\mathbb Z_{2}$, then a gauge algebra $\mathfrak{g} = \mathfrak{su}(k)$ is possible only for $k=2n$ - see the discussion in section \ref{sec_Generalisation} for an explicit construction. Furthermore, the total gauge group is given by 
\bea
G_0 / \mathbb Z_k \times G'.
\eea 
This can be directly understood in terms of the construction of our coweight element $\Xi_k$ in (\ref{def-Xi}).  Exponentiation of $\Xi_k$  generates a $\mathbb Z_k$ subgroup of $Z_{G_0}$. Since  $\Xi_k$ has integer pairing with every representation that is present (\textit{i.e.}\ with every lattice point in the weight lattice $\Lambda$, but not not $\Lambda_0$), the corresponding center element (viewed as an element of $G_0$) acts trivially on every such representation; the actual gauge group is therefore not $G_0 \times G'$, but $G_0/\mathbb Z_k \times G'$.\\\\
Indeed, to construct an element in the center of $G_0$ one exponentiates a linear combination $ \Xi= \sum m_i F_i$
%For $\mathfrak{g}$ the Lie algebra of the Lie group $G$ the center $Z_\mathfrak{g}$ is the ideal $\{X \,|\, [X,Y] = 0 \,\, \forall \,\, Y \in \mathfrak{g}\}$. In particular an element $Z \in Z_\mathfrak{g}$ is in the Cartan subalgebra $\mathfrak{h} \subset \mathfrak{g}$. The Cartan-Weyl basis for $\mathfrak{h}$ is $\{H_i\}$ enumerated by $i = 1,\dots, r$ where $r = \text{rank } \mathfrak{g}$ and hence we have
% \begin{equation}\label{eq:cartan_sum}
% \Xi = \sum m_i F_i
% \end{equation}
of Cartan generators $F_i$ for suitable coefficients $m_i$. We denote by $\rho_d$ a $d$-dimensional representation of $\mathfrak{g}$. A state $|\lambda^n, \rho_d \rangle$ in the representation $\rho_d$ is labeled by the weight $\lambda^n$ in the weight system of $\rho_d$. Letting $\Xi$ act on such a state gives
\begin{equation}
\Xi \cdot|\lambda^n, \rho_d \rangle = \sum_i m_i \lambda_i^n |\lambda^n, \rho_d \rangle, \, \qquad n = 1, \dots, d,\,\qquad i=1,\dots,r,
\end{equation}
where $\lambda_i^n$ is the eigenvalue of $F_i$ on this state vector. An element $c$ in the center $Z_{G_0} \in G_0$ commutes with any element in $G_0$ and is represented as a multiple of the $d\times d$ unit matrix when acting on the state $|\lambda^n, \rho_d \rangle$, i.e
\begin{equation}\label{eq:center_on_group}
c |_{\rho_d} \cdot|\lambda^n, \rho_d \rangle =  a_n\, \one \cdot|\lambda^n, \rho_d \rangle \, 
\end{equation}
for $a_n \in \mathbb C$. To identify $c$ as the exponentiation of $\Xi$ we identify
\begin{equation}
a_n = \text{exp }(2 \pi i \sum m_i \lambda_i^n).
\end{equation}
For $c$ to lie in a $\mathbb Z_k$ subgroup of the center of $G_0$, $c^k$ acts as $\one$ on any representation $\rho_d$, or equivalently  $(a_n)^k = 1$ for all $n$. 
Therefore, if we identify $\Xi$ with the $k$-fractional linear combination $\Xi_k =  - \frac{1}{k} \sum_i a_i F_i$, we see that this does indeed generate a $\mathbb Z_k$ subgroup of $Z_{G_0}$. Moreover, since $\Xi_k$ has integer pairing with all weights in the weight lattice $\Lambda$ of the actual gauge group $G$, the element $c$ acts trivially on every such representation. We can therefore view $G$ as the result of 'gauging' $\mathbb Z_k$, \textit{i.e.}\ $G = G_0 /\mathbb Z_k$, as claimed. Finally, note that all results of this section generalize to more complicated Mordell-Weil torsion groups of the form $\mathbb Z_{k_1} \oplus \ldots \oplus \mathbb Z_{k_n}$.

%%%%%%%%%%%%%%%%%%%%%%%%%%%%%%%%%%%%%%%%%%%%%%%%%%%%%%%%%%%%%%%%%%%%%%%%%%%%%%%%%%%%%%%%%%%%%%%%%%%%%%%%%%%%%%%%%%%%%%%%%%%%%%%%%%%%%%%%%%%%%%%
\section{Mordell-Weil group \texorpdfstring{$\mathbb Z_2$}{Z2}} \label{sec_4}
%%%%%%%%%%%%%%%%%%%%%%%%%%%%%%%%%%%%%%%%%%%%%%%%%%%%%%%%%%%%%%%%%%%%%%%%%%%%%%%%%%%%%%%%%%%%%%%%%%%%%%%%%%%%%%%%%%%%%%%%%%%%%%%%%%%%%%%%%%%%%%%

In the subsequent sections we exemplify the structure of F-theory compactifications on elliptic fibrations with torsional Mordell-Weil group as outlined above.
Ref. \cite{Aspinwall:1998xj} has derived the defining equations describing elliptic fibrations with Mordell-Weil group 
${\mathbb Z}_k$ for $k=2,3,4,5,6$, ${\mathbb Z}_2 \oplus {\mathbb Z}_n$ with $n=2,4$ and ${\mathbb Z}_3 \oplus {\mathbb Z}_3$ as hypersurfaces in $\mathbb P_{2,3,1}[6]$ fibrations.
As it turns out, the restriction of the complex structure moduli of the fibration necessary for the Mordell-Weil group to have torsion induces singularities in the fiber over divisors on the base ${\cal B}$.
To explicitly analyse these singular loci and their resolution we focus in this work on the subset of geometries in the list of \cite{Aspinwall:1998xj} which can be treated torically as certain hypersurfaces.
As noted already, there exist 16 reflexive polygons in two dimensions which describe an elliptic curve as a hypersurface in a toric ambient space.
Of these only three admit torsional sections in the Mordell-Weil group as the intersection of a toric divisor with the generic hypersurface defined by the dual polygon. The Mordell-Weil group of these fibrations has already been provided in \cite{Braun:2013nqa}. 
As we will show, they correspond to the geometries with Mordell-Weil group ${\mathbb Z}_2$ and ${\mathbb Z}_3$ as well as a further specialisation of the ${\mathbb Z}_2$-model in the list of \cite{Aspinwall:1998xj}.
For each of these three fibration types we construct a compact model fibered over a generic base ${\cal B}$
and analyse in detail the interplay between the torsional sections and the global structure of the gauge group.
In addition we implement further non-abelian gauge symmetries by the construction of toric tops \cite{Bouchard:2003bu}.

\subsection{An $SU(2)/{\mathbb Z}_2$-fibration} \label{sec_SU2Z2}

We begin with the simplest example of an elliptic fibration with torsional Mordell-Weil group, which turns out to be $\mathbb Z_2$.
% First we consider the general features of this fibration and its description in terms of reflexive polygons, and then we construct compact four-folds with further fiber degenerations via tops. 
As derived in \cite{Aspinwall:1998xj}, an elliptic fibration with a $\mathbb Z_2$-torsional section
admits a representation as the hypersurface $P=0$ with 
 \begin{equation}\label{eq:hse-Z2-fiber-singular}
 P = -y^2 - a_1 x\,y\,z + x^3+a_2\,x^2\,z^2+a_4\,x\,z^4
\end{equation}
and $[x:y:z]$ fiber coordinates in a $\mathbb P_{2,3,1}$-fibration over some base $\mathcal{B}$.
To ensure that the variety $P=0$ satisfies the Calabi-Yau condition the coefficients $a_i$ must be sections of $\mathcal{\bar K}^{i}_\mathcal{B}$ with    $\mathcal{\bar K}_\mathcal{B}$      the anti-canonical bundle 
of the base ${\cal B}$.
Note that (\ref{eq:hse-Z2-fiber-singular}) corresponds to an otherwise generic Tate model with $a_6 \equiv 0$ and $a_3 \equiv 0$. It can therefore be viewed as a further specialisation of the $U(1)$ restricted Tate model, defined  in \cite{Grimm:2010ez} by setting $a_6 \equiv 0$. The latter has Mordell-Weil group $\mathbb Z$ and in turn represents a special case of the elliptic fibrations with Mordell-Weil group $\mathbb Z$ as described in \cite{Morrison:2012ei}.

\subsubsection{Singularity structure and resolution} \label{subsec_sing}

\begin{figure}[t]
    \centering
\begin{tikzpicture}
   \shadedraw[ball color=purple!40!white] (-0.3,0) .. controls (-0.3,1) and (.3,2) .. (1,2)
               .. controls (1.7,2) and (1.8,1.5) .. (1.8,1.2)
               .. controls (1.8,.5) and (1,.5) .. (1,0)
               .. controls (1,-.5) and (1.8,-.5) .. (1.8,-1.2)
               .. controls (1.8,-1.5) and (1.7,-2) .. (1,-2)
               .. controls (.3,-2) and (-0.3,-1) .. (-0.3,0);
\begin{scope}[xshift=3.6cm]
\shadedraw[ball color=gray!30!white,xscale=-1] (-0.3,0) .. controls (-0.3,1) and (.3,2) .. (1,2)
               .. controls (1.7,2) and (1.8,1.5) .. (1.8,1.2)
               .. controls (1.8,.5) and (1,.5) .. (1,0)
               .. controls (1,-.5) and (1.8,-.5) .. (1.8,-1.2)
               .. controls (1.8,-1.5) and (1.7,-2) .. (1,-2)
               .. controls (.3,-2) and (-0.3,-1) .. (-0.3,0);
\end{scope}
\begin{scope}[xshift=10cm]
% grey ball
\begin{scope}[scale=2]
\shadedraw[ball color=gray!30!white, shading=ball] (1,0) ..controls (1,-.5) and (0,-.5) .. (0,-1) 
.. controls (0,-.5) and (-1,-.5) .. (-1,0) 
.. controls (-1,.5) and (-.5,1) .. (0,1)
.. controls (.5,1) and  (1,.5).. (1,0);
\end{scope}
% green cross
\draw[green, line width=0.5mm, xshift=1cm,yshift=1cm] (0,0)--(0.2,0.2);
\draw[green, line width=0.5mm, xshift=1cm,yshift=1cm] (0,0.2)--(0.2,0);
% blue cross
\draw[blue, line width=0.5mm, xshift=-1cm,yshift=-1cm] (0,0)--(0.2,0.2);
\draw[blue, line width=0.5mm, xshift=-1cm,yshift=-1cm] (0,0.2)--(0.2,0);
\end{scope}

% blue cross
\draw[blue, line width=0.5mm, xshift=2.5cm,yshift=1cm] (0,0)--(0.2,0.2);
\draw[blue, line width=0.5mm, xshift=2.5cm,yshift=1cm] (0,0.2)--(0.2,0);
% green cross
\draw[green, line width=0.5mm, xshift=5mm,yshift=-1cm] (0,0)--(0.2,0.2);
\draw[green, line width=0.5mm, xshift=5mm,yshift=-1cm] (0,0.2)--(0.2,0);
\end{tikzpicture}

\caption{To the left we depict the factorised fiber over the base locus $a_4=0$; the \textcolor{purple}{purple} $\mathbb P^1$ indicates the $s=0$ part while the \textcolor{gray}{grey} $\mathbb P^1$ is the second irreducible part of the elliptic curve. To the right the fiber over the base locus $a_4=\tfrac14(a_2+\tfrac14 a_1^2)^2$ is shown. The multiplicity is one, and  the fiber is singular.
The \textcolor{blue}{blue} and \textcolor{green}{green} crosses indicate the specified points $z=0$ and the $\mathbb Z_2$-point $t=0$ of the elliptic curve, respectively.}\label{fig:fiber_over_discriminant_zeroes}
\end{figure}
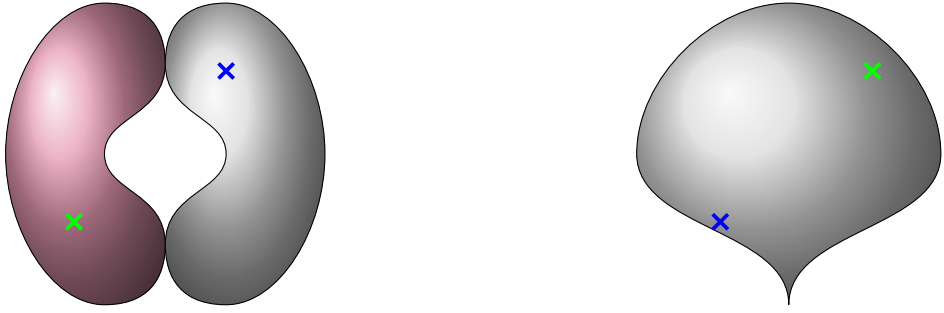

The elliptic fibration  (\ref{eq:hse-Z2-fiber-singular}) is easily brought into Weierstrass form (\ref{eq:weierstrass_equation}) with
\begin{equation*}\label{eq:discriminant-Z2-case}
f = a_4 - \frac{1}{3}\left(a_2+\frac{a_1^2}{4}\right)^2, \quad g = \frac{1}{27}\left(a_2+\frac{a_1^2}{4}\right)\left(2(a_2+\frac{a_1^2}{4})^2 - 9a_4\right).
\end{equation*}
From $f$ and $g$ and the discriminant
\bea \label{discr-su2}
\Delta=\frac{1}{16}a_4^2\left(4\,a_4-\left(a_2+\tfrac14 a_1^2\right)^2\right)
\eea
one infers an $\mathfrak{su}(2)$-singularity at $a_4=0$. Indeed,  the gradient of \eqref{eq:hse-Z2-fiber-singular} in the patch $z\ne 0$,
\begin{equation}
\begin{aligned}
&d P = (-a_1\,y+3\,x^2+2\,a_2\,x+a_4)\,dx-(2\,y+a_1\,x)\,dy\, -xy\,d_B\, a_1+ x^2\, d_B\,a_2 + x\,d_B\,a_4 \,,
%  &(a_1\,y\,z-(3\,x^2+2\,a_2\,x\,z^2+a_4\,z^4))\,dx+(2\,y+a_1\,x\,z)\,dy +(a_1 x\,y+\\
% &-(2\,a_2\,x^2\,z+4\,a_4\,x\,z^3))\,dz+((d_B\, a_1) x\,y\,z-((d_B\,a_2)\,x^2\,z^2+(d_B\,a_4)\,x\,z^4))\,,
\end{aligned}
\end{equation}
with $d_B$  the total derivative with respect to the base coordinates, vanishes together with the hypersurface equation \eqref{eq:hse-Z2-fiber-singular} for $x=y=a_4=0$. 
The situation is similar to the $U(1)$-restricted model with $a_6 \equiv 0$ but $a_3 \neq 0$  \cite{Grimm:2010ez}, in which, however, the singularity appeared over the curve  $\{a_3 = 0\} \cap  \{a_4 = 0\}$ on ${\cal B}$.
Since in (\ref{eq:hse-Z2-fiber-singular}) $a_3$ is set to zero from the very beginning, the $\mathfrak{su}(2)$ locus is promoted to the divisor $ \{a_4 = 0\}$. We will come back to this enhancement of the $\mathfrak{u}(1)$ gauge algebra of the $U(1)$ restricted Tate model to $\mathfrak{su}(2)$ by setting $a_3\equiv 0$  in section \ref{unHiggsing}.
 \\ \\
To resolve the singularity we perform a blow-up in the fiber ambient space
\begin{equation}
 x\rightarrow s\,x\,,\qquad y\rightarrow s\,y\,.
\end{equation}
Since $a_6\equiv 0$, this does not spoil the Calabi-Yau condition of the hypersurface as one can see from the proper transform of \eqref{eq:hse-Z2-fiber-singular} given by
\begin{equation}\label{eq:hse-Z2-fiber-1st-bu}
  \hat{P} = -y^2\,s  -a_1 x\,y\,z\,s + x^3\,s^2+a_2\,x^2\,z^2\,s+a_4\,x\,z^4 \,,
\end{equation}
which is checked to be smooth (see \cite{Grimm:2010ez,Krause:2011xj} for the analogous blow-up if $a_3 \neq 0$). In order to facilitate the  description of the $\mathbb Z_2$-torsional section it turns out useful to 
perform a further ambient space blow-up 
\begin{equation}
 s\rightarrow t\,s\,,\qquad x\rightarrow t\,x\,,
\end{equation}
under which
the proper transform of \eqref{eq:hse-Z2-fiber-1st-bu}  becomes 
\begin{equation}\label{eq:hse-Z2-fiber-2nd-bu}
\hat{P}  = -y^2\,s - a_1 x\,y\,z\,s\,t + x^3\,s^2\,t^4+a_2\,x^2\,z^2\,s\,t^2+a_4\,x\,z^4.
\end{equation}
The Stanley-Reisner ideal relations after the two blow-ups are
\begin{equation}
 \textmd{SR-i}:\quad \{y\,t,\,y\,x,\,s\,x,\,s\,z,\,t\,z\}\,,
\end{equation}
and we observe that the divisor $X: \{x=0\}$ does not intersect the hypersurface. % in this representation of the curve. 
Hence $x$ can be set to one in \eqref{eq:hse-Z2-fiber-2nd-bu} and from now on we will analyse the fibration $\hat P=0$ with
\begin{equation}\label{eq:hse-Z2-fiber-2nd-bub}
\hat{P}  = -y^2\,s - a_1 y\,z\,s\,t + s^2\,t^4+a_2\,z^2\,s\,t^2+a_4 \,z^4 \,
\end{equation}
over a suitable base ${\cal B}$. If ${\cal B}$ is 3-dimensional, this defines an elliptically fibered Calabi-Yau 4-fold $\hat Y_4$.
\begin{figure}[t]
    \centering
%lightgray!30!white
%blue!30!white
  \begin{tikzpicture}[scale=1.5]
%%%% feft polygon %%%%
  \filldraw [ultra thick, draw=black, fill=lightgray!30!white]
      (-2,-1)--(2,-1)--(0,1)--cycle;
    \foreach \x in {-2,-1,...,2}{% Two indices running over each
      \foreach \y in {-1,0,...,1}{% node on the grid we have drawn 
        \node[draw,circle,inner sep=1.3pt,fill] at (\x,\y) {};
            % Places a dot at those points
      }
    }
  \draw[ultra thick, -latex]
       (0,0) -- (0,1) node[above] {$y$};
  \draw[ultra thick, -latex]
       (0,0) -- (1,0) node[above right] {$s$};
  \draw[ultra thick, -latex]
       (0,0) -- (2,-1) node[below right] {$t$};
  \draw[ultra thick, -latex]
       (0,0) -- (-2,-1) node[below left] {$z$};
  \node [below] at (1,-1.05)  {$x$};
%%%% right polygon %%%%
\begin{scope}[xshift=0.33\textwidth]
\filldraw [ultra thick, draw=black, fill=blue!30!white]
      (0,1)--(-1,-1)--(1,-1)--cycle;
    \foreach \x in {-1,0,...,1}{% Two indices running over each
      \foreach \y in {-1,0,...,1}{% node on the grid we have drawn 
        \node[draw,circle,inner sep=1.3pt,fill] at (\x,\y) {};
            % Places a dot at those points
      }
    }
  \node [above] at (0,1)  {$y^2\,s$};
  \node [below right] at (1,-1) {$t^4\,s^2$};
  \node [below left] at(-1,-1)  {$z^4$};
\end{scope}
  \end{tikzpicture}
      \caption{Polygon 13 of \cite{Bouchard:2003bu} together with its dual polygon. The coordinate $x$ is blown-down, and not part of the fan.}\label{fig:polygon13}
\end{figure}
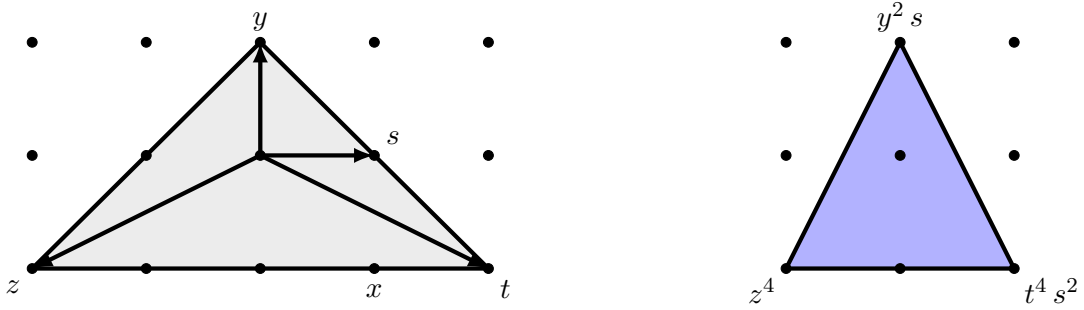
The weight matrix of the homogeneous coordinates
can be taken to be
\begin{equation}\label{eq:weight-matrix_Z2-case}
 \begin{array}{|c|c|c|c||c|}
\hline y & z & s & t & \sum \tabularnewline \hline \hline
2 &  1 &  0 &  1 &  4 \tabularnewline\hline
1 &  1 &  2 &  0 & 4\tabularnewline\hline
\end{array}
\end{equation}
and  the Stanley-Reisner ideal  simplifies to
\begin{equation}
 \{y\,t,\,s\,z\}\,.
\end{equation}
Note that the weight matrix  (\ref{eq:weight-matrix_Z2-case}) coincides with the weight matrix as read off from 
 the toric fan depicted in Figure~\ref{fig:polygon13}, which corresponds to polygon 13 in the list \cite{Bouchard:2003bu} of 16 torically embedded hypersurface elliptic curves. The fibration (\ref{eq:hse-Z2-fiber-2nd-bub}) with $s \equiv 1$, corresponding to the blow-down of the resolution divisor associated with the $\mathfrak{su}(2)$ singularity over $a_4=0$, has been analysed previously in \cite{Berglund:1998va} and shown to correspond to an elliptic fibration with restricted $SL(2,\mathbb Z)$ monodromy. We will analyze this relation in more detail in section \ref{sec_Generalisation}.\\\\
 The advantage of passing to the hypersurface representation (\ref{eq:hse-Z2-fiber-2nd-bub}) is that the $\mathbb Z_2$-torsional point on the elliptic fiber is now explictly given by the intersection of the fiber with the toric divisor 
 \bea
 T: t=0.
\eea
This can be checked via the group law on the elliptic curve.
We will henceforth denote $T$ as the $\mathbb Z_2$ section of the fibration. The holomorphic zero-section is given by $Z: z=0$.\\\\
To study the geometry further we note that the fibration restricted to the $\mathfrak{su}(2)$-sublocus  $\{a_4=0\}$ in the discriminant (\ref{discr-su2}) factorises as
\begin{equation}\label{eq:hse-Z2-over-a4}
 \begin{aligned}
\hat P|_{a_4=0}  =  s\,\left( - y^2 - a_1\,y\,z\,t + \left(s\,t^4+a_2\,z^2\,t^2\right)\right).
 \end{aligned}
\end{equation}
The resolution divisor $S: s=0$ is a $\mathbb P^1$-fibration over the locus $\{a_4=0\}$ on ${\cal B}$ as the coordinate $s$ is just a toric ambient space coordinate.
The other irreducible component of \eqref{eq:hse-Z2-over-a4}  is quadratic in $y$ and must therefore be studied in more detail. Note first that this component does not intersect the $\mathbb Z_2$ section $T$, but only the holomorphic zero-section $Z$. 
Since $z$ and $t$ cannot both vanish along it, we can go to the patch where $y$ and $s$ can vanish simultaneously. Here the second factor of \eqref{eq:hse-Z2-over-a4} becomes
\begin{equation}\label{eq:hse-Z2-over-a4-2nd-part}
 \begin{aligned}
  y^2 + a_1\,y-\left(s+a_2\right)=0\,.
 \end{aligned}
\end{equation}
The discriminant of this quadratic equation is a linear function in $s$ so that we find one branching point in the $s$-plane. Since the point at `$s=\infty$' ($z=0$) is also single valued, we can  take the branch-cut from $s=-(\frac14 a_1^2+a_2)$ to infinity. Gluing the two $\mathbb P^1$s viewed as compactified complex planes
along the branch-cut, we obtain again a $\mathbb P^1$. The two irreducible parts of \eqref{eq:hse-Z2-over-a4} intersect each other in two points, as can be seen from \eqref{eq:hse-Z2-over-a4-2nd-part}. The factorised fiber over the base divisor $\{a_4=0\}$ is depicted on the left in Figure~\ref{fig:fiber_over_discriminant_zeroes}. 
Over the zero set of the second factor of the discriminant (\ref{discr-su2}),
\begin{equation}\label{eq:discriminant_Z2-case_2nd-part}
4\,a_4-(a_2+\tfrac14 a_1^2)^2=0 \,,
\end{equation}
we analyse the fiber structure by substituting \eqref{eq:discriminant_Z2-case_2nd-part} into \eqref{eq:hse-Z2-fiber-2nd-bub}. This gives the hypersurface equation
\begin{equation*}
 \hat P|_{(\ldots=0)} = - y^2\,s - a_1 \,y\,z\,s\,t +s^2\,t^4+a_2\,z^2\,s\,t^2+\tfrac14(a_2+\tfrac14 a_1^2)^2\,z^4\,.
\end{equation*}
To determine the fiber type, we can go to the patch where $y$ and $z$ are allowed to vanish simultaneously. We set $s=1$ since the divisor $\{s=0\}$ does not intersect the elliptic curve away from $\{ a_4=0 \}$ and complete the square as
\begin{equation*}
\begin{aligned}
&y^2 + a_1 \,y\,z=1+a_2\,z^2+\tfrac14(a_2+\tfrac14 a_1^2)^2\,z^4\\
&\Rightarrow (y+\frac{1}{2}a_1\,z)^2=1 +(a_2+\tfrac14 a_1^2)\,z^2+\tfrac14(a_2+ a_1^2)^2\,z^4\\
&\Rightarrow (y+\frac{1}{2}a_1\,z)^2=\left(1 +\tfrac12(a_2+\tfrac14 a_1^2)\,z^2\right)^2\\
&\Rightarrow \left(y+\frac{1}{2}a_1\,z - 1 - \tfrac12(a_2+\tfrac14 a_1^2)\,z^2\right)\left(y+\frac{1}{2}a_1\,z + 1 +\tfrac12(a_2+\tfrac14 a_1^2)\,z^2\right)=0\,.
\end{aligned}
\end{equation*}
Therefore, it appears as if the elliptic curve factorises into two rational curves. However, these two $\mathbb P^1$s are equivalent as follows from the second row of the weight matrix \eqref{eq:weight-matrix_Z2-case} because the equivalence relation $(y,z)\sim(-y,-z)$ is left over after setting $s$ to one\footnote{This can also be seen from the $N$-lattice polygon of Figure~\ref{fig:polygon13} because $y$ and $z$ do not span the lattice. The patch where $y$ and $z$ are allowed to vanish simultaneously is, therefore, $\mathbb{C}^2/\mathbb Z_2$ and not $\mathbb{C}^2$ as one would na\"ively think.}. Thus the fiber is just a single rational curve; moreover, it has a singular point, cf.\ Figure~\ref{fig:fiber_over_discriminant_zeroes}, at $y = - \frac{1}{2} a_1 z$, $s= -\frac{1}{8}(a_1^2 + 4 a_2) z^2$ (and $t=1$ due to the Stanley-Reisner ideal), where the gradient along the fiber coordinates vanishes even though the fibration as such is non-singular.
Thus the fiber is of Kodaira-type $I_1$, and the locus (\ref{eq:discriminant_Z2-case_2nd-part}) does not give rise to any further gauge symmetry. \\\\
Interestingly, apart from the codimension-one splitting of the fiber over $\{a_4=0\}$ no further degeneration of the fiber occurs in higher codimension. In particular, the fiber over the intersection curve $\{a_4=0\} \cap \{a_2 + \frac{1}{4}a_1^2 =0\}$ of the two components of the discriminant does not factorise further. This can be understood by considering the vanishing of $f$ and $g$ along that locus:  $f$ vanishes to order $1$, $g$ vanishes to order $2$ and the discriminant $\Delta$ consequently to order $3$, giving a Kodaira fiber of type $III$.  This type of fiber has two components just like the familiar $A_1$-fiber, but they are tangent to each other rather than meeting at two distinct points, and there is no enhancement or matter (consistent with 
\cite{Bershadsky:1996nh,Grassi:2011hq}).
This is remarkable because naively one might have expected an enhancement from $A_1$ to $A_2$ at the intersection of the $A_1$-locus with the $I_1$-component of the discriminant and thus localised massless matter in the fundamental of $\mathfrak{su}(2)$. The absence of this enhancement and the associated fundamental representation is a typical property of fibrations with torsional Mordell-Weil group. To summarize, the fibration (\ref{eq:hse-Z2-fiber-singular}) gives rise to an F-theory compactification with gauge algebra $\mathfrak{su}(2)$ and no localised
 charged matter.
 % As we will see, the absence of charged localized matter is a consequence of the $\mathbb Z_2$ Mordell-Weill group and the resulting global structure of the gauge group.

%%%%%%%%%%%%%%%%%%%%%%%%%%%%%%%%%%%%%%%%%%%%%%%%%%%%%%%%%%%%%%%%%%%%%%%%%%%%%%%%%%%%%%%%%%%%%%%%%%%%%%%%%%%%%%%%%%%%%%%%%%%%%%%%%%%%%%%%%%%%%%%
\subsubsection{Torsional divisors and free quotient} \label{subsec_freequotient}
%%%%%%%%%%%%%%%%%%%%%%%%%%%%%%%%%%%%%%%%%%%%%%%%%%%%%%%%%%%%%%%%%%%%%%%%%%%%%%%%%%%%%%%%%%%%%%%%%%%%%%%%%%%%%%%%%%%%%%%%%%%%%%%%%%%%%%%%%%%%%%%

The absence of charged localized matter in the fundamental representation is a consequence of the $\mathbb Z_2$ Mordell-Weil group and the resulting global structure of the gauge group.
To see this  let us first exemplify how the torsional Mordell-Weil group of the elliptic fiber induces a torsional element in  $H^{1,1}(\hat{Y}_4, \mathbb{Z})$ modulo the integer lattice spanned by the resolution divisors.
In the present model with gauge algebra $\mathfrak{g} = \mathfrak{su}(2)$ the lattice of resolution divisors is simply $\langle S \rangle_{\mathbb Z}$.
To find the element $\Sigma_2$ of the form (\ref{Sigmadef}) we make an Ansatz and demand that (\ref{pi-condition}) be satisfied. In the present situation this amounts to demanding that 
 $\Sigma_2$ have `one leg in the fiber' and that it be orthogonal to the exceptional divisor $S$, in the sense that for all $\omega_4 \in H^4(B)$ and $\omega_2 \in H^2(B)$
\bea
\int_{\hat Y_4} \Sigma_2 \wedge Z \wedge \pi^*\omega_4 =  \int_{\hat Y_4} \Sigma_2 \wedge \pi^*\omega_2 \wedge \pi^*\omega_4  = \int_{\hat Y_4} \Sigma_2 \wedge S \wedge \pi^*\omega_4 = 0.
\eea
This uniquely determines 
\begin{equation}
\Sigma_2 = T - Z - \bar{\mathcal{K}} + \frac{1}{2} S  \,
\end{equation}
 with $\bar{\mathcal{K}} = \pi^{-1}\bar{\mathcal{K}}_\mathcal{B}$ .
This element is in fact trivial in $H^2(\hat Y_4, \mathbb R)$.
Indeed,
 recall that the fibration $\hat Y_4$ is described as the  hypersurface  (\ref{eq:hse-Z2-fiber-2nd-bu}) in an ambient toric space. 
Consider the toric divisor $X: \{x=0\}$ in this ambient space. Its class is
\begin{equation}
X= 2Z - S - 2T + 2{\bar {\cal K}} = -2  \Sigma_2.
\end{equation}
However, as discussed, $X$ does not intersect the hypersurface $\hat Y_4$ and therefore its class is trivial on the hypersurface. Thus also $\Sigma_2$ is trivial in $H^{1,1}(\hat Y_4, \mathbb R)$.  This  implies that
\begin{equation} \label{xi2def}
\Xi_2 := T-Z- \bar{\mathcal{K}} = -\frac{1}{2}  S,
\end{equation}
thereby identifying $\Xi_2$ as 2-torsion in $H^{1,1}(\hat{Y}_4, \mathbb{Z}) / \langle S \rangle_{\mathbb Z}$. \\\\
According to the discussion in section \ref{tors-sec-gen}, associated with $\Xi_2$ is an extra coweight defined over $\frac{1}{2}\mathbb Z$. Thus, to preserve the pairing with the weights, the weight lattice is forced to be coarser. In particular the representation $\mathbf{2}$ of $\mathfrak{su}(2)$ cannot be present in this model as its weight would have half-integer pairing with the fractional coweight  $\Xi_2 = - \frac{1}{2} S$, in contradiction with the fact that $T-Z- \bar{\mathcal{K}}$  is manifestly integer. 
This is the deeper reason behind the absence of a fundamental representation at the intersection of the $\mathfrak{su}(2)$-divisor $\{a_4=0\}$ with the second discriminant component.
The gauge group of the model is thus
\bea
G = SU(2)/{\mathbb Z}_2
\eea
with $\pi_1(G) = \mathbb Z_2$.\\\\
One can give an intuitive geometric explanation for the appearance of the 2-torsion element  $\Xi_2$ in $H^{1,1}(\hat{Y}_4, \mathbb{Z}) / \langle S \rangle_{\mathbb Z}$ as follows:
Restrict the elliptically fibered Calabi-Yau $ \hat Y_4$ over ${\cal B}$ given by the hypersurface equation \eqref{eq:hse-Z2-fiber-2nd-bu} to ${\cal B} \backslash \{a_4=0\}$. 
As will be discussed momentarily, the resulting space $\hat Y_4'$
is a free $\mathbb Z_2$ quotient,
\begin{eqnarray} \label{Z2quotient}
\hat Y_4' = \widetilde{\hat Y_4'}/\mathbb Z_2,
\end{eqnarray}
with $\widetilde{\hat Y_4'}$ an elliptic fibration over ${\cal B} \backslash \{a_4=0\}$.
Correspondingly 
\begin{eqnarray}
\pi_1(\hat Y'_4) \supset \mathbb Z_2,
\end{eqnarray}
where additional discrete torsion pieces may arise if $\pi_1({\cal B}\backslash \{a_4=0\})$ is non-trivial.
Since the resolution divisor $S$ is fibered over $\{a_4=0\}$ this is in agreement with the appearance of a torsional element   in $H^{1,1}(\hat{Y}_4, \mathbb{Z}) / \langle S \rangle_{\mathbb Z}$.\\\\
The relation (\ref{Z2quotient}) can be seen as follows:
 Consider the fibration over a generic locus on the base ${\cal B}$ where $a_4\neq 0$. Since the resolution divisor $s=0$ intersects the fiber only over $\{ a_4=0 \}$ we can set $s$ to one away from that locus. %In addition, as explained above, we can set $x$ to  as $x=0$ does not lie on the hypersurface.
Then \eqref{eq:hse-Z2-fiber-2nd-bub} becomes
\begin{eqnarray} \label{eq:hse-Z2-fiber-2nd-bu-2}
 y^2 + a_1 y\,z\,t= t^4+a_2 z^2 \, t^2+a_4 \, z^4\,.
\end{eqnarray}
This is a special $\mathbb P_{1,1,2}[4]$ fibration with homogeneous coordinates $[t : z : y]$, which in addition to the equivalence relation $(t,z,y) \sim (\lambda t, \lambda z,\lambda^2 y)$ enjoys a further $\mathbb Z_2$ identification
\begin{eqnarray} \label{eq:Z2symmetry}
t \sim -t, \qquad \quad y \sim -y.
\end{eqnarray}
In fact, the most generic $\mathbb P_{1,1,2}[4]$ representation of an elliptic curve contains the nine terms
\begin{eqnarray}
y^2,\, t^4,\, z^4 ,\, z^2 t^2,\, y z t; \qquad \quad y t^2,\, y z^2,\, z t^3,\, t z^3.
\end{eqnarray}
Precisely the first five terms present in \eqref{eq:hse-Z2-fiber-2nd-bu-2} are compatible with the $\mathbb Z_2$ identification \eqref{eq:Z2symmetry}. Note that by a coordinate redefinition we can set $a_1 \equiv 0$, thereby arriving at the special $\mathbb P_{1,1,2}[4]$-fibration that goes by the name of the L\'egendre family.
In any case, we can view \eqref{eq:hse-Z2-fiber-2nd-bu-2} as the result of starting with  a $\mathbb P_{1,1,2}[4]$ fibration described by the hypersurface equation 
\begin{eqnarray} \label{eq:Z2-cover}
 y^2 + a_1 y\,z\,t= t^4+a_2 z^2 \, t^2+a_4 \, z^4\, + c_1 \, y \, t^2 + c_2 \, y \, z^2 + c_3 \,  z \, t^3 + c_4 \, t \,  z^3,
\end{eqnarray}
enforcing the $\mathbb Z_2$ symmetry by setting $c_i \equiv 0$ (we call the resulting space  $\widetilde{\hat Y_4'}$) and then quotienting by this $\mathbb Z_2$ symmetry. The fact that  ${\hat Y_4'}$ is really the quotient of  $\widetilde{\hat Y_4'}$ by \eqref{eq:Z2symmetry}
is automatically implemented by the toric description because the dual polyhedron exclusively contains monomials invariant under  \eqref{eq:Z2symmetry}.
Importantly, the $\mathbb Z_2$ acts freely as the fixed point sets $\{t=y=0\}$ and $\{z=y=0\}$ do not lie on  ${\hat Y_4'}$ due to the Stanely-Reisner ideal. Note that the role of this $\mathbb Z_2$ quotient symmetry was stressed already in \cite{Berglund:1998va} albeit in a slightly different context.\\\\
This description makes the existence of discrete one-cycles on $\hat Y_4'$ manifest:
Consider the locus $z=0$ on \eqref{eq:hse-Z2-fiber-2nd-bu-2}. On $\widetilde{\hat Y_4'}$ it is given by $y = \pm 1$, where we have used the scaling of $\mathbb P_{1,1,2}$ to set $t=1$ since $t$ and $z$ cannot simultaneously vanish as a consequence of the Stanely-Reisner ideal. A path from $y=-1$ to $y=+1$ on the double cover $\widetilde{\hat Y_4'}$ corresponds to a non-contractible closed loop on ${\hat Y_4'}$. This loop is torsional as going along it twice is contractible again. \\\\
%Note that away from $a_4=0$ the same argument gives a non-contractible loop at $t=0$ corresponding to the path from $y= - \sqrt{a_4}$ to $y= + \sqrt{a_4}$ on the %double cover. However, this loop degenerates to a point at $a_4=0$ and thus does not give rise to a globally defined discrete one-cycle on $Y$.
The existence of a torsion one-cycle implies also a torsion six-cycle because in general 
\begin{eqnarray}
{\rm Tor}_p(Y) \simeq {\rm Tor}_{D-p-1}(Y) 
\end{eqnarray}
with $D$ the real dimension of $Y$. This picture has relied on setting $s=1$ and is thus really valid away from the locus $a_4=0$.
Therefore all we can conclude is the existence of a 2-torsion element in $H^{1,1}(\hat{Y}_4, \mathbb{Z}) / \langle S \rangle_{\mathbb Z}$.

%%%%%%%%%%%%%%%%%%%%%%%%%%%%%%%%%%%%%%%%%%%%%%%%%%%%%%%%%%%%%%%%%%%%%%%%%%%%%%%%%%%%%%%%%%%%%%%%%%%%%%%%%%%%%%%%%%%%%%%%%%%%%%%%%%%%%%%%%%%%%%%
\subsection{An $(SU(2) \times SU(2))/\mathbb Z_2$-fibration} \label{sec_SU2SU2Z2}
%%%%%%%%%%%%%%%%%%%%%%%%%%%%%%%%%%%%%%%%%%%%%%%%%%%%%%%%%%%%%%%%%%%%%%%%%%%%%%%%%%%%%%%%%%%%%%%%%%%%%%%%%%%%%%%%%%%%%%%%%%%%%%%%%%%%%%%%%%%%%%%
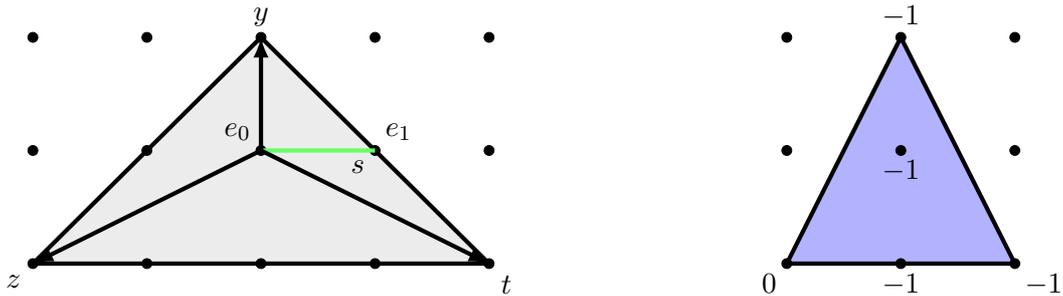
\begin{figure}
    \centering
  \begin{tikzpicture}[scale=1.5]
%%%% feft polygon %%%%
  \filldraw [ultra thick, draw=black, fill=lightgray!30!white]
      (-2,-1)--(2,-1)--(0,1)--cycle;
    \foreach \x in {-2,-1,...,2}{% Two indices running over each
      \foreach \y in {-1,0,...,1}{% node on the grid we have drawn 
        \node[draw,circle,inner sep=1.3pt,fill] at (\x,\y) {};
            % Places a dot at those points
      }
    }
          \filldraw [ultra thick, draw=green!60!white]
      (0,0)--(1,0)--(0,0)--cycle;
  \draw[ultra thick, -latex]
       (0,0) -- (0,1) node[above] {$y$};
  \node [below left] at (1,0) {$s$};
  \draw[ultra thick, -latex]
       (0,0) -- (2,-1) node[below right] {$t$};
  \draw[ultra thick, -latex]
       (0,0) -- (-2,-1) node[below left] {$z$};
  \node [above left] at (0,0) {$e_0$};
  \node [above right] at (1,0) {$e_1$};
%%%% right polygon %%%%
\begin{scope}[xshift=0.33\textwidth]
\filldraw [ultra thick, draw=black, fill=blue!30!white]
      (0,1)--(-1,-1)--(1,-1)--cycle;
    \foreach \x in {-1,0,...,1}{% Two indices running over each
      \foreach \y in {-1,0,...,1}{% node on the grid we have drawn 
        \node[draw,circle,inner sep=1.3pt,fill] at (\x,\y) {};
            % Places a dot at those points
      }
    }
    \node [below] at (0,0)  {$-1$};
  \node [above] at (0,1)  {$-1$};
  \node [below right] at (1,-1) {$-1$};
  \node [below] at (0,-1)  {$-1$};
  \node [below left] at(-1,-1)  {$0$};
\end{scope}
  \end{tikzpicture}
\caption{On the lefthand side the only possible $\mathfrak{su}(2)$-top over polygon 13 of \cite{Bouchard:2003bu} is depicted. The green color indicates the layer at height one, containing the nodes $e_0$ and $e_1$. On the righthand side we give the dual top, bounded from below by the values $z_{min}$, shown next to the nodes. }\label{fig:SU2-top-over_polygon_13}
\end{figure}
\noindent The analysis so far has treated all coefficients $a_i$ appearing in (\ref{eq:hse-Z2-fiber-singular}) as maximally generic.  We now further restrict the coefficients $a_i$ defining the $\mathbb Z_2$-torsional fibration in its singular form (\ref{eq:hse-Z2-fiber-singular}) or its resolution (\ref{eq:hse-Z2-fiber-2nd-bub})
such as to create additional non-abelian singularities in the fiber. %over divisors other than $\{a_4 =0\}$. 
A special class of such restrictions corresponds to specializations $a_i \rightarrow a_{i,j} w^j$ with $W: \{w=0\}$ a base divisor and  $a_{i,j}$ \emph{generic}. 
Since the fibration (\ref{eq:hse-Z2-fiber-singular}) is in global Tate form, the possible enhancements one can obtain via such specialisations can be conveniently determined via Tate's algorithm \cite{Bershadsky:1996nh,Katz:2011qp,Grassi:2011hq} as summarized \textit{e.g.}\ in Table 2 of \cite{Katz:2011qp}.
Another advantage of this class of enhancements is that the corresponding fibrations can be treated torically.  Indeed, the possible enhancements of type $a_i \rightarrow a_{i,j} w^j$ with generic  $a_{i,j}$ which admit a crepant resolution are classified by the 
  tops construction \cite{Candelas:1996su,Candelas:1997pv}, which provides both the possible vanishing patterns $a_{i,j}$ (coinciding with Tate's algorithm) and the toric resolution.
  For a detailed account of how to read off the vanishing orders from the toric data of a top in the present context we also refer to \cite{Borchmann:2013hta}.\\\\
 From the classification of tops by  
 Bouchard and Skarke~\cite{Bouchard:2003bu}  for the 16 hypersurface elliptic fibrations, we 
 note that the only tops possible for the fiber (\ref{eq:hse-Z2-fiber-2nd-bu})  correspond to singularity type $A_{2n+1}$ for $n \geq 0$, $C_n$ and $D_{2n+4}$ for $n \geq 1$, $B_3$ and $E_7$.
 This is indeed in agreement with an analysis via Tate's algorithm as a consequence of $a_3 \equiv 0$ and $a_6 \equiv 0$. 
 The associated gauge algebras have the property that their universal cover groups have a center with a $\mathbb Z_2$-subgroup. Indeed, as we will exemplify below, in all models of this type the Mordell-Weil torsion $\mathbb Z_2$ will be identified with this  $\mathbb Z_2$-subgroup of the center.

 % $A_n$-type tops possible for the fiber (\ref{eq:hse-Z2-fiber-2nd-bu})  are those with $n=2\,i+1$ and $i\in \mathbb N_0$. 
 % This, in fact, follows also straightforwardly from Tate's algorithm applied to the fibration (\ref{eq:hse-Z2-fiber-singular}) as a consequence of $a_3 \equiv 0$.
 % In view of the general relation of the $\mathbb Z_2$ torsion and the global structure of the gauge group
  % the underlying reason is that the total gauge group is forced to be of the form $G_0/\mathbb Z_2$, which excludes in $G_0$ any factor of $SU(n+1)$ with $n$ even.\\\\
 % must have a common $\mathbb Z_2$ factor.
\noindent  To verify this pattern explicitly we begin with an $A_1$ top, corresponding to an affine $\mathfrak{su}(2)$-type fiber over a divisor $W: w=0$ on ${\cal B}$. There is, in fact, only one possible $A_1$ top over this polygon, see Figure \ref{fig:SU2-top-over_polygon_13}. The singular version of the associated fibration is obtained by replacing in (\ref{eq:hse-Z2-fiber-singular}) $a_4$ by $a_{4,1} w$. 
 The discriminant of this fibration,
 \begin{equation}
\Delta \sim w^2 a_{4,1}^2\big((a_1^2-4a_2)^2-64w a_{4,1}\big),
\end{equation}
reflects the gauge algebra $\mathfrak{su}(2) \oplus \mathfrak{su}(2)$.\\\\
The toric resolution of this fibration is described by the hypersurface equation
\begin{equation}\label{eq:SU(2)_top_polygon_13}
\hat P = sy^2 + a_1 s t y z -e_1s^2t^4-a_2 st^2z^2 - a_{4,1}e_0z^4 \,,
\end{equation}
corresponding to the reflexive pair in Figure \ref{fig:SU2-top-over_polygon_13}
(again after scaling $x$ to one, since $X$ does not intersect the hypersurface). For definiteness we choose a triangulation with Stanley-Reisner ideal
\begin{equation}
\{ sz, tz, ty,e_0s, e_1z\}.
 \end{equation}
The extra $\mathfrak{su}(2)$-fiber is found over $W:\{w=0\}$ with $\pi^*w = e_0 e_1$. 
Indeed, over $W$ the two fiber components $\mathbb P^1_0$ and $\mathbb P^1_1$ are given by the intersection of the ambient divisors $E_0: \{e_0=0\}$ and  $E_1: \{e_1=0\}$ with the hypersurface equation and two generic divisors in the base,
\bea
\mathbb P^1_i = E_i \cap \hat P|_{e_i=0} \cap D_a \cap D_b, \qquad i=0,1.
\eea
They intersect as the affine $\mathfrak{su}(2)$ Dynkin diagram. \\\\
The discriminant also suggests three codimension-two enhancement loci, at $W \cap \{a_{4,1} = 0\}$,  $W \cap \{a_1^2 = 4a_2\}$ and $\{a_{4,1} = 0\} \cap \{a_1^2 = 4a_2\}$. Splitting of fiber components only occurs over the first one\footnote{The other two loci are completely analogous to the curve $\{a_4=0\} \cap \{a_1^2 = 4a_2\}$ analysed in the previous section, where no splitting of the fiber was found despite an enhancement of the vanishing order of the discriminant.}, where $\mathbb P^1_1$ factors into the two components
\bea \label{P11su2}
\mathbb P^1_{1s} &=& E_1 \cap \{s=0\}  \cap \{a_{4,1}=0\}  \cap D_a \cap D_b, \\
\mathbb P^1_{1A} &=& E_1 \cap (y+\frac{1}{2}a_1t \pm t\sqrt{\frac{a_1^2}{4}-a_2} )(y+\frac{1}{2}a_1t \mp t\sqrt{\frac{a_1^2}{4}-a_2} ) = 0\} \cap \{a_{4,1}=0\}  \cap D_a \cap D_b. \nonumber
\eea
%\eea
%\begin{equation} \label{P11su2}
%\mathbb P^1_1 = E_1 \cap \{s(y+\frac{1}{2}a_1t \pm t\sqrt{\frac{a_1^2}{4}-a_2} )(y+\frac{1}{2}a_1t \mp t\sqrt{\frac{a_1^2}{4}-a_2} )\} \cap D_a \cap D_b
%\end{equation}
Note that the two factors in brackets appearing in $\mathbb P^1_{1A}$ get exchanged when the sign of the square root changes across a branch cut on ${\cal B}$
so that $\mathbb P^1_{1A}$ really describes a single $\mathbb P^1$.
% and we denote the corresponding fiber components as $\mathbb P^1_{1\pm}$. 
The weight 
\begin{equation}
\mathbb P^1_{1A} \cdot (E_1,S)=(-1,1)
\end{equation}
is in the weight system of the $(\mathbf{2}, \mathbf{2})$ of $\mathfrak{su}(2) \oplus \mathfrak{su}(2)$. This implies massless matter in the $(\mathbf{2}, \mathbf{2})$ representation over  $W \cap \{a_{4,1} = 0\}$. Again, no fundamental matter $({\bf 1},{\bf 2})$ or $({\bf 2},{\bf 1})$ is found. \\\\
Our  derivation of the extra coweight induced by the torsional section $T:\{t=0\}$ is only mildly modified by the extra $\mathfrak{su}(2)$ singularity compared to the previous section.
 The Shioda map $\Sigma_2$ of $T$ takes the form
\begin{equation}
\Sigma_2 = T-Z-\bar{\mathcal{K}} + \frac{1}{2}(S+E_1),
\end{equation}
which is trivial on the hypersurface since the divisor class
\begin{equation}
X = 2Z - S - 2T + 2\bar{\mathcal{K}} - E_1
\end{equation}
does not intersect \eqref{eq:SU(2)_top_polygon_13} due to the Stanley-Reisner ideal. The extra coweight is associated with the class
\begin{equation}
\Xi_2 \equiv T-Z-\bar{\mathcal{K}} = -\frac{1}{2}(S+E_1),
\end{equation}
which is torsion in $H^{1,1}(\hat Y_4, \mathbb Z)/\langle S, E_1\rangle_\mathbb Z$ and manifestly integral on the split curves over $W \cap \{a_{4,1} = 0\}$. This explains why the bifundamental representation is indeed present, whereas fundamental representations of the form $({\bf 1},{\bf 2})$ or $({\bf 2},{\bf 1})$, which for group theoretic reasons would have fractional pairing with the coweight $\Xi_2$, are not possible. \\\\
This refinement of the coweight lattice makes the gauge group non-simply connected and the gauge group is 
\begin{equation}\label{eq:su2_top_polygon_13_gauge_group}
G = \frac{SU(2)\times SU(2)}{\mathbb Z_2}.
\end{equation}
An example of this type was also given in \cite{Morrison:2014era}.

\begin{figure}[h!]
    \centering
  \begin{tikzpicture}[scale=1.5]
%%%% feft polygon %%%%
  \filldraw [ultra thick, draw=black, fill=lightgray!30!white]
      (-2,-1)--(2,-1)--(0,1)--cycle;
    \foreach \x in {-2,-1,...,2}{% Two indices running over each
      \foreach \y in {-1,0,...,1}{% node on the grid we have drawn 
        \node[draw,circle,inner sep=1.3pt,fill] at (\x,\y) {};
            % Places a dot at those points
      }
    }
          \filldraw [ultra thick, draw=black, fill=green, opacity=0.6]
      (0,0)--(2,0)--(1,1)--cycle;
  \draw[ultra thick, -latex]
       (0,0) -- (0,1) node[above] {$y$};
  \node [below left] at (1,0) {$s$};
  \draw[ultra thick, -latex]
       (0,0) -- (2,-1) node[below right] {$t$};
  \draw[ultra thick, -latex]
       (0,0) -- (-2,-1) node[below left] {$z$};
  \node [above left] at (0,0) {$e_0$};
  \node [above] at (1,1) {$e_1$};
  \node [right] at (2,0) {$e_2$};
  \node [above right] at (1,0) {$e_3$};
%%%% right polygon %%%%
\begin{scope}[xshift=0.33\textwidth]
\filldraw [ultra thick, draw=black, fill=blue!30!white]
      (0,1)--(-1,-1)--(1,-1)--cycle;
    \foreach \x in {-1,0,...,1}{% Two indices running over each
      \foreach \y in {-1,0,...,1}{% node on the grid we have drawn 
        \node[draw,circle,inner sep=1.3pt,fill] at (\x,\y) {};
            % Places a dot at those points
      }
    }
    \node [below] at (0,0)  {$-1$};
  \node [above] at (0,1)  {$-1$};
  \node [below right] at (1,-1) {$-1$};
  \node [below] at (0,-1)  {$0$};
  \node [below left] at(-1,-1)  {$1$};
\end{scope}
  \end{tikzpicture}
\caption{The lefthand side shows an $\mathfrak{su}(4)$-top over polygon 13 of \cite{Bouchard:2003bu}. The green layer contains the points at height one. On the righthand side we depict the dual top, bounded from below by the values $z_{min}$, shown next to the nodes. }\label{fig:SU4-top-over_polygon_13}
\end{figure}
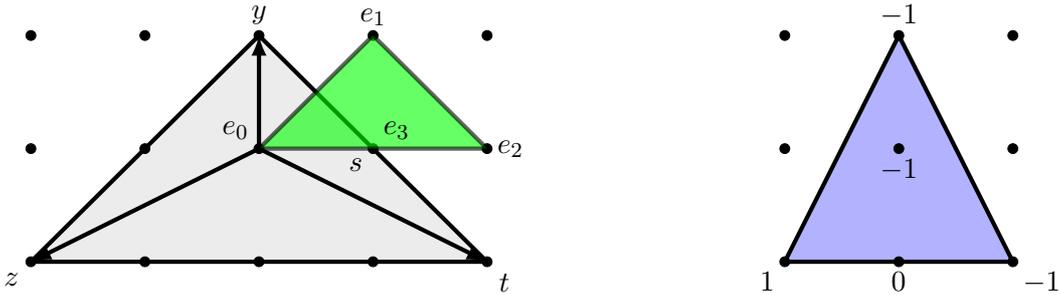
%%%%%%%%%%%%%%%%%%%%%%%%%%%%%%%%%%%%%%%%%%%%%%%%%%%%%%%%%%%%%%%%%%%%%%%%%%%%%%%%%%%%%%%%%%%%%%%%%%%%%%%%%%%%%%%%%%%%%%%%%%%%%%%%%%%%%%%%%%%%%%%
\subsection{An $(SU(4) \times SU(2))/\mathbb Z_2$-fibration} \label{SU4SU2Z2}
%%%%%%%%%%%%%%%%%%%%%%%%%%%%%%%%%%%%%%%%%%%%%%%%%%%%%%%%%%%%%%%%%%%%%%%%%%%%%%%%%%%%%%%%%%%%%%%%%%%%%%%%%%%%%%%%%%%%%%%%%%%%%%%%%%%%%%%%%%%%%%%
In this section we consider the next example in the $A$-series \cite{Bouchard:2003bu}, corresponding to an affine $\mathfrak{su}(4)$-type fiber. This construction yields the unique top of Figure~\ref{fig:SU4-top-over_polygon_13} associated with the hypersurface equation
\begin{equation}\label{eq:hse-SU4-case}
\hat P = -y^2\,s\,e_1 - a_1 \,y\,z\,s\,t + s^2\,t^4\,e_2^2\,e_3+a_{2,1}\,z^2\,s\,t^2\,e_0\,e_2\,e_3 + a_{4,2}\,z^4\,e_0^2\,e_3\,.
\end{equation}
The pullback of the projection of the fibration obeys $e_0 e_1 e_2 e_3  = \pi^* \,w$, defining an affine $\mathfrak{su}(4)$ fiber over $W: \{w=0\}$ in the base. From one of the 16 triangulations of this top we obtain the Stanley-Reisner ideal
\begin{equation}\label{eq:SR-ideal_for_SU4}
 \textmd{SR-i}:\quad \{y\,t,\, y\,e_0,\, y\,e_2,\, y\,e_3,\, s\,z,\, s\,e_0,\, s\,e_2,\, s\,e_3,\, z\,e_2,\, z\,e_3,\, e_0\,e_2,\, t\,z\,e_1,\, t\,e_0\,e_1,\, t\,e_1\,e_3\}\,.
\end{equation}
%\subsubsection{codimension-one}
%In addition to the $A_1$ singularity, with the resolution divisor  $\{s=0\}$ we now have a fiber degeneration over $\{w=0\}$ with the irreducible components
%\begin{equation}
%\mathbb P^1_i = \{e_i = 0\} \cap \{P_W|_{e_i = 0}\} \cap D_a \cap D_b
%\end{equation}
%for $D_a$, $D_b$ generic divisors in the base $\mathcal{B}$. The intersection form of these components are that of an affine $A_3$ Dynkin diagram. 
%%%%%%%%%%%%%%%%%%%%%%%%%%%%%%%%%%%%%%%%%%%%%%%%%%%%%%%%%%%%%%%%%%%%%%%%%%%%%%%%%%%%%%%%%%%%%%%%%%%%%%%%%%%%%%%%%%%%%%%%%%%%%%%%%%%%%%%%%%%%%%%
%\subsubsection{$\mathfrak{su}(4)$-divisor}
%%%%%%%%%%%%%%%%%%%%%%%%%%%%%%%%%%%%%%%%%%%%%%%%%%%%%%%%%%%%%%%%%%%%%%%%%%%%%%%%%%%%%%%%%%%%%%%%%%%%%%%%%%%%%%%%%%%%%%%%%%%%%%%%%%%%%%%%%%%%%%%
The three exceptional divisors $e_1$, $e_2$, $e_3$ and the part of the original fiber $e_0$ are all fibered over $\{w=0\}$ with fiber components 
\begin{equation}
\mathbb P^1_i = \{E_i\} \cap \{\hat P|_{e_i = 0} = 0\} \cap D_a \cap D_b\quad i = 0, \dots, 3.
\end{equation}
The explicit equations are provided in appendix \ref{app:su4poly13}. The irreducible fiber components intersect like the nodes of the affine Dynkin diagram of $\mathfrak{su}(4)$ type. This is also seen in Figure \ref{fig:SU4-top-over_polygon_13}, where the upper layer reproduces this structure by construction. 
%%%%%%%%%%%%%%%%%%%%%%%%%%%%%%%%%%%%%%%%%%%%%%%%%%%%%%%%%%%%%%%%%%%%%%%%%%%%%%%%%%%%%%%%%%%%%%%%%%%%%%%%%%%%%%%%%%%%%%%%%%%%%%%%%%%%%%%%%%%%%%%
%\subsubsection{Matter curves}
%%%%%%%%%%%%%%%%%%%%%%%%%%%%%%%%%%%%%%%%%%%%%%%%%%%%%%%%%%%%%%%%%%%%%%%%%%%%%%%%%%%%%%%%%%%%%%%%%%%%%%%%%%%%%%%%%%%%%%%%%%%%%%%%%%%%%%%%%%%%%%%
To analyze the localised charged matter we infer from  the discriminant of \eqref{eq:hse-SU4-case},
\begin{equation}
 \Delta= 16\, w^4\,  a_{4,2}^2 \left(\left(4\, w\, a_{2,1}+a_1^2\right){}^2-64\, w^2\, a_{4,2}\right)\,,
\end{equation}
%Recall that $\pi^* w=\prod_i e_i$ with $\pi$ the projection of the fibration. From the discriminant we find  
the codimension-two enhancement loci\footnote{All other enhancement loci as read off from the discriminant do not correspond to an extra fiber splitting.}
\begin{equation}
 \{w=a_{4,2}=0\}\qquad\textmd{and}\qquad \{w=a_1=0\}\,.
\end{equation}
The factorization properties of the fiber components (see appendix \ref{app:su4poly13})  identify the split curves in the fiber. At $\{w=a_{4,2}=0\}$ the component $\mathbb P^1_1$ splits into three components, whose intersection numbers with the exceptional divisors from the $\mathfrak{su}(4)$ and $\mathfrak{su}(2)$ singularities are
\begin{equation}
 \begin{aligned}
  &\mathbb P^1_{e_1=s=0} \cdot (E_1,E_2,E_3)=(0,0,0)\,,   & \mathbb P^1_{e_1=s=0} \cdot (S)=(-2)\,,\\
  &\mathbb P^1_{e_1=t=0} \cdot (E_1,E_2,E_3)=(-1,1,0)\,,  & \mathbb P^1_{e_1=t=0} \cdot (S)=(1)\,,\\
  &\mathbb P^1_{e_1=R1=0} \cdot (E_1,E_2,E_3)=(-1,0,0)\,, & \mathbb P^1_{e_1=R1=0} \cdot (S)=(1)\,,\\
 \end{aligned}
\end{equation}
respectively. The $(-1,1,0)$ and $(-1,0,0)$ are weights in the fundamental of $\mathfrak{su}(4)$ and from the right column we find the weights of the fundamental representation of $\mathfrak{su}(2)$ (which is the same as the anti-fundamental). Indeed the full weight system is reproduced by taking linear combinations of fibral curves. Hence the charged matter at this locus transforms in representation $(\mathbf{4},\mathbf{2})$ of $\mathfrak{su}(4) \oplus \mathfrak{su}(2)$.\\ \\
Over $\{w=a_1=0\}$ the relevant intersections are
\begin{equation}
 \begin{aligned}
  &\mathbb P^1_{e_1=e_3=0} \cdot (E_1,E_2,E_3)=(0,1,-2)\,,        & \mathbb P^1_{e_1=e_3=0} \cdot (S)=(0)\,,\\
  &\mathbb P^1_{e_1=R2_1=0} \cdot (E_1,E_2,E_3)=(-1,0,1)\,,         & \mathbb P^1_{e_1=R2_1=0} \cdot (S)=(0)\,,\\
  &\mathbb P^1_{e_1=R2_2=0} \cdot (E_1,E_2,E_3)=(-1,0,1)\,,         & \mathbb P^1_{e_1=R2_2=0} \cdot (S)=(0)\,,
 \end{aligned}
\end{equation}
where $(-1,0,1)$ is  one of the weights in the $\mathbf 6$-representation of $\mathfrak{su}(4)$. States originating from these curves are uncharged under $\mathfrak{su}(2)$. This is as expected since this locus is away from the $\mathfrak{su}(2)$ divisor $\{a_{4,2} = 0\}$. The following table summarizes the matter spectrum:
\begin{equation}
 \begin{array}{cc}
  \multicolumn{2}{c}{\textit{Top over polygon 13: } \mathfrak{su}(4) \times \mathfrak{su}(2)} \\ \\
 \textit{Locus} & \textit{Charged matter}\tabularnewline 
w \cap a_{4,2} &  (\mathbf{4},\mathbf{2})  \tabularnewline
w \cap a_1 &  (\mathbf{6},\mathbf{1})  \tabularnewline
\end{array} \,.
\end{equation}
Again we stress the absence of fundamental representations. The Shioda-type Ansatz for the toric divisor class $T$ yields
\begin{equation}
\Sigma_2 = T - Z - \bar{\mathcal{K}} + \frac{1}{2}\left( S+ E_1 + 2 E_2 + E_3 \right) \, ,
\end{equation}
which for the same reasons as before turns out to be trivial in $H^{1,1}(Y_4, \mathbb R)$.
% Giving the fibration as a hypersurface in an ambient toric space, the toric divisor $X$ do not intersect the hypersurface. The class of $X$ is given by
% \begin{equation}
% X= 2Z - S - 2T + 2 \bar{\mathcal{K}} - E_1 - 2E_2 - E_3
% \end{equation}
% and we see that
% \begin{equation}
% -2 \Sigma_2 = X
% \end{equation}
% and since $X$ is trivial on the hypersurface, so is $\Sigma_2$ in $H^{1,1}(Y_4, \mathbb R)$. Therefore $\Sigma_2$ is not giving rise to an extra $U(1)$ symmetry. Moreover, this triviality implies
The coweight element
\begin{equation}
\Xi_2 = T-Z- \bar{\mathcal{K}} = -\frac{1}{2} ( S+ E_1 + 2 E_2 + E_3),
\end{equation}
which is 2-torsion in $H^{1,1}(\hat{Y}_4, \mathbb{Z}) / \langle S,E_1,E_2,E_3 \rangle_{\mathbb Z}$, forces  the weight lattice to be coarser in order to preserve the integer pairing of coweights and weights. Indeed, the intersection of $\Xi_2$ with all split curves corresponding to weights of the matter representations is integer, and representations such as $({\bf 4},1)$ or $(1,{\bf 2})$ which would violate this integral pairing are absent. This identifies the global gauge group as
\bea
G = \frac{SU(4) \times SU(2)}{\mathbb Z_2}.
\eea

%%%%%%%%%%%%%%%%%%%%%%%%%%%%%%%%%%%%%%%%%%%%%%%%%%%%%%%%%%%%%%%%%%%%%%%%%%%%%%%%%%%%%%%%%%%%%%%%%%%%%%%%%%%%%%%%%%%%%%%%%%%%%%%%%%%%%%%%%%%%%%%
\subsection{A $(Spin(7) \times SU(2))/\mathbb Z_2$-fibration}

\begin{figure}
    \centering
  \begin{tikzpicture}[scale=1.5]
%%%% feft polygon %%%%
  \filldraw [ultra thick, draw=black, fill=lightgray!30!white]
      (-2,-1)--(2,-1)--(0,1)--cycle;
    \foreach \x in {-2,-1,...,2}{% Two indices running over each
      \foreach \y in {-1,0,...,1}{% node on the grid we have drawn 
        \node[draw,circle,inner sep=1.3pt,fill] at (\x,\y) {};
            % Places a dot at those points
      }
    }
          \filldraw [ultra thick, draw=black, fill=green, opacity=0.6]
      (0,0)--(2,0)--(1,1)--cycle;
  \draw[ultra thick, -latex]
       (0,0) -- (0,1) node[above] {$y$};
  \node [below left] at (1,0) {$s$};
  \draw[ultra thick, -latex]
       (0,0) -- (2,-1) node[below right] {$t$};
  \draw[ultra thick, -latex]
       (0,0) -- (-2,-1) node[below left] {$z$};
       
  \draw[dashed]
  	(0,0) -- (2,1);
\draw[dashed]
(2,-1) -- (2,1);
\draw[dashed]
(0,1) -- (2,1);
  \node [above left] at (0,0) {$e_0$};
  \node [above] at (1,1) {$e_3$};
  \node [right] at (2,0) {$e_1$};
  \node [right] at (2,1) {$e_2$};
%%%% right polygon %%%%
\begin{scope}[xshift=0.33\textwidth]
\filldraw [ultra thick, draw=black, fill=blue!30!white]
      (0,1)--(-1,-1)--(1,-1)--cycle;
    \foreach \x in {-1,0,...,1}{% Two indices running over each
      \foreach \y in {-1,0,...,1}{% node on the grid we have drawn 
        \node[draw,circle,inner sep=1.3pt,fill] at (\x,\y) {};
            % Places a dot at those points
      }
    }
    \node [below] at (0,0)  {$-\frac{1}{2}$};
  \node [above] at (0,1)  {$-1$};
  \node [below right] at (1,-1) {$-1$};
  \node [below] at (0,-1)  {$0$};
  \node [below left] at(-1,-1)  {$1$};
\end{scope}
  \end{tikzpicture}
\caption{The lefthand side shows the unique $B_3$-top over polygon 13 of \cite{Bouchard:2003bu}. The green layer contains the points at height one and the node labelled $e_2$ is at height two. On the right side we depict the dual top, bounded from below by the values $z_{min}$, shown next to the nodes. }\label{fig:B3-top-over_polygon_13}
\end{figure}
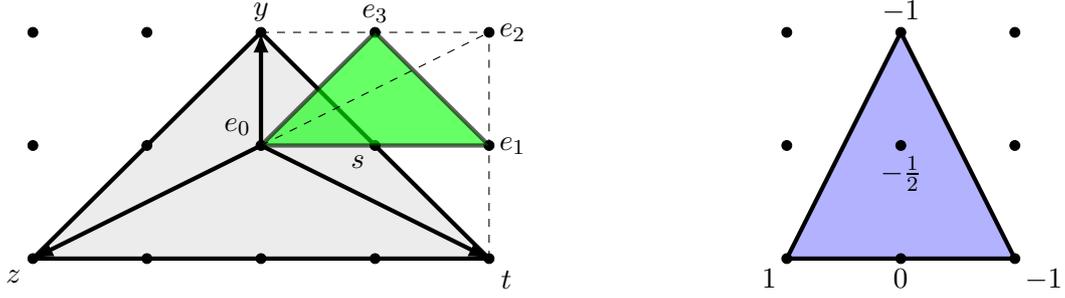

Keeping the same ambient fiber space as in previous section, we now consider a top corresponding to the non-simply laced Lie algebra $B_3$. The top is constructed uniquely from the classification \cite{Bouchard:2003bu} and the corresponding hypersurface equation is
\begin{equation}
\hat P = e_1^2 s^2 t^4 + e_3 s y^2 + a_1 e_0 e_1 e_2 e_3 s t y z +  a_2 e_0 e_1 s t^2 z^2 + a_4 e_0^2 z^4 \,.
\end{equation}
Having a node at $z=2$ in the top defines the divisor $W = \{w = 0\}$ in the base with $\pi^* w = e_0 e_1 e_2^2 e_3$ and gives multiplicity $2$ to the corresponding curve $\mathbb P^1_2$ in the fiber over $W$. The occurrence of the multiplicity of the node in the projection to the base is crucial to make $W$ scale under the scaling relations coming from the $z \geq 1$ layers of the top.  The affine $B_3$ Dynkin diagram is read off along the edges at $z\geq1$ of the top in Figure \ref{fig:B3-top-over_polygon_13}. The non-simply laced structure of this algebra is reflected in the fact that the intersection of the ambient divisor $E_3$ with the hypersurface
\begin{equation}
 E_3 \cap \hat P|_{e_3 = 0} \cap D_a \cap D_b, \qquad D_{a,b} \subset \mathcal{B}
\end{equation}
gives rise to two curves. These are described by the factorization
\begin{equation}
\begin{aligned}
 \{e_3 = 0\} &\cap \{ s^2 + a_2 e_0 s + a_4 e_0^2 = 0\} \qquad \Leftrightarrow \\
 \{e_3 = 0\} &\cap \{ (s + \frac{1}{2} a_2 e_0 \pm e_0\sqrt{\frac{a_2^2}{4}-a_4})(s + \frac{1}{2} a_2 e_0 \mp e_0\sqrt{\frac{a_2^2}{4}-a_4})= 0\} \,.
\end{aligned}
\end{equation}
The two factors on the righthand side give rise to the curves $\mathbb P^1_{3\pm}$ and they get exchanged when the signs of the square roots shift upon travelling along $W$ in the base. As a check, the negative of the Cartan matrix $C_{ij}$ of $B_3$ is reproduced as the intersection numbers
\begin{equation}
E_i \cdot (\mathbb P^1_0, \mathbb P^1_1,\mathbb P^1_2,\mathbb P^1_{3\pm} )_j = -C_{ij} \, .
\end{equation}
By analyzing the codimension-two loci, we find only one curve in $\mathcal{B}$ over which the fiber degenerates further. This happens over $W \cap \{a_4 = 0\}$ and by calculating the charges of the split fiber components weights in the weight system of the $(\mathbf{8}, \mathbf{2})$ of $\mathfrak{so}(7) \oplus \mathfrak{su}(2)$ are found, where $\mathbf{8}$ is the spinor representation. \\ \\
Using that the toric divisor $X$ does not restrict to the hypersurface the Shioda map of the torsional section gives a class
\begin{equation}
\Xi_2 \equiv T -Z - \bar{\mathcal{K}} = -\frac{1}{2}(S + 2E_1 +2E_2 + E_3)
\end{equation}
with integer intersection with all fiber components over the matter curve. 

\noindent Consistently with the appearance of the representation $(\mathbf{8}, \mathbf{2})$ of $\mathfrak{so}(7) \oplus \mathfrak{su}(2)$    the gauge group is 
\begin{equation}
G = \frac{Spin(7)\times SU(2)}{\mathbb Z_2},
\end{equation}
where the $\mathbb Z_2$ is the common center of $Spin(7)$ and $SU(2)$ and $\pi_1(G) = \mathbb Z_2$. Even though not realized in this geometry, all representations $(\mathbf{8}, \mathbf{R_e})$ for $\mathbf{R}_e$ an even-dimensional representation of $SU(2)$ would also be allowed, and also the representations $(\mathbf{7}, \mathbf{R}_o)$ for $\mathbf{R}_o$ an odd-dimensional representation of $SU(2)$.

\subsection{Generalisation to $Sp(n)/\mathbb Z_2$, $SU(2n)/\mathbb Z_2$, $Spin(4n) /\mathbb Z_2$, Type IIB limit and restricted monodromies} \label{sec_Generalisation}

The toric enhancements described in the previous sections involved the specialization $a_4 \rightarrow a_{4,n} w^n$ for $W: \{w=0\}$ some divisor different from the $A_1$-locus $\{a_4=0\}$.  
Clearly one can also identify $w$ with $a_4$, thereby producing a single gauge group factor. According to the general discussion, this single group factor will be strongly constrained by the requirement that the universal cover gauge group $G_0$ contain a $\mathbb Z_2$-subgroup in its center.

\noindent Indeed, if $\bar{{\cal K}}_{\cal B}^{4/n}$ exists as a line bundle with non-trivial sections, we can simply factorise
\bea
a_4 = (\tilde a_4)^n
\eea
with $\tilde a_4 \in H^0({\cal B}, \bar {\cal K}_{\cal B}^{4/n})$. 
Since the analysis of the singular geometry and its resolution has been exemplified in detail in the previous sections, we content ourselves with determining the resulting gauge groups by application of Tate's algorithm \cite{Bershadsky:1996nh,Katz:2011qp} without explicitly constructing the resolution.
%To this end, note first that by a coordinate transformation one can always set $a_1 \equiv 0$ in the defining equation (\ref{eq:hse-Z2-fiber-singular}).
For generic $a_2$, Tate's algorithm in the form of Table 2 of \cite{Katz:2011qp} indicates that the fiber over $\tilde a_4=0$ is of Kodaira type  $I_{2n}^{ns}$, with the superscript denoting the \emph{non-split} type.
The associated gauge algebra is the rank $n$ Lie algebra $\mathfrak{sp}(n)$ (with the convention that $\mathfrak{sp}(1) \simeq \mathfrak{su}(2)$). 
This identifies the gauge group as
\bea
G = \frac{Sp(n)}{\mathbb Z_2}.
\eea
As described in subsection \ref{subsec_sing}, if $n=1$ the global structure of $G$ makes extra massless representations along the curve $\{\tilde a_4 = 0\} \cap \{\frac{1}{4} a_1^2 + a_2 =0\}$ impossible; this is no longer true for $n \geq 2$.
Indeed, in this case Tate's algorithm predicts, as described in detail in \cite{Grassi:2011hq}, for the fiber type over this curve Kodaira type $I^{*s}_{2n-4}$ (with the superscript standing for \emph{split} type), corresponding to gauge algebra $\mathfrak{so}(4n)$. 
From the branching rule of the adjoint of $SO(4n)$ along  $SO(4n) \rightarrow SU(2n) \times U(1) \rightarrow Sp(n) \times U(1)$ one deduces matter in the 2-index antisymmetric representation of $Sp(n)$ of dimension $2 n^2 - n -1$ along $\{\tilde a_4 = 0\} \cap \{\frac{1}{4} a_1^2 + a_2 =0\}$ (see in particular Table 9 of  \cite{Grassi:2011hq}). This is compatible with the gauge group $G = \frac{Sp(n)}{\mathbb Z_2}$. \\\\
Next, one can engineer a gauge algebra $\mathfrak{su}(2n)$ by factoring $a_4 = (\tilde a_4)^n$ and in addition restricting $a_2 = a_{2,1} \tilde a_4$ for suitable $a_{2,1} \in H^0({\cal B}, \bar {\cal K}_{\cal B}^{2 - 4/n} )$ (if existent). In this case the gauge group is 
\bea
G = \frac{SU(2n)}{\mathbb Z_2}, \qquad \quad n \geq 2.
\eea
Note that the Mordell-Weil torsion group $\mathbb Z_2$ appears here as a proper subgroup of the center $\mathbb Z_{2n}$ of the universal cover $G_0 = SU(2n)$.
The same argument as above predicts massless matter  in the antisymmetric representation of $SU(2n)$ localised on the curve $\{\tilde a_4 = 0\} \cap \{ a_{1}=0\}$.
The appearance of this matter distinguishes $G = SU(2n) / {\mathbb Z}_2$ as realized here from $SU(2n)/{\mathbb Z}_{2n}$. The possibility that the Mordell-Weil torsion appears as a proper subgroup of the center of the universal cover $G_0$
had previously been noted in eight-dimensional F-theory compactifications on K3 in \cite{Aspinwall:1998xj, Ganor:1996pc}. \\

\noindent The only remaining chain of enhancements of this type which is possible according to Tate's algorithm leads to gauge algebra $\mathfrak{so}(4n)$ with  $n \geq 4$ and corresponds to
 $a_4 = (\tilde a_4)^n$, $a_2 = a_{2,1} \tilde a_4$ and $a_1 = a_{1,1} \tilde a_4$. The restriction to $n \geq 4$ comes about as a necessary condition for a section $a_1 \in H^0({\cal B},\bar {\cal K}_{\cal B}^{1 - 4/n}) $  to exist. According to the analysis in \cite{Grassi:2011hq} we expect matter in the vector representation along the curve $\{\tilde a_4=0\} \cap \{ a_{2,1} =0 \}$. Note that the universal cover group $G_0=Spin(4n)$ has center $\mathbb Z_2 
\times \mathbb Z_2$. The appearance of the vector representation (but not the spinor) is in perfect agreement with the gauge group being
\bea
G = \frac{Spin(4n)}{\mathbb Z_2} = SO(4n), \qquad \quad n \geq 4 .
\eea
\noindent The observed pattern has a natural interpretation in the weak coupling Type IIB orientifold limit.
This Sen limit \cite{Sen:1997gv} is realized as the limit $\epsilon \rightarrow 0$ after rescaling $a_3 \rightarrow  \epsilon \, a_3$, $a_4 \rightarrow \epsilon \, a_4$, $a_6 \rightarrow \epsilon^2 \, a_6$ \cite{Donagi:2009ra}. 
The discriminant locus can be brought into the form
\bea \label{Deltaepsilon}
\Delta \simeq  \epsilon^2 h^2 (\eta^2 - h \chi) + {\cal O}(\epsilon^3),
\eea
and the Type IIB  Calabi-Yau 
\bea
 X_\textmd{IIB}: \xi^2 = h
\eea
is a double cover of the F-theory base ${\cal B}$  branched over the orientifold plane  localised at $h=0$. The orientifold action on $ X_\textmd{IIB}$ acts as $\xi \rightarrow - \xi$. The locus $\eta^2 - h \chi = 0$ on ${\cal B}$ and its uplift to the Calabi-Yau double cover  $ X_\textmd{IIB}$ represents the D7-brane locus.
In the configuration at hand, due to the restriction $ a_3 \equiv 0$ and $a_6 \equiv 0$, one finds
\bea
 h=  - \frac{1}{12} ( a_1^2 + 4 a_2), \qquad \chi =0, \qquad \eta = a_4 = (\tilde a_4)^n.
\eea
For generic $a_2$ the D7-brane system is given by a stack of D7-branes on the uplift of the divisor $\{ a_4=0 \}$ to the double cover $ X_\textmd{IIB}$; since this locus is invariant under the orientifold projection, the D7-brane stack supports gauge algebra $\mathfrak{sp}(n)$. The antisymmetric matter appears at the intersection with the O7-plane at $h=0$.
If $a_2 = a_{2,1} \tilde a_4$, then the analysis of \cite{oai:arXiv.org:1202.3138} shows that the D7-branes wrap a divisor on the Calabi-Yau double cover which is not mapped to itself under the orientifold action. Its corresponding non-abelian gauge algebra is therefore indeed $\mathfrak{su}(n)$ with antisymmetric matter at the intersection of the  D7-brane stack with its image  on top of the O7-plane.
For completeness, note that the further specialization  $a_1 = a_{1,1} \tilde a_4$, corresponding to the $Spin(4n)/\mathbb Z_2$ series in F-theory, has an ill-defined weak-coupling limit with two O7-planes  intersecting over a curve of conifold singularities. 
 \\\\
Apart from reproducing the F-theory predictions, this weak coupling analysis exemplifies how the global structure of the gauge group in the Type IIB limit can be understood from the specific D7-brane configuration and the absence (or presence) of certain matter representations. In the situation under consideration, what changes the gauge group from $Sp(n)$ or $SU(2n)$ to $Sp(n)/\mathbb Z_2$ and $SU(2n)/\mathbb Z_2$ is that in the discriminant (\ref{Deltaepsilon})  no extra single D7-brane arises in addition to the non-abelian brane stack at $\{\tilde a_4 = 0\}$; if present the intersection curve of such a brane with the D7-brane stack would lead to matter in the fundamental representation of $Sp(n)$ or $SU(2n)$ and thus change the global structure of the gauge group. \\\\
Finally, let us point out that the elliptic fibration (\ref{eq:hse-Z2-fiber-2nd-bub}) with $s\equiv 1$, \textit{i.e.}\ the singular model corresponding to the blow-down of the $A_1$-fiber at $\{a_4 =0\}$, was considered in \cite{Berglund:1998va} from a related, but slightly different perspective: In this work it was shown that this class of elliptic fibrations does not exhaust the full $SL(2,\mathbb Z)$ monodromy group, but only the subgroup $\Gamma_0(2) \subset SL(2,\mathbb Z)$.\footnote{Recall that $\Gamma_0(k)$ is defined as the subgroup of $SL(2,\mathbb Z)$-matrices $\begin{pmatrix}
a & b \\
c & d
\end{pmatrix}
$
with $c\equiv0 \quad {\rm mod} \,\,  k$.} In fact, restricted $\Gamma_0(k)$-monodromy is a consequence of the existence of an order $k$ point on the elliptic fiber \cite{Berglund:1998va}, which, in the language of our analysis, is equivalent to Mordell-Weil $k$-torsion.
There are a number of geometric consequences of this \cite{Silverman:2008}.  
For example, the 
modular curve $\mathfrak{h}/\Gamma_0(2)$ has {\em two}\/ ``cusp'' points
at which $j=\infty$, corresponding to the two irreducible factors $a_4$ and
$\left(4a_4-(a_2+\frac14a_1^2)^2\right)$ of the discriminant \eqref{discr-su2}.
As we have seen in examples, it is the factor $a_4$ which vanishes when 
the corresponding gauge group factor is related to $\mathbb{Z}_2$ torsion. By contrast, one can in principle also engineer additional gauge group factors by factorising $\left(4a_4-(a_2+\frac14a_1^2)^2\right)$ without factorising $a_4$ as such. Such non-toric enhancements would lead to what we called the `spectator' gauge group $G'$ in section \ref{sec_global} and which is unconstrained by the $\mathbb Z_2$ torsion.
Indeed, while all gauge algebras that can be  engineered torically are easily checked to lead to Kodaira monodromies contained in $\Gamma_0(2)$, this set does not exhaust the list of $\Gamma_0(2)$-compatible singularities (\textit{e.g.}\ it misses $A_{2k}$ - see appendix B of \cite{Berglund:1998va}). Such algebras would have to come from a non-toric enhancement involving the second factor of the discriminant. 
We will see an example of an abelian spectator group $G'=U(1)$ in the next section.

%%%%%%%%%%%%%%%%%%%%%%%%%%%%%%%%%%%%%%%%%%%%%%%%%%%%%%%%%%%%%%%%%%%%%%%%%%%%%%%%%%%%%%%%%%%%%%%%%%%%%%%%%%%%%%%%%%%%%%%%%%%%%%%%%%%%%%%%%%%%%%%
\section{Mordell-Weil group $\mathbb Z \oplus \mathbb Z_2$} \label{sec_5}
%%%%%%%%%%%%%%%%%%%%%%%%%%%%%%%%%%%%%%%%%%%%%%%%%%%%%%%%%%%%%%%%%%%%%%%%%%%%%%%%%%%%%%%%%%%%%%%%%%%%%%%%%%%%%%%%%%%%%%%%%%%%%%%%%%%%%%%%%%%%%%%

\subsection{An $(SU(2) \times SU(2))/{\mathbb Z_2} \times U(1)$ fibration}\label{sec_SU2SU2Z2U1}

\begin{figure}
    \centering
  \begin{tikzpicture}[scale=1.5]
%%%% left polygon %%%%
  \filldraw [ultra thick, draw=black, fill=lightgray!30!white]
      (-1,-1)--(1,-1)--(1,1)--(-1,1)--cycle;
    \foreach \x in {-1,...,1}{% Two indices running over each
      \foreach \y in {-1,0,...,1}{% node on the grid we have drawn 
        \node[draw,circle,inner sep=1.3pt,fill] at (\x,\y) {};
            % Places a dot at those points
      }
    }
  \draw[ultra thick, -latex]
       (0,0) -- (-1,-1) node[below left] {$z$};
  \draw[ultra thick, -latex]
       (0,0) -- (1,1) node[above right] {$v$};
  \draw[ultra thick, -latex]
       (0,0) -- (-1,1) node[above left] {$w$};
  \draw[ultra thick, -latex]
       (0,0) -- (1,-1) node[below right] {$u$};
   \draw[ultra thick, -latex]
       (0,0) -- (0,1) node[above] {$d$};
   \draw[ultra thick, -latex]
       (0,0) -- (1,0) node[right] {$c$};

\begin{scope}[xshift=0.33\textwidth]
\filldraw [ultra thick, draw=black, fill=blue!30!white]
      (-1,0)--(0,1)--(1,0)--(0,-1)--cycle;
    \foreach \x in {-1,0,...,1}{% Two indices running over each
      \foreach \y in {-1,0,...,1}{% node on the grid we have drawn 
        \node[draw,circle,inner sep=1.3pt,fill] at (\x,\y) {};
            % Places a dot at those points
      }
    }
%  \node [below] at (0,0) {$-1$};
  \node [left] at (-1,0)  {$d w^2 z^2$};
  \node [above] at (0,1) {$c d^2 v^2 w^2$};
  \node [right] at (1,0)  {$c^2 d u^2 v^2$};
    \node [below] at (0,-1)  {$c u^2 z^2$};
\end{scope}

  \end{tikzpicture}
      \caption{Polygon 15 of \cite{Bouchard:2003bu} together with its dual polygon.}\label{fig:polygon15}
\end{figure}
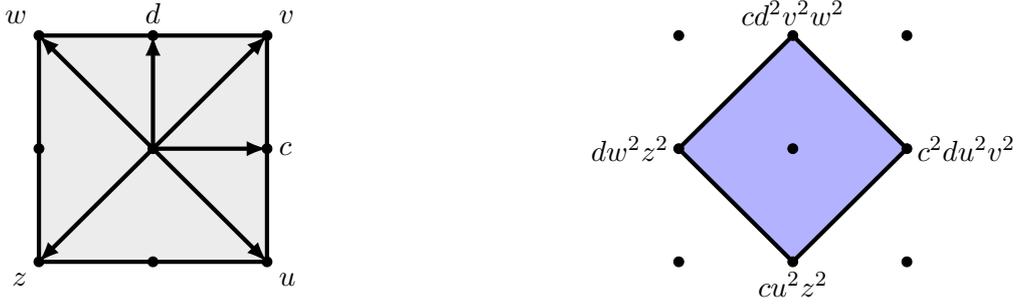

The generic elliptic fibration with $\mathbb Z_2$-torsional Mordell-Weil group admits an interesting specialization such as to enhance the Mordell-Weil group to $\mathbb Z_2 \oplus \mathbb Z$. As it turns out the generator of the free part of the Mordell-Weil group can be described again very conveniently as a toric section. \\\\
In fact, the specialization we have in mind gives rise to the second of the three elliptic fibrations realized as hypersurfaces in a toric ambient space with  Mordell-Weil torsion \cite{Braun:2013nqa}.
The fiber is defined by the reflexive pair in Figure \ref{fig:polygon15}, which corresponds to polygon 15 and its dual in the classification of ~\cite{Bouchard:2003bu}.
The associated elliptic curve is the vanishing locus of a biquadric in a blow-up of $\mathbb P^1 \times \mathbb P^1$. The hypersurface equation defined via the dual polygon is
\begin{equation}\label{eq:hypersurf_poly_15}
 \hat P = cd^2v^2w^2 + c^2d u^2 v^2 + \gamma_1 c d u v w z + \gamma_2 d w^2 z^2   + \delta_2  c u^2 z^2,
\end{equation}
where we have set the coefficients of the first two terms to one since they are sections of the trivial bundle over the base\footnote{If we  had chosen a fibration such that these two coefficients are sections of non-trivial bundles, $z=0$ would not be a holomorphic section but a birational one.}.
%\footnote{This is why two vectors in the fan are omitted: After setting the two first coefficients in \eqref{eq:hypersurf_poly_15} to one, the corresponding toric divisors doesn't intersect the hypersurface and can thus be scaled away.}
The coefficients $\gamma_i$ and $\delta_i$ are sections of ${\bar{\cal K}}^i$. %, making the hypersurface equation homogeneous of degree two. 
A choice for the scaling relations of the fiber coordinates is
\begin{equation}\label{eq:weight-matrix_polygon15}
 \begin{array}{|c|c|c|c|c|c||c|}
\hline
u & v & w & z & c & d &\sum  \tabularnewline \hline \hline
1 &  0 & 1 &  0 &  0 & 0 & 2 \tabularnewline\hline
0 &  1 &  0 &  1 & 0  & 0 & 2\tabularnewline\hline
0 &  0 &  0 &  1 & 1  & 1 & 3\tabularnewline\hline
0 &  0 &  1 &  1 & 2  & 0 & 4\tabularnewline\hline
\end{array}\,,
\end{equation}
which is consistent with the degree of homogeneity of \eqref{eq:hypersurf_poly_15}. % under the respective scalings.
The Stanley-Reisner ideal of the toric ambient space of the fiber takes the form
$\{u v, uw, ud, vz, zc, zd,   
 wc, cd, vw\}$.\\\\
The biquadric \eqref{eq:hypersurf_poly_15} can be brought into Weierstrass form, where it can be compared with the Weierstrass model associated with the fibration \eqref{eq:hse-Z2-fiber-singular} analysed in the previous section.
This identifies
\begin{equation}\label{eq:specialization_of_z2_model}
	a_1 = \gamma_1\,, \qquad a_2 = -(\gamma_2 + \delta_2)\,, \qquad a_4 = \gamma_2 \delta_2 \,,
\end{equation}
where $a_i$ are the coefficients of the generic $\mathbb Z_2$-torsion fibration (\ref{eq:hse-Z2-fiber-singular}).
As we will show, the result of this specialization of $a_2$ and $a_4$ is the enhancement of the Mordell-Weil group from $\mathbb Z_2$ to $\mathbb Z_2 \oplus \mathbb Z$ (as computed previously in \cite{Braun:2013nqa}) and the appearance of an extra $\mathfrak{su}(2)$ factor.\\\\
%
% This model is a restricted case of the equation \eqref{eq:hse-Z2-fiber-singular}. As constructed in \cite{Aspinwall:1998xj} this equation is a general representation of a curve with $\mathbb Z_2$ torsion. We thus expect the existence of an extra section in the model treated in this section to be due to imposing extra conditions on the sections $a_i$. Indeed, by mapping \eqref{eq:hypersurf_poly_15} to Weierstrass form one finds the identification
% \begin{equation}\label{eq:specialization_of_z2_model}
% 	a_1 = \gamma_1\,, \qquad a_2 = -(\gamma_2 + \delta_2)\,, \qquad a_4 = \gamma_2 \delta_2 \,.
% \end{equation}
To analyse the non-abelian sector, we first note that the discriminant of equation (\ref{eq:hypersurf_poly_15})  takes the form
\begin{equation}\label{eq:discriminant_poly15}
\Delta \sim \gamma_2^2 \delta_2^2 [\gamma_1^4 -8 \gamma_1^2(\gamma_2 + \delta_2) + 16(\gamma_2 -\delta_2)^2].
\end{equation}
Together with the Weierstrass functions $f$ and $g$ of the associated Weierstrass model this suggests
an $A_1$ singularity at $\{\gamma_2 = 0\}$ and $\{\delta_2 = 0\}$ respectively. Indeed, the hypersurface equation factorises over these loci as
\begin{equation}
\begin{aligned}
   \{\gamma_2 = 0\} : & \qquad c \left(c d u^2 v^2+d^2 v^2 w^2+\gamma _1 d u v w 
   z+\delta_2 u^2 z^2\right),  \\
\{\delta_2 = 0\} : & \qquad d \left(c d v^2 w^2+c^2 u^2 v^2+\gamma _1 c u v w 
   z+\gamma _2 w^2 z^2\right),
\end{aligned}   
\end{equation}
and we identify the irreducible components $\mathbb P^1_c$ and $\mathbb P^1_d$ as the restriction to the fiber of the resolution divisors $C : \{c = 0\}$ and $D: \{d=0\}$ of these singularities.\\ \\
On general grounds \cite{GrassiPerduca,Braun:2013nqa}, the intersection of the toric divisors $U:\{u=0\}$,  $V:\{v=0\}$, $W: \{w=0\}$, $Z:\{z=0\}$ with the hypersurface give rise to sections of the fibration, not all of which are independent.
Since $Z: \{z=0\}$ is a holomorphic section we choose it as the zero-section. Then, the Mordell-Weil group is generated by differences of sections $U-Z$, $V-Z$, $W-Z$, which are not all independent. Let us first consider the Shioda map for the section $U:\{u=0\}$. 
 %The intersection numbers of the resulting divisor with the split fiber components at codimension-two enhancement loci gives us the $U(1)$ charges of these states. 
Requiring, as usual, one leg in the fiber as well as orthogonality with the exceptional divisors gives 
\begin{equation} \label{WUequ}
W_U = 2(U - Z - \bar{\mathcal{K}}) + C \,,
\end{equation}
which is unique up to an overall normalization, here chosen such as to arrive at integer charges below. 
We take this non-trivial element $W_U$ as the generator of the free part of the Mordell-Weil group, and physically identify it with the generator of the associated, suitably normalized $U(1)$ part of the gauge group.\\\\
On the other hand, the intersection of the section $V: \{v=0\}$ with the elliptic curve describes a 2-torsion point, as noted already in \cite{Braun:2013nqa}. The Shioda map for $V: \{v=0\}$ yields the element
\begin{equation}
\Sigma_2 = V-Z-\bar{\mathcal{K}} + \frac{1}{2}(C+D) \, .
\end{equation}
However, $V$ is not an independent toric divisor class, but may be expressed as
\begin{equation}\label{eq:toric_divisor_V}
V = Z+\bar{\mathcal{K}} -\frac{1}{2}(C+D),
\end{equation}
which makes $\Sigma_2$ a trivial class. Since the model we consider here is a restriction of the model with just a $\mathbb Z_2$ section we have the analogous situation that $\Sigma_2$ is given by a divisor in the ambient space which restricts to a trivial class on the hypersurface. The integer class
\begin{equation}
\Xi_2 \equiv V-Z-\bar{\mathcal{K}} = \frac{1}{2}(C+D)
\end{equation}
 is 2-torsion in $H^{1,1}(\hat{Y}_4,\mathbb Z)$ modulo resolution classes and to be identified with a coweight element momentarily.\\\\
 Having established the gauge algebra $\mathfrak{su}(2) \oplus \mathfrak{su}(2) \oplus \mathfrak{u}(1)$ we turn to the matter representations in codimension 2.
 From the discriminant \eqref{eq:discriminant_poly15} the three potential enhancement loci which could host matter charged under the non-abelian gauge groups are identified  as
\begin{equation}\label{eq:matter_curves_without_top_polygon_15}
\begin{aligned}
\{\gamma_2 = \delta_2 = 0\}, \quad &\{\gamma_1 = \gamma_2 = 0\}, \quad \{\gamma_1 = \delta_2 = 0\}\,. \\
\end{aligned}
\end{equation}
At the loci $\{\gamma_1 = \gamma_2 = 0\}$ and $\{\gamma_1 = \delta_2 = 0\}$, which would naively give rise to fundamental matter, the equation does not factorize further, and hence no extra matter is found there.
But at the locus $\{\gamma_2 = \delta_2 = 0\}$ the equation factorizes as
\begin{equation} \label{fiberbifunda}
cdv\underbrace{(c u^2 v + d v w^2 + \gamma_1 u w z)}_{R},
\end{equation}
where the curves $\mathbb P^1_c,\, \mathbb P^1_d,\, \mathbb P^1_v$ and the last component $\mathbb P^1_R$ intersect as the affine $A_3$ Dynkin diagram. We calculate the charges of the split component $\mathbb P^1_{v=0}$ as
\begin{equation}
\begin{aligned}
\mathbb P^1_{v=0} &\cdot (C,D)=(1,1), \\
%\mathbb P^1_{R} &\cdot (C,D)=(1,1)\,,
\end{aligned}
\end{equation}
giving the highest weights of the bifundamental $(\mathbf{2},\mathbf{2})$. 
By acting on this with the respective roots the entire  $(\mathbf{2},\mathbf{2})$ is reproduced.
%We interpret these as two bifundamentals, since the $\mathbf{2} = \bar{\mathbf{2}}$ for $\mathfrak{su}(2)$. 
With the normalization (\ref{WUequ}) the $U(1)$ charge of this state is 
\bea
W_U \cdot \mathbb P^1_{v=0}   = 1.
\eea  
%(\mathbf{2},\mathbf{2})_{1}$.
% Computing the $U(1)$ charges differentiate between the two split curves and under $ \mathfrak{su}(2)_C \times \mathfrak{su}(2)_D \times U(1)$ the representations are $(\mathbf{2},\mathbf{2})_{2}$ and $(\mathbf{2},\mathbf{2})_{-2}$ \\ \\
%
Extra massless matter is localized at the singlet curve $\{\gamma_2 = \delta_2\} \cap \{\gamma_1 = 0\}$. This is an $I_2$ locus over which the hypersurface equation factorizes as 
\begin{equation}
(cu^2 + dw^2)(cdv^2 + \delta_2 z^2),
\end{equation}
and we denote the fiber components by $\mathbb P^1_-$ and $\mathbb P^1_+$ respectively. These have zero intersection with the Cartan divisors $C, D$ (and are thus invariant also under the center of the gauge group) and their $U(1)$-charges are computed as
\begin{equation}
W_U \cdot \mathbb P^1_\pm = \pm 2.
\end{equation}
Hence we find a representation $(\mathbf{1},\mathbf{1})_{\pm 2}$ with respect to $\mathfrak{su}(2)_C \oplus \mathfrak{su}(2)_D \oplus U(1)$. \\\\
At the intersection points  $\gamma_1 = \gamma_2 = \delta_2 = 0$ of the two matter curves the fiber type changes to form a non-affine Dynkin diagram of $D_4$. This is because the component $(c u^2 v + d v w^2 + \gamma_1 u w z)_{R}$  in the fiber over the curve   $\{\gamma_2 = \delta_2 = 0\}$ splits off a factor of $v$ as $\gamma_1=0$, corresponding to a factorisation
\bea
c \, d \, v^2 \, (c u^2  + d  w^2 ).
\eea
At those points a Yukawa coupling $(\mathbf{2},\mathbf{2})_1 \,  (\mathbf{2},\mathbf{2})_1 \, (\mathbf{1},\mathbf{1})_{- 2}$ is localised.\\\\
 As is manifest, the divisor $\Xi_2$ has integer pairing with all split curves  associated with the representations $(\mathbf{2},\mathbf{2})_1$ and $(\mathbf{1},\mathbf{1})_{\pm 2}$
 and is therefore identified with a coweight. With coefficients in $\frac{1}{2}\mathbb Z$ the coweight lattice is made finer by this extra coweight, and only weights in representations integer paired with $\Xi_2$ are allowed. 
 Again this is the reason for the absence of for example a fundamental representation at the loci $\{\gamma_1 = \gamma_2 = 0\}$ and $\{\gamma_1 = \delta_2 = 0\}$.
 Note that the expression for $\Sigma_2$ does not include a term proportional to the $U(1)$-generator $W_U$, but only the generators $C$ and $D$ of the $\mathfrak{su}(2)_C \oplus \mathfrak{su}(2)_D$ Cartan $U(1)$s.
In particular, integrality of the pairing of $\Xi_2$ does therefore not constrain the allowed $U(1)$ charges, but only the non-abelian part of the representation. 
We conclude that the gauge group is
 \begin{equation}\label{eq:gauge_group_polygon_15}
 G= \frac{SU(2)_C \times SU(2)_D }{\mathbb Z_2}\times U(1),
 \end{equation}
 whose first fundamental group $\pi_1(G) = \mathbb Z \oplus \mathbb Z_2$ coincides with the Mordell-Weil group as expected.

 \subsection{A chain of fibrations via Higgsing} \label{unHiggsing}

The elliptic fibrations described in sections \ref{sec_SU2Z2}, \ref{sec_SU2SU2Z2} and \ref{sec_SU2SU2Z2U1} can be viewed as a successive specialization of a  Tate model 
\bea
P = y^2 - x^3 + a_1 x y z + a_2 x^2 z^2 + a_3 y z^3 + a_4 x z^4 + a_6 z^6,
\eea
which for generic $a_i \in H^0({\cal B}, {\bar{\cal K}}^i)$ has trivial Mordell-Weil and gauge group. %$G = \emptyset$.
If $a_6 \equiv 0$, the fibration corresponds to a  $U(1)$ restricted Tate model \cite{Grimm:2010ez} with Mordell-Weil group $\mathbb Z$, gauge group $G=U(1)$ and a massless singlet ${\bf 1}_{\pm 1}$ localized at the curve $\{a_3 = 0\} \cap \{a_4 = 0\}$.  
The extra section degenerates to a $\mathbb P^1$ over this matter curve \cite{Grimm:2010ez,Braun:2011zm,oai:arXiv.org:1202.3138}. 
From this, one reaches the fibration (\ref{eq:hse-Z2-fiber-singular}) with Mordell-Weil group $\mathbb Z_2$ and $G= SU(2)/\mathbb Z_2$ by setting in addition $a_3 \equiv 0$. This promotes the $U(1)$ generator of the $U(1)$ restricted model to the $\mathfrak{su}(2)$ Cartan generator, which is $\mathbb P^1$ fibered over the $\mathfrak{su}(2)$-divisor $\{a_4=0\}$.
Since the $U(1)$ restricted model has only one type of charged singlet, which becomes part of the $\mathfrak{su}(2)$ adjoint multiplet, the specialization to $a_3 \equiv 0$ does not give rise to any extra matter states. This way the gauge group $G= SU(2)/\mathbb Z_2$ could in fact have been anticipated even without any knowledge of the torsional Mordell-Weil group.
The reverse process corresponds to the Higgsing of $G= SU(2)/\mathbb Z_2$ to $U(1)$ via a Higgs in the adjoint of $SU(2)$, more precisely the component with zero Cartan charge. \\\\
A further factorisation $a_4 = a_{4,1} w$ enhances, as described, the gauge group to $G= (SU(2) \times SU(2))/\mathbb Z_2$ (cf. \ref{eq:SU(2)_top_polygon_13}) without changing the Mordell-Weil group. Finally, if $w \in H^0({\cal B}, \bar{\cal K}^2)$, specialising in addition to $a_2 = - (w + a_{4,1})$ enhances the Mordell-Weil group to $\mathbb Z_2 \oplus \mathbb Z$ and the gauge group to $G = (SU(2) \times SU(2))/\mathbb Z_2 \times U(1)$ - see (\ref{eq:specialization_of_z2_model}) with $\gamma_2 = a_{4,1}$ and $\delta_2 = w$.
The reversed chain of Higgsing thus relates all these fibrations as 
\begin{equation} \label{chain1}
\frac{SU(2)\times SU(2)}{\mathbb Z_2}\times U(1) \rightarrow\, \frac{SU(2)\times SU(2)}{\mathbb Z_2}\,\rightarrow  \,\frac{SU(2)}{\mathbb Z_2}\,\rightarrow\,U(1)\,\rightarrow\, \mathbb \emptyset.
\end{equation}
Note that the fibration (\ref{eq:SU(2)_top_polygon_13}) with $G= (SU(2) \times SU(2))/\mathbb Z_2$ can be shown to coincide with a model that was recently considered in \cite{Morrison:2014era}. In this paper, a different chain of Higgsing was considered which takes the form 
\begin{equation} \label{chain2}
\frac{SU(2)\times SU(2)}{\mathbb Z_2}\,\rightarrow\, {SU(2)}\,\rightarrow\,U(1)\,\rightarrow\, \mathbb Z_2.
\end{equation}
The chain (\ref{chain1}) is a specialization of the deformations involved in (\ref{chain2}). In particular,
the fibration with Mordell-Weil group $\mathbb Z$ and $G=U(1)$ reached in (\ref{chain2}) is described as a special $\mathbb P_{1,1,2}[4]$-fibration \cite{Morrison:2012ei} and can in general not be represented as a global Tate model.
However, a specialization of this family of fibrations corresponds to the $U(1)$ restricted Tate model appearing in (\ref{chain1}). The endpoint of the Higgsing process (\ref{chain2}) with gauge group $\mathbb Z_2$ is a genus-one fibration \cite{Braun:2014oya} which is not an elliptic fibration. The absence of a $\mathbb Z_2$ remnant in the last step in our chain (\ref{chain1}) can be viewed as a consequence of the division by the $\mathbb Z_2$ center in the $G= SU(2)/\mathbb Z_2$ model.

%%%%%%%%%%%%%%%%%%%%%%%%%%%%%%%%%%%%%%%%%%%%%%%%%%%%%%%%%%%%%%%%%%%%%%%%%%%%%%%%%%%%%%%%%%%%%%%%%%%%%%%%%%%%%%%%%%%%%%%%%%%%%%%%%%%%%%%%%%%%%%%%%%%%%%%%%%%%%%%%%%%%%%%%%%%%%%%%%%%%%%%%%%%%%%%%%%%%%%%%%%%%%%%%%%
\subsection{An $(SU(4) \times SU(2) \times SU(2))/{\mathbb Z_2} \times U(1)$ fibration} \label{SU4SU2SU2}
%%%%%%%%%%%%%%%%%%%%%%%%%%%%%%%%%%%%%%%%%%%%%%%%%%%%%%%%%%%%%%%%%%%%%%%%%%%%%%%%%%%%%%%%%%%%%%%%%%%%%%%%%%%%%%%%%%%%%%%%%%%%%%%%%%%%%%%%%%%%%%%%%%%%%%%%%%%%%%%%%%%%%%%%%%%%%%%%%%%%%%%%%%%%%%%%%%%%%%%%%%%%%%%%%%
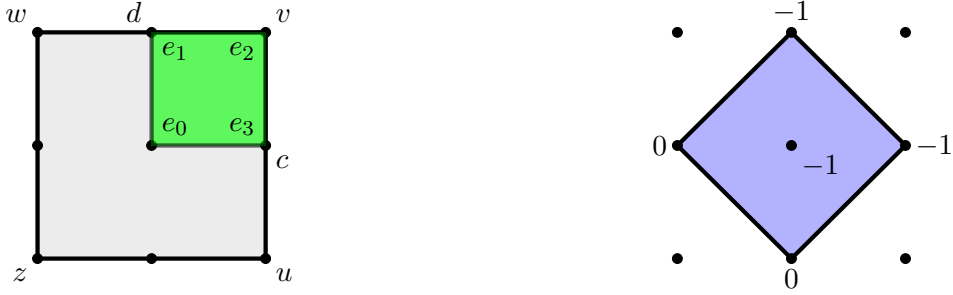
\begin{figure}
    \centering
  \begin{tikzpicture}[scale=1.5]
%%%% left polygon %%%%
  \filldraw [ultra thick, draw=black, fill=lightgray!30!white] %means: 30% lightgray and 70% white 
      (-1,-1)--(1,-1)--(1,1)--(-1,1)--cycle;
    \foreach \x in {-1,...,1}{% Two indices running over each
      \foreach \y in {-1,0,...,1}{% node on the grid we have drawn 
        \node[draw,circle,inner sep=1.3pt,fill] at (\x,\y) {};
            % Places a dot at those points
      }
    }
   \filldraw [ultra thick, draw=black, fill=green, opacity=0.6]
      (0,0)--(1,0)--(1,1)--(0,1)--cycle;
%lightgray!30!white
%blue!30!white
\node[below left] at (-1,-1) {$z$};
\node[above right] at (1,1) {$v$};
\node[above left] at (-1,1) {$w$};
\node[below right] at (1,-1) {$u$};
\node [above left] at (0,1)  {$d$};
\node [below right] at (1,0)  {$c$};
%exceptionals
\node [above right] at (0,0)  {$e_0$};
\node [below right] at (0,1)  {$e_1$};
\node [below left] at (1,1)  {$e_2$};
\node [above left] at (1,0)  {$e_3$};

\begin{scope}[xshift=0.33\textwidth]
\filldraw [ultra thick, draw=black, fill=blue!30!white]
      (-1,0)--(0,1)--(1,0)--(0,-1)--cycle;
    \foreach \x in {-1,0,...,1}{% Two indices running over each
      \foreach \y in {-1,0,...,1}{% node on the grid we have drawn 
        \node[draw,circle,inner sep=1.3pt,fill] at (\x,\y) {};
            % Places a dot at those points
      }
    }
 \node [below right] at (0,0) {$-1$};
  \node [left] at (-1,0)  {$0$};
  \node [above] at (0,1) {$-1$};
  \node [right] at (1,0)  {$-1$};
    \node [below] at (0,-1)  {$0$};
\end{scope}

  \end{tikzpicture}
      \caption{To the left a $\mathfrak{su}(4)$ top over polygon 15 of \cite{Bouchard:2003bu}. The green layer contains the points at height one. To the right the dual top, bounded from below by the values $z_{min}$ shown next to the nodes.}\label{fig:su4top_polygon15}
\end{figure}
We now exemplify the implementation of a further non-abelian singularity by constructing a top. According to the classification in \cite{Bouchard:2003bu} the only A-type singularities admissible over this polygon are $A_{3+2\,n}$, in agreement with Tate's algorithm. %$A_3, A_5, A_7, \dots$. 
We consider here the $A_3 = \mathfrak{su}(4)$ case, with a unique top corresponding to the dual on the righthand side in Figure \ref{fig:su4top_polygon15}. The hypersurface equation is given by
\begin{equation}
\hat P = e_2 e_3 c^2 d  u^2 v^2 + e_1 e_2 c d^2 v^2 w^2+ \gamma_1 c d u v w z+\gamma_2 e_0 e_1 d w^2 z^2+ \delta_2 e_0 e_3 c u^2 z^2
\end{equation}
with  discriminant
\begin{equation}\label{eq:su4top_discriminant}
\Delta \sim \varpi^4 \gamma _2^2 \delta_2^2 \left[\gamma _1^4 -8 \varpi \gamma_1^2
   (\delta_2 + \gamma_2)+16\varpi^2( \gamma_2-\delta_2 )^2\right]
\end{equation}
for $\pi^* \varpi = e_0 e_1 e_2 e_3$. We see that imposing the factorization
\begin{equation}
\gamma_1 \rightarrow \gamma_1, \quad \gamma_2 \rightarrow \varpi \gamma_2, \quad \delta_2 \rightarrow \varpi\delta_2
\end{equation}
on the coefficients of \eqref{eq:hypersurf_poly_15} gives the same behaviour as the top construction. This pattern is just the standard factorisation deduced by the Tate algorithm. For the chosen triangulation of this top we obtain a Stanley-Reisner ideal generated by
\begin{center}
$\{u v, uw, ud, vz, zc, zd,   
 wc, cd, vw, $\\ 
 $ce_0, de_0,ve_0 , ce_1, ue_1, ze_1, de_2, we_2, ze_2, ce_3,de_3, ve_3,we_3, ze_3,e_1e_3,ue_0 e_2
\}$.
\end{center}
In addition to the $A_1$ singularities, with resolution divisors  $C$ and  $D$, we have a fiber degeneration over $\{\varpi=0\}$ with  irreducible components
\begin{equation}
\mathbb P^1_i = \{E_i\} \cap \{\hat P|_{e_i = 0} = 0\} \cap D_a \cap D_b\quad i = 0, \dots, 3,
\end{equation}
where $D_a$ and $D_b$ are some generic divisors in $\mathcal{B}$. These are intersecting as the affine $\mathfrak{su}(4)$ Dynkin diagram, as can be read off from the top in Figure \ref{fig:su4top_polygon15}. For the explicit expressions we refer to appendix \ref{app:su4poly15}. The $U(1)$-generator from the previous section gets corrected by the exceptional divisors from the extra $\mathfrak{su}(4)$ locus and takes the form
\begin{equation}
W_U = 4(U - Z - \bar{\mathcal{K}}) + 2 C + E_1 + 2E_2 + 3 E_3 \,.
\end{equation}
The normalization is chosen such as to give integer charges of all matter states. In the same way we get additional contributions to the Shioda map $\Sigma_2$ of the torsion section, which is a trivial class since $V$ can be written as the linear combination 
\begin{equation}\label{eq:toric_divisor_with_su4}
V = Z+\bar{\mathcal{K}} -\frac{1}{2}(C+D+E_1+2E_2+E_3) \,.
\end{equation}
We identify with the new coweight the integer class
\begin{equation}
\Xi_2 \equiv V-Z-\bar{\mathcal{K}} = \frac{1}{2}(C+D+E_1+2E_2+E_3),
\end{equation}
which is 2-torsion in $H^{1,1}(\hat{Y}_4,\mathbb Z)$ modulo resolution classes. Repeating the analysis of the previous section we find that the extra coweight class $\Xi_2$ is independent of the $U(1)$-generator. \\\\
In what follows we compute the additional charged matter representations localized at codimension-two loci in the base, \textit{i.e.}\  the matter curves that lie in the $\mathfrak{su}(4)$ divisor $\{\varpi=0\}$. The full equations are omitted here and are found appendix \ref{app:su4poly15}.  By inspection of the discriminant \eqref{eq:su4top_discriminant} the potential enhancement loci are
\begin{equation}
\begin{aligned}
\{\varpi = \gamma_1 = 0\}, \quad &\{\varpi = \gamma_2 = 0\}, \quad \{\varpi = \delta_2 = 0\}\,, \\
\end{aligned}
\end{equation}
in addition to the curves considered in the previous section.  At $\{\varpi = \gamma_1 = 0\}$ the fiber components $\mathbb P^1_0$ and $\mathbb P^1_2$ split and the total fiber has the intersection structure of the affine $D_4$ Dynkin diagram. The weights of the split curves are 
\begin{equation}
 \begin{aligned}
  \mathbb P^1_{e_0=e_2=0} &\cdot (E_1,E_2,E_3)=(1,-1,1)\,,   & \mathbb P^1_{e_0=e_2=0} \cdot (C,D)=(0,0)\,,\\
  \mathbb P^1_{e_0=e_3u^2 + e_1w^2=0} &\cdot (E_1,E_2,E_3)=(0,1,0)\,,  & \mathbb P^1_{e_0=e_3u^2 + e_1w^2=0} \cdot (C,D)=(0,0)\,, \\
  \mathbb P^1_{e_2=\gamma_2 e_1 + \delta_2 e_3 cu^2=0}, &\cdot (E_1,E_2,E_3)=(0,-1,0)\,,   & \mathbb P^1_{e_2=\gamma_2 e_1 + \delta_2 e_3 cu^2=0} \cdot (C,D)=(0,0)\,.
 \end{aligned}
\end{equation}
The $(0,1,0)$ is the highest weight of the $\mathbf{6}$ of $\mathfrak{su}(4)$. Including the $U(1)$ charges  we therefore find the representation $(\mathbf{6},\mathbf{1},\mathbf{1})_2 + c.c.$.\\\\
At $\{\varpi = \gamma_2 = 0\}$ the curve $\mathbb P^1_2$ splits into three components and the full fiber has the structure of an affine $A_5$ Dynkin diagram. We expect to find matter charged under the $\mathfrak{su}(4)$ and the $\mathfrak{su}(2)_C$ factors along this curve in the base. Indeed the split curves have charges
\begin{equation}
 \begin{aligned}
  \mathbb P^1_{e_2=c=0} &\cdot (E_1,E_2,E_3)=(0,0,0)\,,   & \mathbb P^1_{e_2=c=0} \cdot (C,D)=(-2,0)\,,\\
  \mathbb P^1_{e_2=u=0} &\cdot (E_1,E_2,E_3)=(0,-1,1)\,,  & \mathbb P^1_{e_2=u=0} \cdot (C,D)=(1,0)\,, \\
  \mathbb P^1_{e_2=\gamma_1 v + \delta_2 e_0 e_3 u=0} &\cdot (E_1,E_2,E_3)=(1,-1,0)\,,   & \mathbb P^1_{e_2=\gamma_1 v + \delta_2 e_0 e_3 u=0} \cdot (C,D)=(1,0)\,,
 \end{aligned}
\end{equation}
where the $(0,-1,1)$ and the $(1,-1,0)$ are weights in the fundamentals $\mathbf{4}$ and $\bar{\mathbf{4}}$ respectively. Including the $U(1)$ charges we have the $(\mathbf{4},\mathbf{2},\mathbf{1})_1 + c.c. $  along this matter curve. \\ \\
Along $\varpi = \delta_2 = 0$ the configuration is completely analogous to that along $\varpi=\gamma_2=0$ and gives rise to massless matter in representation
$(\mathbf{4},\mathbf{1},\mathbf{2})_1 + c.c.$.  The massless matter spectrum is summarized in the following table:
\begin{equation}
 \begin{array}{cc}
  \multicolumn{2}{c}{\textit{Top over polygon 15: } \mathfrak{su}(4) \times \mathfrak{su}(2)_C \times \mathfrak{su}(2)_D \times U(1)}\\ \\
 \textit{Locus} & \textit{Charged matter}\tabularnewline 
 \gamma_1 \cap \{ \gamma_2 = \delta_2 \}& (\mathbf{1},\mathbf{1},\mathbf{1})_{4} \, ,\,(\mathbf{1},\mathbf{1},\mathbf{1})_{-4}\tabularnewline
\gamma_2 \cap \delta_2 &  (\mathbf{1},\mathbf{2},\mathbf{2})_2 \,,\, (\mathbf{1},\mathbf{2},\mathbf{2})_{-2} \tabularnewline
\varpi \cap \gamma_1 &  (\mathbf{6},\mathbf{1},\mathbf{1})_2\,,\,(\mathbf{6},\mathbf{1},\mathbf{1})_{-2} \tabularnewline
\varpi \cap \gamma_2 &  (\mathbf{4},\mathbf{2},\mathbf{1})_1 \, ,\,(\bar{\mathbf{4}},\mathbf{2},\mathbf{1})_{-1} \tabularnewline
\varpi \cap \delta_2 & (\mathbf{4},\mathbf{1},\mathbf{2})_1 \, ,\,(\bar{\mathbf{4}},\mathbf{1},\mathbf{2})_{-1} \tabularnewline
\end{array} 
\end{equation}
It is confirmed that the coweight element $\Xi_2$ is integer-valued on all split curves responsible for the matter representations. We finally conclude that the gauge group is
 \begin{equation}
 \frac{SU(4)\times SU(2)_C \times SU(2)_D }{\mathbb Z_2}\times U(1) \,.
 \end{equation}

%%%%%%%%%%%%%%%%%%%%%%%%%%%%%%%%%%%%%%%%%%%%%%%%%%%%%%%%%%%%%%%%%%%%%%%%%%%%%%%%%%%%%%%%%%%%%%%%%%%%%%%%%%%%%%%%%%%%%%%%%%%%%%%%%%%%%%%%%%%%%%%
\section{Mordell-Weil group \texorpdfstring{$\mathbb Z_3$}{Z3}} \label{sec_6}
%%%%%%%%%%%%%%%%%%%%%%%%%%%%%%%%%%%%%%%%%%%%%%%%%%%%%%%%%%%%%%%%%%%%%%%%%%%%%%%%%%%%%%%%%%%%%%%%%%%%%%%%%%%%%%%%%%%%%%%%%%%%%%%%%%%%%%%%%%%%%%%

As a further illustration we now analyze elliptic fibrations with $\mathbb Z_3$ torsional Mordell-Weil group. The general form of such fibrations was derived in \cite{Aspinwall:1998xj}. As we will show this fibration allows for a toric representation,
which in fact coincides with the last of the 3 reflexive pairs of polygons admitting a torsional Mordell-Weil group \cite{Braun:2013nqa}. The fan is given by the 16th reflexive polygon in the enumeration by  \cite{Bouchard:2003bu}.
We first present the toric representation of this fibration, its singularity structure and impose further non-abelian degenerations of the fiber to analyse the resulting matter spectrum and global structure of the gauge group.

\subsection{An $SU(3)/\mathbb Z_3$-fibration}

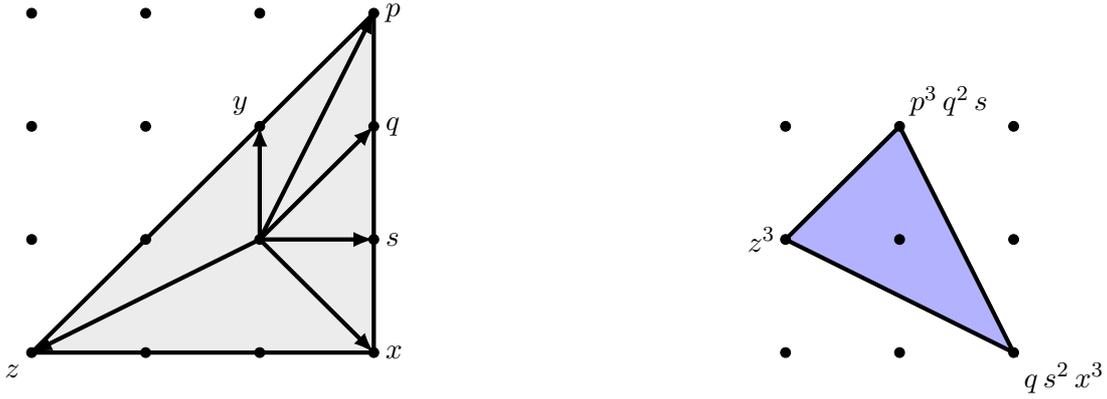
\begin{figure}
    \centering
  \begin{tikzpicture}[scale=1.5]
%%%% left polygon %%%%
  \filldraw [ultra thick, draw=black, fill=lightgray!30!white]
      (-2,-1)--(1,-1)--(1,2)--cycle;
    \foreach \x in {-2,-1,...,1}{% Two indices running over each
      \foreach \y in {-1,0,...,2}{% node on the grid we have drawn 
        \node[draw,circle,inner sep=1.3pt,fill] at (\x,\y) {};
            % Places a dot at those points
      }
    }
    %lightgray!30!white
%blue!30!white
  \draw[ultra thick, -latex]
       (0,0) -- (-2,-1) node[below left] {$z$};
  \draw[ultra thick, -latex]
       (0,0) -- (0,1) node[above left] {$y$};
  \draw[ultra thick, -latex]
       (0,0) -- (1,2) node[right] {$p$};
  \draw[ultra thick, -latex]
       (0,0) -- (1,1) node[right] {$q$};
  \draw[ultra thick, -latex]
       (0,0) -- (1,0) node[right] {$s$};
  \draw[ultra thick, -latex]
       (0,0) -- (1,-1) node[right] {$x$};
%%%% right polygon %%%%
\begin{scope}[xshift=0.33\textwidth]
\filldraw [ultra thick, draw=black, fill=blue!30!white]
      (-1,0)--(0,1)--(1,-1)--cycle;
    \foreach \x in {-1,0,...,1}{% Two indices running over each
      \foreach \y in {-1,0,...,1}{% node on the grid we have drawn 
        \node[draw,circle,inner sep=1.3pt,fill] at (\x,\y) {};
            % Places a dot at those points
      }
    }
  \node [left] at (-1,0)  {$z^3$};
  \node [above right] at (0,1) {$p^3\,q^2\,s$};
  \node [below right] at(1,-1)  {$q\,s^2\,x^3$};
\end{scope}
  \end{tikzpicture}
      \caption{Polygon 16 of \cite{Bouchard:2003bu} together with its dual polygon. The coordinate $y$ is scaled to one and does not contribute to the monomials.}\label{fig:polygon16}
\end{figure}

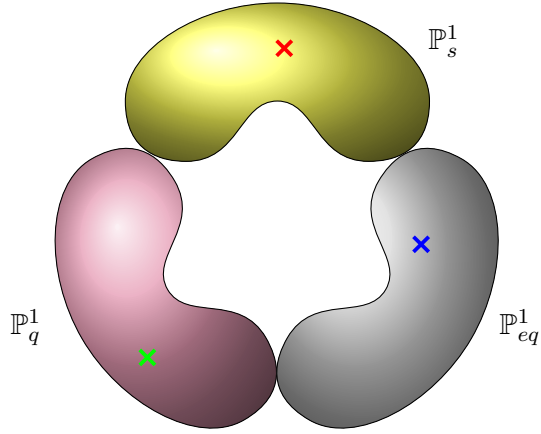
\begin{figure}
    \centering
\begin{tikzpicture}
\begin{scope}[rotate=30]
   \shadedraw[ball color=purple!40!white] (-0.3,0) .. controls (-0.3,1) and (.3,2) .. (1,2)
               .. controls (1.7,2) and (1.8,1.5) .. (1.8,1.2)
               .. controls (1.8,.5) and (1,.5) .. (1,0)
               .. controls (1,-.5) and (1.8,-.5) .. (1.8,-1.2)
               .. controls (1.8,-1.5) and (1.7,-2) .. (1,-2)
               .. controls (.3,-2) and (-0.3,-1) .. (-0.3,0);
\end{scope}
\begin{scope}[xshift=4.6cm,rotate=-30]
\shadedraw[ball color=gray!30!white,xscale=-1] (-0.3,0) .. controls (-0.3,1) and (.3,2) .. (1,2)
               .. controls (1.7,2) and (1.8,1.5) .. (1.8,1.2)
               .. controls (1.8,.5) and (1,.5) .. (1,0)
               .. controls (1,-.5) and (1.8,-.5) .. (1.8,-1.2)
               .. controls (1.8,-1.5) and (1.7,-2) .. (1,-2)
               .. controls (.3,-2) and (-0.3,-1) .. (-0.3,0);
\end{scope}
\begin{scope}[xshift=2.31cm,yshift=4cm,rotate=-90]
\shadedraw[ball color=yellow!65!white,xscale=1] (-0.3,0) .. controls (-0.3,1) and (.3,2) .. (1,2)
               .. controls (1.7,2) and (1.8,1.5) .. (1.8,1.2)
               .. controls (1.8,.5) and (1,.5) .. (1,0)
               .. controls (1,-.5) and (1.8,-.5) .. (1.8,-1.2)
               .. controls (1.8,-1.5) and (1.7,-2) .. (1,-2)
               .. controls (.3,-2) and (-0.3,-1) .. (-0.3,0);
\end{scope}
%labels
\begin{scope}
\node at (-1,0) {$\mathbb P^1_q$}; 
\node at (5.5,0) {$\mathbb P^1_{eq}$};
\node at (4.5,3.8) {$\mathbb P^1_s$};
\end{scope}
% blue cross
\draw[blue, line width=0.5mm, xshift=4.1cm,yshift=1cm] (0,0)--(0.2,0.2);
\draw[blue, line width=0.5mm, xshift=4.1cm,yshift=1cm] (0,0.2)--(0.2,0);
% green cross
\draw[green, line width=0.5mm, xshift=5mm,yshift=-0.5cm] (0,0)--(0.2,0.2);
\draw[green, line width=0.5mm, xshift=5mm,yshift=-0.5cm] (0,0.2)--(0.2,0);
% red cross
\draw[red, line width=0.5mm, xshift=2.3cm,yshift=3.6cm] (0,0)--(0.2,0.2);
\draw[red, line width=0.5mm, xshift=2.3cm,yshift=3.6cm] (0,0.2)--(0.2,0);
\end{tikzpicture}
\caption{The factorised fiber over the base locus $a_3=0$; The \textcolor{blue}{blue} cross indicates the zero point $z=0$ and the \textcolor{green}{green} and \textcolor{red}{red} crosses indicates the points $p=0$ and $x=0$, respectively.}\label{fig:fiber_over_a3}
\end{figure}

The generic form of an elliptic fibration with a $\mathbb Z_3$-section is given by the vanishing locus of the hypersurface equation \cite{Aspinwall:1998xj} 
\begin{equation}\label{eq:generic_z3_equation}
P = y^2 + a_1 xyz + a_3 y z^3 - x^3 
\end{equation}
in weighted projective space $\mathbb P_{[2, 3, 1]}$. 
Such fibrations therefore fit again into the class of global Tate models, but with $a_6 \equiv 0$ and in addition $a_2 \equiv 0$ and $a_4 \equiv 0$.
The equivalent Weierstrass model is defined by
\begin{equation} \label{WZ3}
f = \frac{1}{2}a_1a_3 -\frac{1}{48}a_1^4 \,, \qquad \quad g= \frac{1}{4}a_3^2 + \frac{1}{864}a_1^6-\frac{1}{24}a_1^3a_3
\end{equation}
with discriminant
\bea
\Delta = \frac{1}{16}a_3^3(27a_3 - a_1^3).
\eea
The vanishing order of $\Delta$ at $\{a_3=0\}$, where neither $f$ nor $g$ vanish, signals an $A_2$-singularity over this locus. The singularity at $x=y=a_3=0$ is resolved by two blow-ups 
\begin{equation}
(x,y) \rightarrow (sx,sy), \qquad (s,y) \rightarrow (qs,qy) \, 
\end{equation}
with proper transform
\begin{equation}\label{eq:2ndblowup}
\hat{P} = s q^2 y^2 + a_1 q s x y z + a_3 y z^3 - q s^2 x^3
\end{equation}
as the resulting equation. The Stanley-Reisner ideal after these two blow-ups is $\{qx,qy,qz,xy,sz\}$ (see Fig \ref{fig:polygon16}).
%%%%%%%%%%%%%%%%%%%%%%%%%%%%%%%%%%%%%%%%%%%%%%%%%%%%%%%%%%%%%%%%%%%%%%%%%%%%%%%%%%%%%%%%%%%%%%%%%%%%%%%%%%%%%%%%%%%%%%%%%%%%%%%%%%%%%%%%%%%%%%%
%\subsection{Toric description}
%%%%%%%%%%%%%%%%%%%%%%%%%%%%%%%%%%%%%%%%%%%%%%%%%%%%%%%%%%%%%%%%%%%%%%%%%%%%%%%%%%%%%%%%%%%%%%%%%%%%%%%%%%%%%%%%%%%%%%%%%%%%%%%%%%%%%%%%%%%%%%%
The hypersurface equation \eqref{eq:generic_z3_equation} has an equivalent toric description as a generic hypersurface which makes the vanishing of the coefficients $a_2, a_4$ and $a_6$ manifest. To see this we perform yet another blow-up by
\begin{equation}
q \rightarrow pq, \qquad y \rightarrow py \, ,
\end{equation}
under which the proper transform of equation \eqref{eq:2ndblowup} is
\begin{equation}
\hat P = s p^3 q^2 y^2 + a_1 p q s x y z + a_3 y z^3 - q s^2 x^3. 
\end{equation}
The Stanley-Reisner ideal now extends to $\{sz, qz,pz,xy,sy,qy,ps,px,qx\}$ and implies that the locus $\{y=0\}$ does not intersect the hypersurface any more. Hence we can use one scaling relation to set $y=1$. After this step we arrive at the hypersurface equation
\begin{equation}\label{eq:3rdblowup}
\hat P = p^3 q^2 s + a_1 p q s x z + a_3 z^3 - q s^2 x^3
\end{equation}
defined in the ambient space with scaling relations
\begin{equation}\label{eq:weight-matrix_Z3-case}
 \begin{array}{|c|c|c|c|c||c|}
\hline
x & z & s & q & p &\sum  \tabularnewline \hline \hline
1 &  1 &  0 &  0 &  1 & 3 \tabularnewline\hline
1 &  2 &  0 &  3 & 0  & 6\tabularnewline\hline
0 &  1 &  1 &  1 & 0  & 3\tabularnewline\hline
\end{array}
\end{equation}
and SR ideal $\{sz,qz,px,ps,qx\}$. A blow-down of this fibration was also considered in \cite{Berglund:1998va}, where it was shown that the structure group of the elliptic fibration is the subgroup $\Gamma_0(3)$ of $SL(2,\mathbb Z)$.
As we will see, the structure of admissible gauge groups is in agreement with the appearance of such restricted monodromy.\\\\
Over the locus $\{a_3 = 0 \}$ the hypersurface equation \eqref{eq:3rdblowup} factors as
\begin{equation} \label{a3P}
\hat P|_{a_3=0} = qs(p^3 q- sx^3 -a_1 pxz) 
\end{equation}
with three irreducible factors. The intersection pattern of the irreducible parts of the fiber, denoted by $\mathbb P^1_s$, $\mathbb P^1_q$ and $\mathbb P^1_{eq}$, is shown in Fig. \ref{fig:fiber_over_a3}. 
The two resolution divisors $Q: \{q=0\}$ and $S: \{s=0\}$ are $\mathbb P^1$-fibrations over $\{a_3 = 0 \}$ and are associated with the two Cartan generators of $\mathfrak{su}(3)$.\\\\
The vanishing order of the discriminant increases by 1 on the curve $\{a_3=0\} \cap \{a_1 =0\}$, naively suggesting an enhancement of the singularity type from $A_2$ to $A_3$ and thus localised matter in the fundamental ${\bf 3}$ of $\mathfrak{su}(3)$. In actuality, however, no higher degeneration of the fiber structure occurs over this curve because none of the three components in (\ref{a3P}) factorises further.
This can be seen directly by considering the Weierstrass coefficients $f$
and $g$ (\ref{WZ3}):  along $\{a_3=0\} \cap \{a_1 =0\}$, each coefficient
vanishes to order $2$, which implies that the Kodaira type of the
degenerate fibers is type $IV$.  This is very similar to the familiar $A_2$,
except that the three components of the fibers meet in a single point rather
than meeting pairwise at three different points.  There is no enhancement
or matter (consistent with \cite{Bershadsky:1996nh,Grassi:2011hq}).
The absence of the fundamental representation, which would be expected to be present in generic fibrations with $\mathfrak{su}(3)$ gauge algebra, will be understood momentarily from the global structure of the gauge group.  \\\\
The toric Mordell-Weil group is generated by the differences $P-Z$ or $X-Z$ with $P, X, Z$ corresponding to the vertices of polygon 16 \cite{Braun:2013nqa} with coordinates as in Fig. \ref{fig:polygon16}. Using the SR-ideal, we conclude that each of these sections intersects only one of the $\mathbb P^1$'s, and each $\mathbb P^1$ intersects only one of the sections. \\\\
The divisor class $Y: \{y=0\}$ does not intersect the hypersurface and may be expressed as
\begin{equation}
Y = 3Z - S -
    2Q - 3P + 3\bar{\mathcal{K}}.
\end{equation}
Hence we can define the integer class 
\bea
\Xi_3 \equiv P-Z-\bar{\mathcal{K}} = -\frac{1}{3}(S+2Q)
\eea
associated with a new coweight. Any weight of a charged matter representation has to have integer pairing with $\Xi_3$, making the weight lattice an order three coarser lattice. 
In particular, this forbids the fundamental representation of $SU(3)$, in agreement with our findings above.
Note also that the fundamental representation would be transforming under the center $\mathbb Z_3$ of $SU(3)$. Thus the gauge group is $SU(3)/\mathbb Z_3$. \\\\
\noindent Note that the specialization $a_3 = (\tilde a_3)^n$, if admissible, modifies the gauge group to 
\bea
G = SU(3 n) /\mathbb Z_3,
\eea
corresponding to a fiber structure of split Kodaira type $I^s_{3n}$.
For $n=2$, the fiber over the curve $\{\tilde a_3=0\} \cap \{a_1=0\}$ degenerates further to Kodaira type  $IV^*$, as reflected in the vanishing orders $(3,4,8)$ of $(f,g,\Delta)$ in the Weierstrass model.
This signals an enhancement of the singularity type from $A_{5} \simeq \mathfrak{su}(6)$ to $E_6$. From the branching rules of  the adjoint representation of $E_6$ under the decomposition to $\mathfrak{su}(6)$ one infers  massless matter in the  triple-antisymmetric representation ${\bf 20 }$ of $\mathfrak{su}(6)$, in agreement with the gauge group $SU(6)/\mathbb Z_3$. 
However, for  $n \geq 3$ the Kodaira type fiber over $\{\tilde a_3=0\} \cap \{a_1=0\}$ is beyond $E_8$ according to Kodaira's list. This means that no crepant resolution of the fibration exists whenever the locus $\{\tilde a_3=0\} \cap \{a_1=0\}$  is non-trivial, and F-theory on such spaces is ill-defined. 
This complication does not arise for eight-dimensional F-theory compactifications on K3, where the codimension-one loci are points on the base ${\cal B}=\mathbb P^1$ and thus no problematic enhancement of this type arises. Indeed, the case $n=6$ corresponds to the $SU(18)/\mathbb Z_3$ model presented in equ. (5.4) of \cite{Ganor:1996pc} for F-theory on a K3 surface. \\\\
Finally, let us note that the F-theory model does not possess a well-defined weak coupling Type IIB limit, at least not  of the usual type \`a la Sen: Since $a_2 \equiv 0$ (in addition to $a_4\equiv 0$ and $a_6 \equiv 0$), the quantity $h$ defining the Type IIB Calabi-Yau $ X_\textmd{IIB}$ as the hypersurface $\xi^2 =  h$ factorises, $h = - \frac{1}{12} a_1^2$. Thus the locus $\xi=0=a_1$ is singular.

%%%%%%%%%%%%%%%%%%%%%%%%%%%%%%%%%%%%%%%%%%%%%%%%%%%%%%%%%%%%%%%%%%%%%%%%%%%%%%%%%%%%%%%%%%%%%%%%%%%%%%%%%%%%%%%%%%%%%%%%%%%%%%%%%%%%%%%%%%%%%%%
\subsection{An $(SU(6) \times SU(3))/\mathbb Z_3$-fibration} \label{SU6SU3Z3}
%%%%%%%%%%%%%%%%%%%%%%%%%%%%%%%%%%%%%%%%%%%%%%%%%%%%%%%%%%%%%%%%%%%%%%%%%%%%%%%%%%%%%%%%%%%%%%%%%%%%%%%%%%%%%%%%%%%%%%%%%%%%%%%%%%%%%%%%%%%%%%%
\begin{figure}
    \centering
  \begin{tikzpicture}[scale=1.5]
%%%% left polygon %%%%
  \filldraw [ultra thick, draw=black, fill=lightgray!30!white]
      (-2,-1)--(1,-1)--(1,2)--cycle;
    \foreach \x in {-2,-1,...,1}{% Two indices running over each
      \foreach \y in {-1,0,...,2}{% node on the grid we have drawn 
        \node[draw,circle,inner sep=1.3pt,fill] at (\x,\y) {};
            % Places a dot at those points
      }
    }
      \filldraw [ultra thick, draw=black, fill=green, opacity=0.6]
      (0,0)--(2,2)--(2,0)--cycle;
%    \foreach \x in {-2,-1,...,1}{% Two indices running over each
%      \foreach \y in {-1,0,...,2}{% node on the grid we have drawn 
%        \node[draw,circle,inner sep=1.3pt,fill] at (\x,\y) {};
%            % Places a dot at those points
%      }
%    }
  \draw[ultra thick, -latex]
       (0,0) -- (-2,-1) node[below left] {$z$};
  \draw[ultra thick, -latex]
       (0,0) -- (0,1) node[above left] {$y$};
  \draw[ultra thick, -latex]
       (0,0) -- (1,2) node[right] {$p$};
 % \draw[ultra thick, -latex]
   %    (0,0) -- (1,1) node[right] {$q$};
%  \draw[ultra thick, -latex]
%       (0,0) -- (1,0) node[below right] {$s$};
  \draw[ultra thick, -latex]
       (0,0) -- (1,-1) node[right] {$x$};
\node [below] at (0,0)  {$e_0$};
\node [below right] at (1,1)  {$e_1$};
\node [right] at (2,2)  {$e_2$};
\node [right] at (2,1)  {$e_3$};
\node [right] at (2,0)  {$e_4$};
\node [below right] at (1,0)  {$e_5$};
\begin{scope}[xshift=0.33\textwidth]
\filldraw [ultra thick, draw=black, fill=blue!30!white]
      (-1,0)--(0,1)--(1,-1)--cycle;
    \foreach \x in {-1,0,...,1}{% Two indices running over each
      \foreach \y in {-1,0,...,1}{% node on the grid we have drawn 
        \node[draw,circle,inner sep=1.3pt,fill] at (\x,\y) {};
            % Places a dot at those points
      }
    }
        %lightgray!30!white
%blue!30!white
  \node [below] at (0,0) {$-1$};
  \node [left] at (-1,0)  {$1$};
  \node [above right] at (0,1) {$-1$};
  \node [below right] at (1,-1)  {$-1$};
\end{scope}
  \end{tikzpicture}
      \caption{The $\mathfrak{su}(6)$ top over polygon 16 is shown to the left. The green layer contains the points at height one. The right side defines the dual top, bounded from below by the values $z_{min}$ shown next to the nodes. }\label{fig:su6top}
\end{figure}
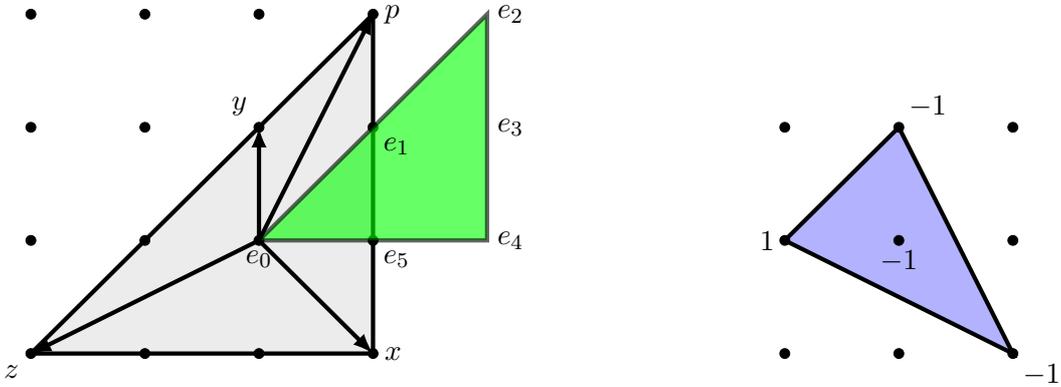
To further illustrate this relation between the $\mathbb Z_3$ Mordell-Weil group and the global structure of the gauge group we implement an additional non-abelian fiber degeneration in codimension-one. This results in an F-theory compactification with a richer matter spectrum. As we will see, only matter representations occur which are compatible with the extra coweight induced by the torsion generator of the Mordell-Weil group. To implement an extra non-Abelian singularity in the hypersurface \eqref{eq:3rdblowup} we construct a top. According to the classification in \cite{Bouchard:2003bu} the only possible tops encoding A-type degenerations are the affine $A_2$, $A_5$, $A_8$ etc. Here we construct the single top corresponding to the affine $A_5$, realizing an $\mathfrak{su}(6)$ theory along a divisor in the base. The hypersurface equation in the ambient space defined by the top is now given by
\begin{equation}
\hat P = e_1 e_2^2 e_3 p^3 q^2 s + a_1 p q s x z + a_3e_0^2 e_1 e_5 z^3 - e_3 e_4^2 e_5 q s^2 x^3,
\end{equation}
where the coefficients of the monomials are chosen to match \eqref{eq:3rdblowup}. The discriminant takes the form
\begin{equation}\label{eq:su6discriminant}
\Delta \sim w^6 a_3^3 (a_1^3 - 27 w^2 a_3) \,,
\end{equation}
where $\pi^* w = e_0 e_1 e_2 e_3 e_4 e_5$ defines the $\mathfrak{su}(6)$-divisor as $W : \{w=0\}$ in the base $\mathcal{B}$. For the chosen triangulation of the top we obtain the Stanley-Reisner ideal 
\begin{equation}
\begin{aligned}
\{&ps, px, qx,qz, sz, p e_3, pe_4, pe_5, qe_0, qe_1, qe_3, qe_4, qe_5, ze_1, ze_2, ze_3,ze_4,ze_5,\\ 
& se_0,se_1,se_3, se_4,se_5, xe_1, x e_0e_3, x e_3 e_5, e_0 e_2, e_0 e_4, e_1 e_4, e_1 e_5, e_2 e_4, e_2 e_5 \} .
\end{aligned}
\end{equation}
In addition to the $A_2$ singularity with resolution divisors  $S$ and  $Q$ one finds a fiber degeneration over $W:\{w=0\}$ with irreducible components
\begin{equation}
\mathbb P^1_i = \{E_i\} \cap \{P_W|_{e_i = 0} = 0\} \cap D_a \cap D_b\quad i = 0, \dots, 5,
\end{equation}
where $D_a$ and $D_b$ are some generic divisors in $\mathcal{B}$. These are intersecting as the affine $A_5$ Dynkin diagram as can also be read off from the top in Figure \ref{fig:su6top}. For the explicit expressions we refer to appendix \ref{app:su6poly16}.  \\\\
%%%%%%%%%%%%%%%%%%%%%%%%%%%%%%%%%%%%%%%%%%%%%%%%%%%%%%%%%%%%%%%%%%%%%%%%%%%%%%%%%%%%%%%%%%%%%%%%%%%%%%%%%%%%%%%%%%%%%%%%%%%%%%%%%%%%%%%%%%%%%%%
We next compute the charged matter representations at enhancement loci in codimension-two. By inspection of the discriminant \eqref{eq:su6discriminant} we see that there are three potentially interesting loci,
\begin{equation}
 \{w=a_1=0\}\, ,\qquad \{w=a_3=0\}\qquad\textmd{and}\qquad \{a_1=a_3=0\}\,.
\end{equation}
The locus $\{a_1=a_3=0\}$, despite the increased vanishing order of $\Delta$, does not give rise any massless matter, as discussed already in the previous section. Thus, no massless states in representation $({\bf 1},{\bf 3})$ of $\mathfrak{su}(6) \oplus \mathfrak{su}(3)$ exist.
The enhancement over the remaining two loci is determined by calculating the factorization of the fiber components over these loci. The explicit equations are presented in appendix \ref{app:su6poly16}.\\\\
At $\{w = a_1 = 0\}$ the fiber components $\mathbb P^1_0$ and $\mathbb P^1_3$ factorize, resulting in six distinct fiber components. They intersect as the \emph{non-affine} $E_6$ Dynkin diagram. The weights at this locus are obtained by computing the intersection numbers of the split fiber components with the resolution divisors $E_i$ and $S,Q$ of the $\mathfrak{su}(6)$ and $\mathfrak{su}(3)$ singularities, respectively. As an example we consider the split curves arising from $\mathbb P^1_0$ and compute the weights
\begin{equation}
 \begin{aligned}
  \mathbb P^1_{e_0=e_3=0} &\cdot (E_1,E_2,E_3,E_4,E_5)=(1,0,-1,0,1)\,,   & \mathbb P^1_{e_0=e_3=0} \cdot (S,Q)=(0,0)\,,\\
  \mathbb P^1_{e_0=e_1 p^3 + e_5 x^3=0} &\cdot (E_1,E_2,E_3,E_4,E_5)=(0,0,1,0,0)\,,  & \mathbb P^1_{e_0=e_1 p^3 + e_5 x^3=0} \cdot (S,Q)=(0,0)\,,
 \end{aligned}
\end{equation}
which are in the $(\mathbf{20},{\bf 1})$ of $\mathfrak{su}(6) \oplus \mathfrak{su}(3)$. \\\\
Over $\{w = a_3 = 0\}$ the component $\mathbb P^1_3$ factorizes. This results in 9 distinct curves, intersecting as the affine $\mathfrak{su}(9)$ Dynkin diagram. We compute the charges
\begin{equation}
 \begin{aligned}
\mathbb P^1_{e_2=x=0} &\cdot (E_1,E_2,E_3,E_4,E_5)=(0,-1,1,0,0)\,,  & \mathbb P^1_{e_2=x=0} \cdot (S,Q)=(1,0)\,,\\
\mathbb P^1_{e_2=a_1 p + e_3 sx^2=0} &\cdot (E_1,E_2,E_3,E_4,E_5)=(1,-1,0,0,0)\,,  & \mathbb P^1_{e_2=a_1 p + e_3 sx^2=0} \cdot (S,Q)=(0,1)\,,
 \end{aligned}
\end{equation}
recognizing the $(0,-1,1,0,0)$ and $(1,-1,0,0,0)$ as a weight of the $\mathbf{6}$ and $\bar{\mathbf{6}}$ of $\mathfrak{su}(6)$, respectively. Taking into account also the $\mathbf{3}$ and $\bar{\mathbf{3}}$ weights of $\mathfrak{su}(3)$ on the right one deduces along $\{w=a_3=0\}$  matter in the bifundamental $(\mathbf{6}, \mathbf{3} )$ (plus its conjugate). \\ 
The matter spectrum is summarized in the following table:
\begin{equation}
 \begin{array}{cc}
  \multicolumn{2}{c}{\textit{Top over polygon 16: } \mathfrak{su}(6) \times \mathfrak{su}(3)} \\ \\
 \textit{Locus} & \textit{Charged matter}\tabularnewline 
w \cap a_3 &  (\mathbf{6},\mathbf{3}) \, , \, (\bar{\mathbf{6}},\bar{\mathbf{3}}) \tabularnewline
w \cap a_1 &  (\mathbf{20},\mathbf{1})  \tabularnewline
\end{array} \,
\end{equation}
Finally we remark that the fibration is non-flat  at the codimension-three points $w=a_1 = a_3 = 0$, where one of the defining equations of the fiber components vanishes identically. 
This is precisely the intersection locus of the matter curves supporting the $ (\mathbf{6},\mathbf{3})$ and    $(\mathbf{20},\mathbf{1})$ representations.
The severe degeneration of the fibration at this locus reflects the fact no triple Yukawa coupling can be constructed out of the $\mathbf{20}$ (antisymmetric in three indices) together with the $\mathbf{6}$ and the $\bar{\mathbf{6}}$. 
Thus, in order to make sense out of F-theory compactified on the associated Calabi-Yau 4-fold the matter curves in question must not meet, which is a strong constraint on the base space ${\cal B}$. This constraint does not arise for F-theory on lower-dimensional Calabi-Yau $n$-folds.\\ \\
We are now in a position to discuss the global structure of the gauge group.
The Shioda-type map for the generator of the $\mathbb Z_3$-torsional Mordell-Weil group reads
\begin{equation}
\Sigma_3 = P - Z - \bar{\mathcal{K}} + \frac{1}{3}\left( S + 2Q + 2E_1 + 4E_2 + 3E_3 +2E_4+E_5 \right)
\end{equation}
 with $\bar{\mathcal{K}} = \pi^* \bar{\mathcal{K}}_\mathcal{B}$. Here $P = \{p=0\}$, whose intersection with the fiber is the $\mathbb Z_3$ torsion point. From the $\mathfrak{su}(6)$ top we infer that the toric divisor class $\{y=0\}$ in the ambient space is expressed as
\begin{equation}
Y = 3Z - S -
    2Q - 3P + 3\bar{\mathcal{K}} - 2E_1 - 4E_2 - 3E_3 - 2E_4 - E_5\,.
\end{equation}
We thus see that
\begin{equation}
-3 \Sigma_3 = Y
\end{equation}
and $Y$ does not intersect the hypersurface. Hence $\Sigma_3$ is trivial in $H^{1,1}(Y_4, \mathbb R)$ and
\begin{equation}
\Xi_3 \equiv P-Z-\bar{\mathcal{K}} = -\frac{1}{3}\left(S+2Q+ 2E_1 + 4E_2 + 3E_3 +2E_4+E_5 \right).
\end{equation}
% Since
% \begin{equation}
% 3(P-Z-\bar{\mathcal{K}}) = \sum_{i=1}^5 k_i E_i + k_6 S + k_7 Q \subset \langle E \rangle_{\mathbb Z} 
% \end{equation}
Again, $P-Z-\bar{\mathcal{K}}$ is a 3-torsion element of $H^{1,1}(\hat{Y}_4, \mathbb Z) / \langle F_i \rangle_{\mathbb Z} $ for $\langle F_i \rangle_{\mathbb Z}$ the lattice spanned by all the exceptional divisors. 
Furthermore, it is easy to check that $\Xi_3$ has integer intersection with all weights computed computed above. Due to the refinement of the coweight lattice the gauge group for this model is thus 
\begin{equation}
G=\frac{SU(6)\times SU(3)}{\mathbb Z_3}
\end{equation}
with $\pi_1(G) = \mathbb Z_3$. The correspondingly coarser weight lattice implies that the center $\Lambda/Q $ of the gauge group is trivial. 
%%%%%%%%%%%%%%%%%%%%%%%%%%%%%%%%%%%%%%%%%%%%%%%%%%%%%%%%%%%%%%%%%%%%%%%%%%%%%%%%%%%%%%%%%%%%%%%%%%%%%%%%%%%%%%%%%%%%%%%%%%%%%%%%%%%%%%%%%%%%%%%
%\subsubsection{$\mathfrak{su}(6)$-divisor}
%%%%%%%%%%%%%%%%%%%%%%%%%%%%%%%%%%%%%%%%%%%%%%%%%%%%%%%%%%%%%%%%%%%%%%%%%%%%%%%%%%%%%%%%%%%%%%%%%%%%%%%%%%%%%%%%%%%%%%%%%%%%%%%%%%%%%%%%%%%%%%%

%%%%%%%%%%%%%%%%%%%%%%%%%%%%%%%%%%%%%%%%%%%%%%%%%%%%%%%%%%%%%%%%%%%%%%%%%%%%%%%%%%%%%%%%%%%%%%%%%%%%%%%%%%%%%%%%%%%%%%%%%%%%%%%%%%%%%%%%%%%%%%%

\section{Conclusions}

In this work we have analyzed F-theory compactifications  on elliptic fibrations with torsional Mordell-Weil group. While non-torsional rational sections give rise to massless $U(1)$ gauge symmetries, the torsional subgroup affects the global structure of the gauge group. In general, the gauge group is of the form $G \times G'$, where $G$ is affected by the Mordell-Weil torsion and $G'$ may or may not be trivial. 
As we have argued, the presence of $\mathbb Z_k$-torsional sections guarantees the existence of a $k$-fractional linear combination of resolution divisors associated with the Cartan generators of $G$ which has integer intersection number with every fiber component. 
This fractional linear combination can be identified with an element of the coweight lattice of $G$, which is rendered finer by a factor $k$ compared to the universal cover $G_0$ of $G$. This enhances the first fundamental group of $G$ by $\mathbb Z_k$ (compared to $G_0$), yielding non-simply connected gauge groups. %If  the torsion subgroup of the Mordell-Weil group is $\mathbb Z_{k_1} \oplus \ldots  \oplus \mathbb Z_{k_n}$, the first fundamental group of the F-theory gauge group is thus $\mathbb Z_{k_1} \oplus \ldots  \oplus \mathbb Z_{k_n}$.
Consistently, the spectrum of allowed matter representations is constrained to the extent that only those elements in the weight lattice are allowed which have an integer pairing with the coweights associated with the Mordell-Weil torsion. 
An equivalent way of putting this is that the torsional subgroup  $\mathbb Z_{k_1} \oplus \ldots  \oplus \mathbb Z_{k_n}$ of the Mordell-Weil group can be identified with a subgroup of the center of the universal cover group $G_0$, and the gauge group of the F-theory compactification is $G_0/({\mathbb Z_{k_1} \oplus \ldots \oplus \mathbb Z_{k_n}}) \times G'$. \\\\
It might be worthwhile pointing out that the torsional Mordell-Weil group has no particular  effect on the structure of the Yukawa couplings between the matter states as such, which is encoded in the fiber type in codimension-three. Contrary to naive expectations, it is thus not relevant to produce \textit{e.g.}\ discrete selection rules in the effective action of an F-theory compactification. \\\\
We have exemplified this picture for elliptic fibrations with torsional Mordell-Weil group $\mathbb Z_2$ and $\mathbb Z_3$, whose defining equation had already been presented in \cite{Aspinwall:1998xj}.  
These fibrations can be analysed torically as hypersurfaces in toric ambient spaces, and, as we have seen, coincide with two out of the 16 possible hypersurface torus fibrations, whose Mordell-Weil group has been computed also in \cite{Braun:2013nqa}.
The third possible hypersurface elliptic fibration with Mordell-Weil torsion, with Modell-Weil group $\mathbb Z \oplus \mathbb Z_2$  \cite{Braun:2013nqa},  turns out to be a further specialization of the $\mathbb Z_2$-model. All these fibrations are related to a special class of elliptic fibrations with Mordell-Weil group $\mathbb Z$ \cite{Grimm:2010ez} by a chain of (un)Higgsings.\\\\
%These coincide with the most general fibrations with Mordell-Weil group $\mathbb Z_2$ and $\mathbb Z_3$ as listed in \cite{Aspinwall:1998xj} as well as a specialization of the $\mathbb Z_2$-model.
A possible next step would be to study also fibrations with Mordell-Weil group $\mathbb Z_4$ and higher. The defining Tate model for examples of such fibrations has been given in \cite{Aspinwall:1998xj}. It would be interesting to express these fibrations as complete intersections (as opposed to hypersurfaces) or even determinantal varieties and to study their properties at the same level of detail as achieved for the hypersurface models in this article. \\\\
An exciting aspect of gauge theories with non-simply connected gauge groups is the physics of non-local operators such as the spectrum of dyonic Wilson line operators.  As studied \textit{e.g.}\ in \cite{Aharony:2013hda},
the spectrum of such dyonic operators depends on the weight lattice of the gauge group $G$ and of its Langlands dual $G^*$.
As we have seen, the weight lattice $\Lambda$ of an F-theory compactification on an elliptic fibration is intimately related to the geometry of torsional sections. %These in turn imply the existence of torsion in $H^{1,1}(\hat Y,\mathbb Z)$ modulo the set of resolution divisors. 
It would be interesting to investigate further the relation between this geometric picture, the spectrum of dyonic Wilson line operators and the global structure of the gauge group in F-theory.  \\ \\ 
{\bf Acknowledgements}

\noindent We are indebted to Eran Palti for initial collaboration and for many important discussions. We also thank Ling Lin for many discussions.
This work was partially funded by DFG under Transregio 33 `The Dark Universe',
and by the US National Science Foundation under grant PHY-1307513.

\newpage
\begin{appendix}

\section{More on fiber structures}\label{app:toric_data}
%%%%%%%%%%%%%%%%%%%%%%%%%%%%%%%%%%%%%%%%%%%%%%%%%%%%%%%%%%%%%%%%%%%%%%%%%%%%%%%%%%%%%%%%%%%%%%%%%%%%%%%%%%%%%%%%%%%%%%%%%%%%%%%%%%%%%%%%%%%%%%%
\subsection{$\mathfrak{su}(4)$ top over polygon 13}\label{app:su4poly13}
%%%%%%%%%%%%%%%%%%%%%%%%%%%%%%%%%%%%%%%%%%%%%%%%%%%%%%%%%%%%%%%%%%%%%%%%%%%%%%%%%%%%%%%%%%%%%%%%%%%%%%%%%%%%%%%%%%%%%%%%%%%%%%%%%%%%%%%%%%%%%%%
Here provide the explicit equations for the fiber components of the $(SU(4) \times SU(2))/\mathbb Z_2$-model discussed in section \ref{SU4SU2Z2}.

\subsubsection{Codimension one}
%%%%%%%%%%%%%%%%%%%%%%%%%%%%%%%%%%%%%%%%%%%%%%%%%%%%%%%%%%%%%%%%%%%%%%%%%%%%%%%%%%%%%%%%%%%%%%%%%%%%%%%%%%%%%%%%%%%%%%%%%%%%%%%%%%%%%%%%%%%%%%%
The equations for the fiber components over $\{w = 0\} \subset \mathcal{B}$ are
\begin{equation}
 \begin{aligned}
  e_0=0:&\quad e_1 + a_1 t z- e_3 t^4 = 0   & (y=s=e_2=1)\,,\\ 
  e_1=0:&\quad a_1 s t y z - e_0 e_2 e_3 s t^2 z^2 a_{2,1} - e_0^2 e_3 z^4 a_{4,2} - e_2^2 e_3 s^2 t^4 = 0    \,,   &\\
  e_2=0:&\quad e_1- e_3 a_{4,2} + a_1 t = 0    & (y=s=z=e_0=1)\,,\\
  e_3=0:&\quad e_1+a_1 t = 0    & (y=s=z=1)\,.
 \end{aligned}
\end{equation}
Here we impose the SR-ideal \eqref{eq:SR-ideal_for_SU4}. The four curves $\mathbb P^1_i$ of these divisors intersect like the nodes of the affine Dynkin diagram of $A_3$.
%%%%%%%%%%%%%%%%%%%%%%%%%%%%%%%%%%%%%%%%%%%%%%%%%%%%%%%%%%%%%%%%%%%%%%%%%%%%%%%%%%%%%%%%%%%%%%%%%%%%%%%%%%%%%%%%%%%%%%%%%%%%%%%%%%%%%%%%%%%%%%%
\subsubsection{Codimension two}
%%%%%%%%%%%%%%%%%%%%%%%%%%%%%%%%%%%%%%%%%%%%%%%%%%%%%%%%%%%%%%%%%%%%%%%%%%%%%%%%%%%%%%%%%%%%%%%%%%%%%%%%%%%%%%%%%%%%%%%%%%%%%%%%%%%%%%%%%%%%%%%
Over $\{w=a_{4,2}=0\}$ we obtain:
\begin{equation}
 \begin{aligned}
  e_0=0:&\quad  a_1 t z-e_3 t^4+e_1 = 0   & (y=s=e_2=1)\,,\\ 
  e_1=0:&\quad  s \,t \underbrace{\left(e_0 e_2 e_3 t z^2 a_{2,1}-a_1 y z+e_2^2 e_3 s t^3\right)}_{\hypertarget{eq:R1}{R1}} = 0   \,,&\\
  e_2=0:&\quad  a_1 t+e_1 = 0   & (y=s=z=e_0=1)\,,\\
  e_3=0:&\quad  a_1 t+e_1 = 0   & (y=s=z=1)\,,
 \end{aligned}
\end{equation}
and over $\{w=a_1=0\}$:
\begin{equation}
 \begin{aligned}
  e_0=0:&\quad e_1-e_3 t^4 = 0    & (y=s=e_2=1)\,,\\ 
  e_1=0:&\quad e_3 \underbrace{\left(e_0 e_2 s t^2 z^2 a_{2,1}+e_0^2 z^4 a_{4,2}+e_2^2 s^2 t^4\right)}_{\hypertarget{eq:R2}{R2}} = 0    \,,&\\
  e_2=0:&\quad e_1-e_3 a_{4,2} = 0    & (y=s=z=e_0=1)\,,\\
  e_3=0:&\quad e_1 = 0    & (y=s=z=1)\,.
 \end{aligned}
\end{equation}
Before calculating the weights we analyse the parts \hyperlink{eq:R1}{$R1$} and \hyperlink{eq:R2}{$R2$} in detail. For  \hyperlink{eq:R1}{$R1$} one can check that the divisors $\{e_2=0\}$, $\{e_3=0\}$, $\{t=0\}$ and $\{z=0\}$ do not intersect the divisor given by \hyperlink{eq:R1}{$R1$} in the toric variety given by the projection along $e_1$. Therefore we can rewrite it as
\begin{equation}\label{eq:R1-reduced}
 e_0\,  a_{2,1}-y \,a_1 +  s =0
\end{equation}
with $e_0$, $y$ and $s$ the homogeneous coordinates of $\mathbb P^2$. Since \eqref{eq:R1-reduced} is a linear equation, we obtain a $\mathbb P^1$ for the curve given by $e_1=0=R1$. In the case of \hyperlink{eq:R2}{$R2$}, we find that $\{e_0=0\}$, $\{e_2=0\}$, $\{z=0\}$ and $\{t=0\}$ does not intersect the divisor \hyperlink{eq:R2}{$R2$} in the toric variety given by the projection along $e_1$. Hence, we rewrite \hyperlink{eq:R2}{$R2$} as
\begin{equation}\label{eq:R2-reduced}
s\, a_{2,1} +  a_{4,2}+ s^2=0,
\end{equation}
where $s$ is now the affine coordinate parametrising  $\mathbb C$ and the remaining homogeneous coordinates $y$ and $e_3$ parametrise a $\mathbb P^1$. Therefore, we obtain two $\mathbb P^1$s from \hyperlink{eq:R2}{$R2$} which are, however, exchanged when going along the matter curve. Around the branch  points $\{w=a_1= a_{4,2}-\frac14 a_{2,1}^2=0\}$ the solutions of $s$ to \eqref{eq:R2-reduced} are exchanged.

%%%%%%%%%%%%%%%%%%%%%%%%%%%%%%%%%%%%%%%%%%%%%%%%%%%%%%%%%%%%%%%%%%%%%%%%%%%%%%%%%%%%%%%%%%%%%%%%%%%%%%%%%%%%%%%%%%%%%%%%%%%%%%%%%%%%%%%%%%%%%%%
\subsection{$\mathfrak{su}(4)$ top over polygon 15}\label{app:su4poly15}
%%%%%%%%%%%%%%%%%%%%%%%%%%%%%%%%%%%%%%%%%%%%%%%%%%%%%%%%%%%%%%%%%%%%%%%%%%%%%%%%%%%%%%%%%%%%%%%%%%%%%%%%%%%%%%%%%%%%%%%%%%%%%%%%%%%%%%%%%%%%%%%
This appendix contains more information on the $(SU(4) \times SU(2) \times SU(2))/{\mathbb Z_2} \times U(1)$ fibration presented in section \ref{SU4SU2SU2}.
%%%%%%%%%%%%%%%%%%%%%%%%%%%%%%%%%%%%%%%%%%%%%%%%%%%%%%%%%%%%%%%%%%%%%%%%%%%%%%%%%%%%%%%%%%%%%%%%%%%%%%%%%%%%%%%%%%%%%%%%%%%%%%%%%%%%%%%%%%%%%%%
\subsubsection{Codimension one}
%%%%%%%%%%%%%%%%%%%%%%%%%%%%%%%%%%%%%%%%%%%%%%%%%%%%%%%%%%%%%%%%%%%%%%%%%%%%%%%%%%%%%%%%%%%%%%%%%%%%%%%%%%%%%%%%%%%%%%%%%%%%%%%%%%%%%%%%%%%%%%%
The irreducible fiber components over $\{\varpi=0\}$ are:
\begin{equation}
 \begin{aligned}
  e_0=0:&\quad e_2 e_3 u^2 + e_1 e_2 w^2 + \gamma_1 uwz=0  & (c = d = v = 1)\,,\\ 
  e_1=0:&\quad e_2 d v^2 + \gamma_1 dvw + \delta_2 e_0=0 & (c=u=z=e_3 = 1)  \,,   &\\
  e_2=0:&\quad \gamma_1 cuv+\gamma_2 e_0 e_1 + \delta_2 e_0 e_3 cu^2= 0    & (d=w=z=1)\,,\\
  e_3=0:&\quad e_2+\gamma_1 u+\gamma_2 e_0 = 0    & (c=d=v=w=z=e_1=1)\,. \\
 \end{aligned}
\end{equation}
The resolution $\mathbb P^1$'s is the intersection of above equations with two generic and independent divisors in the base and they intersect in the pattern of the affine $A_3$ Dynkin diagram.
%%%%%%%%%%%%%%%%%%%%%%%%%%%%%%%%%%%%%%%%%%%%%%%%%%%%%%%%%%%%%%%%%%%%%%%%%%%%%%%%%%%%%%%%%%%%%%%%%%%%%%%%%%%%%%%%%%%%%%%%%%%%%%%%%%%%%%%%%%%%%%%
\subsubsection{Codimension two}
%%%%%%%%%%%%%%%%%%%%%%%%%%%%%%%%%%%%%%%%%%%%%%%%%%%%%%%%%%%%%%%%%%%%%%%%%%%%%%%%%%%%%%%%%%%%%%%%%%%%%%%%%%%%%%%%%%%%%%%%%%%%%%%%%%%%%%%%%%%%%%%
Over $\{\varpi = \gamma_1 = 0\}$ the components of the fiber factorizes as
\begin{equation}
 \begin{aligned}
  e_0=0:&\quad e_2 (e_3 u^2 + e_1 w^2 )=0  & (c = d = v = 1)\,,\\ 
  e_1=0:&\quad e_2 d v^2 + \delta_2 e_0=0 & (c=u=z=e_3 = 1)  \,,   &\\
  e_2=0:&\quad e_0 ( \gamma_2 e_1 + \delta_2 e_3 cu^2)= 0    & (d=w=z=1)\,,\\
  e_3=0:&\quad e_2+\gamma_2 e_0 = 0    & (c=d=v=w=z=e_1=1)\, \\
 \end{aligned}
\end{equation}
and the components intersect as the affine $D_4$ Dynkin diagram.\\ \\
Over $\{\varpi = \gamma_2 = 0\}$ the components of the fiber factorizes as
\begin{equation}
 \begin{aligned}
  e_0=0:&\quad e_2 e_3 u^2 + e_1 e_2 w^2 + \gamma_1 uwz=0  & (c = d = v = 1)\,,\\ 
  e_1=0:&\quad e_2 d v^2 + \gamma_1 dvw + \delta_2 e_0=0 & (c=u=z=e_3 = 1)  \,,   &\\
  e_2=0:&\quad cu(\gamma_1 v + \delta_2 e_0 e_3 u)= 0    & (d=w=z=1)\,,\\
  e_3=0:&\quad e_2+\gamma_1 u = 0    & (c=d=v=w=z=e_1=1)\, \\
 \end{aligned}
\end{equation}
with the intersection structure given by the affine $A_5$ Dynkin diagram. \\ \\
Over $\{\varpi = \delta_2 = 0\}$ the components of the fiber factorizes as
\begin{equation}
 \begin{aligned}
  e_0=0:&\quad e_2 e_3 u^2 + e_1 e_2 w^2 + \gamma_1 uwz=0  & (c = d = v = 1)\,,\\ 
  e_1=0:&\quad dv(e_2 v + \gamma_1 w =0) & (c=u=z=e_3 = 1)  \,,   &\\
  e_2=0:&\quad \gamma_1 cuv+\gamma_2 e_0 e_1 = 0    & (d=w=z=1)\,,\\
  e_3=0:&\quad e_2+\gamma_1 u+\gamma_2 e_0 = 0    & (c=d=v=w=z=e_1=1)\, \\
 \end{aligned}
\end{equation}
intersecting as the affine $A_5$ Dynkin diagram.
%%%%%%%%%%%%%%%%%%%%%%%%%%%%%%%%%%%%%%%%%%%%%%%%%%%%%%%%%%%%%%%%%%%%%%%%%%%%%%%%%%%%%%%%%%%%%%%%%%%%%%%%%%%%%%%%%%%%%%%%%%%%%%%%%%%%%%%%%%%%%%%
\subsection{$\mathfrak{su}(6)$ top over polygon 16}\label{app:su6poly16}
%%%%%%%%%%%%%%%%%%%%%%%%%%%%%%%%%%%%%%%%%%%%%%%%%%%%%%%%%%%%%%%%%%%%%%%%%%%%%%%%%%%%%%%%%%%%%%%%%%%%%%%%%%%%%%%%%%%%%%%%%%%%%%%%%%%%%%%%%%%%%%%
The fiber structure of the $(SU(6) \times SU(3))/\mathbb Z_3$-fibration of section \ref{SU6SU3Z3} can be summarized as follows:
%%%%%%%%%%%%%%%%%%%%%%%%%%%%%%%%%%%%%%%%%%%%%%%%%%%%%%%%%%%%%%%%%%%%%%%%%%%%%%%%%%%%%%%%%%%%%%%%%%%%%%%%%%%%%%%%%%%%%%%%%%%%%%%%%%%%%%%%%%%%%%%
\subsubsection{Codimension one}
%%%%%%%%%%%%%%%%%%%%%%%%%%%%%%%%%%%%%%%%%%%%%%%%%%%%%%%%%%%%%%%%%%%%%%%%%%%%%%%%%%%%%%%%%%%%%%%%%%%%%%%%%%%%%%%%%%%%%%%%%%%%%%%%%%%%%%%%%%%%%%%
The irreducible fiber components over $\{w=0\}$ take the form 
\begin{equation}
 \begin{aligned}
  e_0=0:&\quad e_1e_3 p^3 + e_3 e_5 x^3 + a_1 pxz =0  & (s = q = e_2 = e_4 = 1)\,,\\ 
  e_1=0:&\quad e_3 + a_1 p =0 & (x=s = q = z=e_4 = e_5 = 1)  \,,   &\\
  e_2=0:&\quad a_3 e_1+ a_1 pqsx + e_3 q s^2 x^3= 0    & (z=e_0=e_4=e_5=1)\,,\\
  e_3=0:&\quad a_3 e_0^2 e_1 e_5 + a_1 x = 0    & (s=q=p=z=1)\, \\
  e_4=0:&\quad e_3 + a_3 e_5 + a_1 x = 0    & (s=q=p=z=e_0 = e_1 = e_2 = 1)\, \\
  e_5=0:&\quad e_3 + a_1 x = 0    & (s=q=p=z=e_1=e_2=1)\, .
 \end{aligned}
\end{equation}
The resolution $\mathbb P^1$'s is the intersection of above equations with two generic and independent divisors in the base and they intersect in the pattern of the affine $A_5$ Dynkin diagram.
%%%%%%%%%%%%%%%%%%%%%%%%%%%%%%%%%%%%%%%%%%%%%%%%%%%%%%%%%%%%%%%%%%%%%%%%%%%%%%%%%%%%%%%%%%%%%%%%%%%%%%%%%%%%%%%%%%%%%%%%%%%%%%%%%%%%%%%%%%%%%%%
\subsubsection{Codimension two}
%%%%%%%%%%%%%%%%%%%%%%%%%%%%%%%%%%%%%%%%%%%%%%%%%%%%%%%%%%%%%%%%%%%%%%%%%%%%%%%%%%%%%%%%%%%%%%%%%%%%%%%%%%%%%%%%%%%%%%%%%%%%%%%%%%%%%%%%%%%%%%%
Over $\{w = a_1 = 0\}$ the components of the fiber takes the form
\begin{equation}
 \begin{aligned}
  e_0=0:&\quad e_3 (e_1 p^3 + e_5 x^3 ) =0  & (s = q = e_2 = e_4 = 1)\,,\\ 
  e_1=0:&\quad e_3  =0 & (x=y=s = q = z=e_4 = e_5 = 1)  \,,   &\\
  e_2=0:&\quad a_3 e_1 + e_3 q s^2 x^3= 0    & (y=z=e_0=e_4=e_5=1)\,,\\
  e_3=0:&\quad a_3 e_0^2 e_1 e_5  = 0    & (y=s=q=p=z=1)\, \\
  e_4=0:&\quad e_3 + a_3 e_5 = 0    & (y=s=q=p=z=e_0 = e_1 = e_2 = 1)\, \\
  e_5=0:&\quad e_3  = 0    & (y=s=q=p=z=e_1=e_2=1)\, \\
 \end{aligned}
\end{equation}
resulting in 6 distinct $\mathbb P^1$'s, intersecting as the $E_6$ Dynkin diagram ({\it not affine}).\\ \\
Over $\{w = a_3 = 0\}$ the components of the fiber takes the form
\begin{equation}
 \begin{aligned}
  e_0=0:&\quad e_1e_3 p^3 + e_3 e_5 x^3 + a_1 pxz =0  & (s = q = e_2 = e_4 = 1)\,,\\ 
  e_1=0:&\quad e_3 + a_1 p =0 & (x=y=s = q = z=e_4 = e_5 = 1)  \,,   &\\
  e_2=0:&\quad qsx(a_1 p + e_3 sx^2)= 0    & (y=z=e_0=e_4=e_5=1)\,,\\
  e_3=0:&\quad  a_1 x = 0    & (y=s=q=p=z=1)\, \\
  e_4=0:&\quad e_3 + a_1 x = 0    & (y=s=q=p=z=e_0 = e_1 = e_2 = 1)\, \\
  e_5=0:&\quad e_3 + a_1 x = 0    & (y=s=q=p=z=e_1=e_2=1)\, \\
 \end{aligned}
\end{equation}
resulting in 9 distinct $\mathbb P^1$'s, intersecting as the affine $A_8$ Dynkin diagram.

\end{appendix}

\newpage
\bibliography{papers}  
\bibliographystyle{custom1}

\end{document}